\documentclass[twocolumn,rmp,amsmath,amssymb,showpacs]{revtex4-1}
\usepackage{graphicx,color}
\usepackage{amsmath,amssymb,latexsym,amsthm}
\usepackage{pgfplots}
\usepackage{mathtools}
\usepackage{mhchem}
\usepackage[LGR,T1]{fontenc}
\usepackage[cal=boondoxo]{mathalfa}
\usepackage{setspace}
\usepackage{comment}

\DeclareSymbolFont{ttgreek}{LGR}{cmtt}{m}{n}
\DeclareMathSymbol{\ttsigma}{\mathord}{ttgreek}{`s}

\usepackage{tikz}
\usepgfplotslibrary{fillbetween}

\usetikzlibrary{chains,shapes,fit,calc,arrows}
\usepackage{pgfplots, pgfplotstable}
\usepgfplotslibrary{groupplots}
\usepackage{color}
\newcommand{\mean}[1]{{\left< #1 \right>}}

\usepackage{hyperref}
\hypersetup{%
    colorlinks=true, linktocpage=true, pdfstartpage=1, pdfstartview=FitV,%
    breaklinks=true, pdfpagemode=UseNone, pageanchor=true, pdfpagemode=UseOutlines,%
    plainpages=false, bookmarksnumbered, bookmarksopen=true, bookmarksopenlevel=1,%
    hypertexnames=true, pdfhighlight=/O,%nesting=true,%frenchlinks,%
    urlcolor=brown, linkcolor=red, citecolor=blue, %pagecolor=RoyalBlue,%
    pdftitle={macroscopic ST},%
    pdfauthor={Gianmaria Falasco, Massimiliano Esposito},%
    pdfsubject={},%
    pdfkeywords={},%
    pdfcreator={pdfLaTeX},%
    pdfproducer={LaTeX REVTeX}%
}

\begin{document}

\title{Macroscopic stochastic thermodynamics}
\author{Gianmaria Falasco}
\email{gianmaria.falasco@unipd.it}
\affiliation{Department of Physics and Astronomy, University of Padova, Via Marzolo 8, I-35131 Padova, Italy.}
\author{Massimiliano Esposito}
\email{massimiliano.esposito@uni.lu}
\affiliation{Complex Systems and Statistical Mechanics, Department of Physics and Materials Science, University of Luxembourg, L-1511 Luxembourg, Luxembourg}
%\date{\today}
\begin{abstract}
%We derive the macroscopic limit  of stochastic thermodynamics assuming a Markov jump process as the fundamental mesoscopic description. First, we identify the asymptotic scaling of thermodynamic and kinetic variables to ensure an extensive thermodynamic limit and a deterministic behavior at the macroscopic level. 
%We identify classes of systems (e.g. chemical reaction networks, electronic circuits) whose large deviations properties cannot be described by the Fokker-Planck approximation.
%We then derive the thermodynamic consistent equations describing asymptotic fluctuations, i.e. which preserve the fluctuation theorem. The first and second law of thermodynamics are given, both on average and along fluctuating trajectories. Importantly, the macroscopic entropy of the systems is identified with the average internal entropy of mesoscopic states, while the Shannon entropy associated to the probability distribution over states is proven to asymptotically nullify. Finally, we show how extensivity leads to the emergence of Gibbs-Duhem relations at equilibrium, and how away from equilibrium the free energy is replaced by the dynamically generated quasi-potential as a Lyapunov function for the macroscopic dynamics.
%The latter forms the basis for coarse-graining the macroscopic stochastic dynamics into a Markov jump process on deterministic attractors. We formulate the stochastic thermodynamics that emerges at this new level: currents do not respect detailed balance and thus do not carry all the information about dissipation, but close to equilibrium. 

Starting at the mesoscopic level with a general formulation of stochastic thermodynamics in terms of Markov jump processes, we identify the scaling conditions that ensure 
the emergence of a (typically nonlinear) deterministic dynamics and an extensive thermodynamics at the macroscopic level. 
We then use large deviations theory to build a macroscopic fluctuation theory around this deterministic behavior, which we show preserves the fluctuation theorem. 
For many systems (e.g. chemical reaction networks, electronic circuits, Potts models), this theory does not coincide with Langevin-equation approaches (obtained by adding Gaussian white noise to the deterministic dynamics) which, if used, are thermodynamically inconsistent.  
Einstein-Onsager theory of Gaussian fluctuations and irreversible thermodynamics
are recovered at equilibrium and close to it, respectively. 
Far from equilibirum, the free energy is replaced by the dynamically generated quasi-potential (or self-information) which is a Lyapunov function for the macroscopic dynamics.
%In large but still finite systems the dynamics can be pictured as successions of rare fluctuations between attractors and subsenquent determinitic relaxation within them. 
%Thermodynamics provides remarkable connections between this quasi-potential and the dissipation along deterministic trajectories as well as with the rare fluctuations inducing transitions between attractors.
Remarkably, thermodynamics connects the dissipation along deterministic and escape trajectories to the Freidlin-Wentzell quasi-potential, thus constraining the transition rates between attractors induced by rare fluctuations.
A coherent perspective on minimum and maximum entropy production principles is also provided. For systems that admit a continuous-space limit, we derive a nonequilibrium fluctuating field theory with its associated thermodynamics.
Finally, we coarse grain the macroscopic stochastic dynamics into a Markov jump process describing transitions among deterministic attractors and formulate the stochastic thermodynamics emerging from it. 
%Except close to equilibrium, currents do not respect detailed balance and thus do not carry the full dissipation in the system. 

\end{abstract}

\maketitle

%\tableofcontents
\begin{spacing}{0.9}
\tableofcontents
\end{spacing}

\section{Introduction}

\subsection{The challenge}

\emph{Thermodynamics} is a theory describing energy transfers among systems and energy transformations from one form into another.  
It constitutes a remarkable scientific achievement which started more than three centuries ago with the study of vacuum pumps. 
During the 19th century its developments were spectacular and instrumental to master the technology of steam engines that literally powered the industrial revolution.
In its traditional formulation thermodynamics applies to macroscopic systems at thermodynamic equilibrium. 
The microscopic foundation of the theory, \emph{statistical mechanics}, was established during the late 19th and early 20th century. 
It explains how the observed macroscopic behavior results from very large numbers of microscopic degrees of freedom giving rise to sharply peaked probability distributions.  
The 20th century witnessed the first development of \textit{irreversible thermodynamics} to explain transport phenomena, which is based on the assumption that the bulk properties of macroscopic systems remain close to thermodynamic equilibrium \cite{Prigogine1961Jul,Groot1984}. 

Since the end of the 20th century, a statistical formulation of nonequilibrium thermodynamics, called \textit{stochastic thermodynamics} (ST), was established to describe small fluctuating systems \cite{Peliti2021,Seifert2012,VandenBroeck2015Jan,Gaspard2022,shiraishi2023introduction}. 
In its general formulation \cite{Rao2018a}, ST constructs thermodynamic observables (e.g. heat, work, dissipation) on top of the stochastic dynamics of open systems. The key assumption is that all the degrees of freedom which are not explicitly described by the dynamics -- i.e. the internal structure of the system states and the reservoirs -- must be at equilibrium. The link between observables and dynamics is then provided by the \textit{local detailed balance} property \cite{Lebowitz55new,Esposito2012,Bauer2014,Maes2021,falasco21local}: when an elementary process (e.g. an elementary chemical reaction) induces a transition between two states (e.g. the number of molecules before and after the reaction), the Boltzmann constant $k_B$ times the log ratio of the forward and backward current of that process between the two states is the entropy production (or dissipation) of the process, i.e. the sum of the entropy changes in the environment and in the system. 
The successes of ST over recent years have been impressive both theoretically and experimentally.
Since the theory predicts the temporal changes of thermodynamic quantities, it can be used to address typical \textit{finite-time thermodynamics} questions such as identifying optimal driving protocols leading to maximum power extraction \cite{Seifert2012,Benenti2017Jun}.   
{Driven colloidal particles}, molecular motors, Brownian ratchets and electrical circuits have for instance been considered \cite{Blickle2006time, Mehl2012hidden,Brown2020,Pekola2015Feb,Ciliberto2017Jun}.
ST also predicts the nonequilibrium fluctuations of thermodynamic observables. 
The central discovery of the field is the \textit{fluctuation theorem} asserting that the probability to observe a given positive dissipation is exponentially larger than the probability of its negative counterpart ~\cite{Seifert2012,Rao2018Sep,Gaspard2022,Jarzynski2011Feb}. 
%~\cite{jar97b,cro98,cro99,kur98,mae99,leb99,maes03,esp10,Seifert2012,sag12,lee13,rah14,Rao2018Sep}.
This relation generalizes the second law, which only states that on average the dissipation cannot decrease. 
Previous results such as Onsager reciprocity relations or fluctuation-dissipation relations can be recovered from it.
%close to equilibrium
ST also provides very natural connections with information theory \cite{Wolpert2019Apr}: The system entropy contains a Shannon entropy contribution; dissipation can be expressed as a Kullback-Leibler entropy quantifying how different probabilities of forward trajectories are from their time-reversed counterpart; the difference between the nonequilibrium and equilibrium thermodynamic potential takes the form of a Kullback-Leibler entropy between the system nonequilibrium and equilibrium probability distributions. 
These results provide a rigorous ground to assess, both theoretically \cite{Horowitz14,Parrondo2015Feb} and experimentally \cite{toyabe2010, Berut2012Mar, Jun2014Nov, Koski2015Dec}, the cost of various information processing operations such as Landauer erasure or Maxwell's demons.
More recently, ST has also been used to show that dissipation sets universal bounds -- called \textit{thermodynamic uncertainty relations} and \textit{speed limits} -- on the precision \cite{Barato2015,Horowitz2020,falasco2019unifying} as well as on the duration \cite{Shiraishi2018,Vo2020speed,falasco2020dissipationtime,VanVu2023optimal} of nonequilibrium processes. 

Despite these remarkable achievements, the focus of ST has been the study of small and simple systems far-from-equilibrium, and important questions remain open when considering larger and more complex systems.
In particular, (i) What happens in the thermodynamic limit of ST?  (ii) How should ST be modified when detailed balance is broken? These questions remain poorly explored except for specific model-systems studies.

(i) More specifically, what happens when considering systems whose number of degrees of freedom becomes very large? Can one derive from ST a macroscopic thermodynamics valid far away from equilibrium and connect it close-to-equilibrium to irreversible thermodynamics? These general questions remain largely unanswered. This is surprising given that a motivation for the early developments \cite{hat01,har05,sek10} that led to the present form of ST was the derivation of a nonequilibrium version of traditional thermodynamics \cite{keizer1978thermodynamics,oon98}.
In recent years, only specific classes of systems that exhibit a macroscopic limit have been considered, such as chemical reaction networks \cite{Rao2016,Falasco2018a,Falasco2019a,Avanzini2021a}, electronic circuits \cite{Freitas2020Jul,Freitas2021ec}, and mean-field Ising and Potts models \cite{herpich2018collective,Herpich2019,Herpich2020Jun}.
These results suggest that for these macroscopic systems, one can derive not only a deterministic formulation of nonequilibrium thermodynamics, but also -- with the help of large deviations theory \cite{Touchette2009} -- a theory of macroscopic fluctuations around the deterministic behavior, similar in spirit to the one valid for diffusive systems \cite{Bertini2015Jun}. Importantly, the macroscopic fluctuation theories derived for most of the model systems studied are incompatible with a naive approach consisting of adding Gaussian white noise to the deterministic dynamics. A general theory is needed to clarify these crucial questions.     

(ii) Most dynamics used to describe complex phenomena are highly coarse-grained and do not resolve all the out-of-equilibrium degrees of freedom, thus compromising the detailed balance assumption essential to build a consistent ST. This situation is ubiquitous in active and biological systems, for instance \cite{Marchetti2013Jul, Bechinger2016Nov}. While the dynamics of some motile cells can be modeled by active Brownian dynamics \cite{fodor16howfar}, their energetic cost cannot be deduced from the motion of the cells alone as a large number of hidden sub-cellular processes are also dissipating energy. The practical relevance of a thermodynamics of complex systems is huge. It is necessary to address question such as: To what extend are biology and ecology shaped by the energy flows sustaining them \cite{Garvey2016,Yang2021}? What are energy efficient design principles for information, computation and communication technologies \cite{Auffeves2022,Wolpert2019,lange2020digitalization}?

%It is important to find ways to overcome these limitations.
%Another class of problem results from the fact that most dynamics used to describe complex phenomena are highly coarse-grained and do not resolve all the out-of-equilibrium degrees of freedom, thus compromising the  detailed balance assumption essential to build a consistent ST. This situation is for instance ubiquitous in active and biological systems \cite{Marchetti2013Jul, Bechinger2016Nov}. While the dynamics of some motile cells can be modeled by active Brownian dynamics, their energetic cost cannot be deduced from the motion of the cells alone as a huge number of hidden sub-cellular processes are also burning energy. Given the importance of developing a thermodynamics of complex systems -- e.g. in biology (how does energy conservation constrain biological evolution?), in ecology (how are we disrupting biogeochemical cycles?), in information and communication technologies (how do we keep their energetic cost under control?) -- it is crucial to find ways to overcome these limitations.
A crucial difference between these complex systems and those typically considered in irreversible thermodynamics is that the nonequilibrium constrains or thermodynamic driving forces are not only imposed at the boundaries of the system but throughout the system itself. In biology, for instance, the energy input is chemical (e.g. from ATP hydrolysis) and arises at the molecular scale. This fact a priori jeopardizes notions of  equilibrium used to build traditional macroscopic fluctuation theories \cite{dezarate06,Bertini2015Jun}.
Despite much ongoing work on coarse-graining stochastic dynamics \cite{rah07,pug10,Esposito2012,bo14,esp15,pol17,Bo2017,str19,Herpich2020,maes20}, beside few exceptions in simple models \cite{pietzonka2017entropy,speck2022efficiency,bebon2024thermodynamics} and in the linear regime \cite{GaspardKapral2020}, little is known about how to establish a thermodynamically consistent theory of active systems, in particular at a macroscopic level where nonequilibrium field theoretical descriptions (formulated in physical space) would be very useful.

\subsection{What we achieve}

Starting from a general formulation of ST, we identify the scaling conditions ensuring that a \emph{deterministic} (typically nonlinear) dynamics and a corresponding \emph{extensive} thermodynamics emerge in the macroscopic limit. We find that the same conditions ensure the validity of a thermodynamically consistent macroscopic fluctuation theory -- based on large deviations theory -- describing fluctuations around the deterministic behavior. 
%In general, this theory is shown to be incompatible with approaches based on adding Gaussian white noise to the macroscopic dynamics.   
The importance of these results is manifold.
First, ST is proved to be compatible with macroscopic thermodynamics and equilibrium statistical mechanics, thus resolving some controversy concerning the definition of entropy on the basis of information theory \cite{Goldstein2019entropy}. 
%The macroscopic entropy of the system is identified with the average internal entropy of its mesoscopic states, while the Shannon entropy associated to the probability distribution over states is proven to asymptotically nullify. 
Second, irreversible thermodynamics and the Einstein-Onsager theory of small fluctuations \cite{landau1959fluid,rytov1958correlation,rytov1989priniciples,de2006hydrodynamic,henkel2017nanoscale} are derived from our unified framework for close-to-equilibrium conditions, thus providing a more microscopic foundation to these two originally phenomenological theories. 
Third, we explain how thermodynamic quantities shape the far-from-equilibrium behavior, both the deterministic relaxation and the exponentially (in the system size) rare fluctuations. By doing so, we manage to connect  Freidlin–Wentzell theory of large deviations \cite{freidlin,graham1987macroscopic} to thermodynamics.
This allows us to retrieve the minimum entropy production principle close to equilibrium, relate the life-time of nonequilibrium metastable states to dissipation, and rigorously discuss a maximum entropy production principle. 
Fourth, we identify the pitfalls of commonly used approximations (e.g. nonlinear Langevin equations with multiplicative noise \cite{gillespie2000chemical}) and offer as a result a systematic framework to derive thermodynamic consistent fluctuating field theories, e.g. for active matter.
Finally, we show that ST allows one to perform a physically-motivated coarse graining tailored to reveal emergent states and transitions between them in complex nonequilibrium systems.

\subsection{Plan of the paper}

In Sec. \ref{sec:me} we introduce the dynamical description of the systems, based on the classical master equation for occupation-like variables, e.g. the number of particles in the physical space, of electrons on a conductor, of spins with a certain orientation, or of excitations in some mode space.
At such level of description, which we called mesoscopic, all unresolved degrees of freedom (internal and environmental) are assumed to be thermalized. Detailed balance thus ensues, allowing us to introduce a complete thermodynamic superstructure, given by standard ST, which we briefly recapitulate.

\begin{figure}
\center
\includegraphics[width=0.5\textwidth]{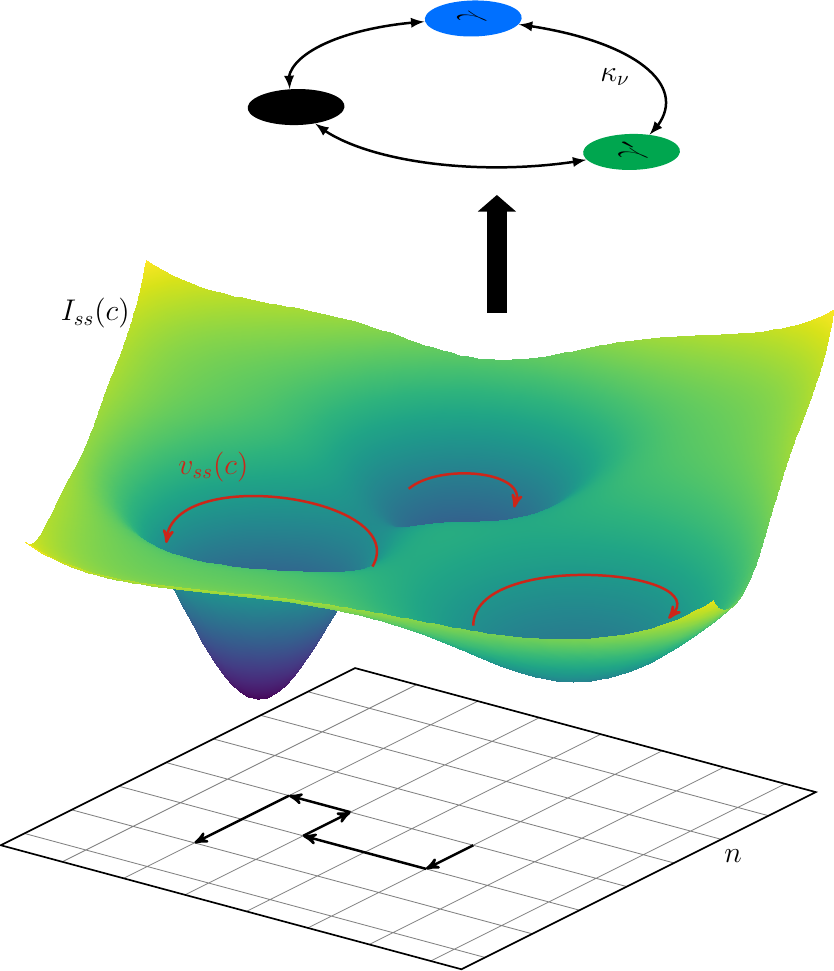}
\caption{Sketch of the different levels of description emerging with growing size $V$ and time scale $\tau$. The mesoscopic level (bottom) consists of a Markov jump process in discrete space $n$ that enjoys local detailed balance. For large $V$ the dynamics of the intensive variable $c=n/V$ is dictated by the quasi-potential $I_\text{ss}$ and the nonconservative drift $v_\text{ss}$ (middle). For large $\tau$ the dynamics reduces again to a Markov jump process on the space of the attractors, regarded as emergent states with internal dissipation (top). \label{fig:scales}}
\end{figure}

We then introduce the macroscopic limit in Sec. \ref{sec:large_scale}, akin to a nonequilibrium thermodynamic limit. We derive the conditions on the transition rate and thermodynamic forces of the mesoscopic dynamics that ensure asymptotically a deterministic dynamics (Sec. \ref{sec:deterministic_dyn}) and an extensive thermodynamics (Sec. \ref{sec:deterministic_thermo}). 
Large deviations theory, the appropriate language to describe the concentration of  probability in the macroscopic limit, is utilized. 
Thanks to it we explain that the macroscopic entropy of a system is independent of the probability distribution at the mesoscopic level and only made up of the internal entropy of the mesoscopic states. We link the phenomenon of dissipative metastability in the stochastic dynamics with the ergodicity breaking and multistability of the macroscopic deterministic dynamics (Sec. \ref{sec:multi_meta}).
In Sec. \ref{subsec:orthogonal} we consider the macroscopic nonlinear dynamics, identifying the weak-noise quasi-potential as the Lyapunov function and connecting it to thermodynamics. We split the deterministic relaxation in its gradient part and the nonequilibrium circulation due to state-space probability currents (see Fig. \ref{fig:scales}). 
In Sec. \ref{sec:macro_linear_dyn} we examine the linear deterministic relaxation close to nonequilibrium fixed points. Close to equilibrium we retrieve linear irreversible thermodynamics and the minimum entropy production principle.

Next, we obtain in Sec. \ref{sec:hamiltonian_dyn}  the macroscopic fluctuations from the extremum action principle. In Sec. \ref{sec:macro_thermo} we formulate the thermodynamics along macroscopic fluctuating trajectories, deriving an emergent second law and connecting the quasi-potential to dissipation close to equilibrium.
We focus on the asymptotic form of both small Gaussian fluctuations and rare large ones. The former are described by linear Langevin equations that reduce to the Onsager-Machlup theory at equilibrium (Sec. \ref{sec:Onsager_Machlup}). We highlight the failure in capturing the rare events (Sec. \ref{sec:KM_truncation}) and the thermodynamics (Sec. \ref{sec:path_langevin}) of the unsystematic truncation of the mesoscopic master equation that gives rise to a nonlinear Langevin equation with multiplicative noise.
In Sec. \ref{sec:large_dev} we quantify the time-reversal asymmetry between relaxation path within an attractor and the escape path out of it (called instanton). Also, the likelihood of rare large fluctuations, encoded into the quasi-potential, allows us to quantify the transition rate out of a basin of attraction. For them we derive in Sec. \ref{sec:bounds_kappa} upper and lower bounds in terms of the entropy produced in relaxation and escape trajectories -- which constitute a novel maximum entropy production principle under certain conditions.
%%%%
%%%
In Sec. \ref{sec:macro_ft} we discuss the asymptotic full counting statistics of thermodynamic currents and the fluctuation theorems showing how the multiplicative Gaussian noise approximations break them.

In Sec. \ref{subsec:continuous} we consider a continuous-space limit, making contact with macroscopic fluctuation theory. We explain how fluxes become linear functions of the thermodynamic forces and the notion of local equilibrium appears.
The associated thermodynamics is discussed in Sec. \ref{sec:entropy_continuous}, showing that the continuous-space limit of the mesoscopic entropy production differs from the apparent one defined in configuration space only, unless all processes are diffusive.

In Sec. \ref{sec:emergent} we formulate ST for the long-time jump dynamics between attractors. This provides a natural way to coarse-grain the original system into emergent states sustained by an internal dissipation. We discuss the conditions of the underlying mesoscopic dynamics required to obtain the standard form of ST for this emergent jump process (see Fig. \ref{fig:scales}).%, i.e. null internal dissipation of states and  detailed balance for transitions.

In Sec. \ref{sec:applications} we provide examples of the general theory, including chemical reaction networks, electronic circuits and driven Potts models. In Sec. \ref{sec:discussion} we conclude by discussing the strength of our framework, its limitations, and open issues.

\section{Stochastic Thermodynamics}\label{sec:me}

\begin{figure}
\includegraphics[width=0.5\textwidth]{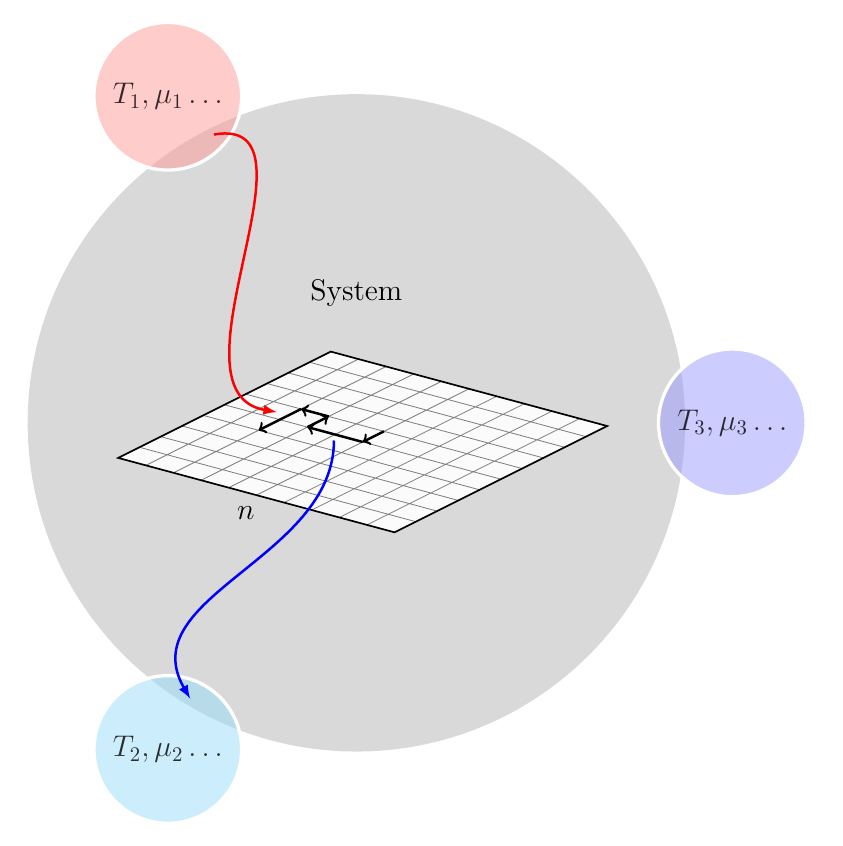}
\caption{Schematic representation of a system coupled to three reservoirs with fixed temperature $T$, (electro)chemical potential $\mu$, etc. Each of them can exchange energy, (charged) particles, etc. through various transition mechanisms which result in jumps of the system state $n$. \label{fig:mesosystem}}
\end{figure}

We consider a small system in contact with several reservoirs, each characterized by prescribed intensive thermodynamic variables (temperature, chemical potential, electric voltage, etc.). We assume that the system's microscopic degrees of freedom are equilibrated and can be grouped into $i=\{1, \dots, N \}$ mesoscopic states, with occupation number $n_i$ evolving under a stochastic dynamics (see Fig. \ref{fig:mesosystem}).
Namely, the occupation number vector $n=\{n_i\}_{i=1}^N$ changes at random times $t_k$ by a finite vector $\Delta_{\rho}=\{\Delta_{\rho}^i\}_{i=1}^N$ due to transitions of type $\rho$ induced by the coupling with the reservoirs: 
\begin{align}\label{eq_ndot}
d_t n(t) = \sum_\rho \Delta_\rho J_\rho(t)=\sum_{\rho >0} \Delta_\rho I_\rho(t).
\end{align}
Here, $J_\rho(t)=\sum_k \delta_{\rho \rho(t)} \delta(t-t_k)$ is the rate of transitions $\rho$ occurring at time $t$ (see, e.g. \cite{Rao2018b}) and $I_\rho(t)=J_\rho(t)-J_{-\rho}(t)$ is the net current. We have used microscopic reversibility, which imposes that for each transition $+\rho$ there exists an inverse transition denoted $-\rho$ and satisfying $\Delta_\rho=-\Delta_{-\rho}$.
The process \eqref{eq_ndot} is taken to be Markovian by choosing for  $J_\rho(t)$ a Poissonian distribution \footnote{Or, equivalently, an exponential distribution for the waiting times $t_{k+1}-t_{k}$.} conditioned on the present state $n(t)$ with average value $R_\rho(n(t))$ \footnote{We chose a compact notation so that the argument of $R_\rho(n)$ (and $\Gamma_{\rho}(n)$, $\Sigma_\rho(n)$, etc.) denotes the starting state of the transition $\rho$ but the function can depend as well on the final state $n +\Delta_\rho$. }.
%, namely,
%\begin{align}
%\mean{J_\rho(t)}_{n(t)=n}=R_\rho(n).
%\end{align}
%Hereafter $\mean{\dots}$ denotes the expectation value with respect to the noise statistics.
We might also consider systems driven by reservoirs whose intensive thermodynamic properties are externally changed in time. This corresponds to nonautonomous dynamics with rates $R_\rho(n(t), \vartheta(t) )$ that carry an explicit dependence on time via a protocol $\vartheta(t)$ -- although we will not write this dependence to avoid clutter. 
For concreteness, we can think of $n_i$ as the number of molecules of chemical species $i$, or the number of charges of conductors $i$, or the number of excitations of modes $i$ (see \ref{sec:applications}).

A trajectory $\mathcal{X}$ is the ordered
set of states $n(t)$ and jumps $\rho(t_k)$ in the time span $ t \in [0,\tau]$. The entire dynamical description of the system, equivalent to \eqref{eq_ndot}, is given by the probability density of a trajectory \cite{sun06,mae08b}
\begin{align}\label{prob_traj}
P[\mathcal{X}] &= P(n(0),0) e^{-\sum_\rho \int_0^\tau dt R_\rho(n(t))}   \prod_k R_{\rho(t_k)}(n(t_k^-))  .
\end{align}
Here appear in order, the probability density of the initial state $n(0)$, the product of probabilities for remaining in a certain state and the product of probability densities for changing state.

The dynamics in the configuration space corresponding to the mesoscopic description \eqref{eq_ndot}, or \eqref{prob_traj}, is given by the master equation for the probability density of occupation numbers,
\begin{align}
%\begin{aligned}
 \partial_t P(n,t) &= \sum_\rho[ R_\rho(n - \Delta_\rho) P(n- \Delta_\rho,t) -R_\rho(n) P(n,t) ] \nonumber \\
&= \sum_{n'} \mathcal{R}_{n n'} P(n',t), \label{me}
%\end{aligned}
\end{align}
with $\mathcal{R}_{n n'}=\sum_\rho R_\rho(n')(\delta_{n' , n-\Delta_\rho}-\delta_{n' , n})$ the transition rate matrix, i.e. the generator of the Markovian dynamics.
We assume that the Markov jump process has a connected graph, thus is ergodic. 
For transition rates that carry an explicit time dependence, the instantaneous stationary probability density is defined by
\begin{align}\label{stationary_p}
\sum_{n'} \mathcal{R}_{n n'}  P^{\text{ss}}_t(n') =0.
\end{align}
For transition rates that do not have any explicit time dependence, the solution of \eqref{stationary_p} is the unique stationary probability density function $P^{\text{ss}}(n) =\lim_{t \to \infty}P(n,t)$ reached by the system at long times.

The transition rates satisfy the condition of local detailed balance \cite{Lebowitz55new,Esposito2012,Bauer2014,Maes2021,falasco21local},
\begin{align}\label{micro_ldb}
&\Sigma_\rho(n):=\ln \frac{R_\rho(n)}{R_{-\rho}(n+\Delta_\rho)} =-\Phi(n+\Delta_\rho)+\Phi(n)+ a_\rho,
\end{align}
which states that the logarithm of the ratio between forward and backward rates of a given transition $\rho$ is the entropy production $\Sigma_\rho$ of such process -- the Boltzmann constant $k_B$ is set to 1 hereafter. Physically, \eqref{micro_ldb} holds when the transitions $\pm \rho$ are caused by reservoirs in thermodynamic equilibrium and all microscopic (internal) degrees of freedom of a mesostate $n$ are as well equilibrated.
In \eqref{micro_ldb}, $\Phi(n)$ is the negative of the Massieu potential, also called free entropy \cite{callen1991thermodynamics}, 
defined by subtracting the internal entropy $S_\text{int}(n)$ of the mesostate $n$ from the energetic contribution of conserved quantities exchanged with the baths \footnote{In case of nonautonomous dynamics $\Phi(n,t)$ carries an explicit time dependence due to $\vartheta(t)$ that we avoid to write, though.}. 
For example, in the particular case of a system exchanging only energy with a single thermal bath at temperature $T$, 
\begin{align}\label{helmoltz}
T \Phi(n) = U(n)-TS_\text{int}(n)
\end{align}
corresponds to the Helmholtz free energy, with $U(n)$ the internal energy of state $n$. 
The quantity 
\begin{align}\label{affinitydef}
a_\rho= \mathbb{X}_\rho \cdot  \mathcal{f}_{\text{nc}}=-a_{-\rho}
\end{align}
in \eqref{micro_ldb} denotes the nonconservative entropy flow  in the transition $\rho$. It is the projection on $\rho$ of the fundamental nonconservative forces $\mathcal{f}_{\text{nc}}$, i.e. the minimal set of independent forces generated by the reservoirs, 
with $\mathbb{X}_\rho$ the variation in the reservoirs of the associated physical quantity due to the transition $\rho$. 
For these reasons, $a_\rho$ is expressed solely in terms of differences of intensive quantities of the reservoirs (e.g. inverse temperatures, chemical and electrical potentials), i.e. it is independent of the state of the system $n$.
Such decomposition into Massieu potential and nonconservative forces can always be performed as outlined in \cite{Rao2018a}.
We note that $
\Sigma_\rho(n) = - \Sigma_{-\rho}(n+\Delta_{\rho})$.

The total entropy production rate \cite{Seifert2005Jul} along a fluctuating trajectory solution of \eqref{eq_ndot},
\begin{align}\label{epr}
\dot \Sigma(t)= \dot \Sigma_e(t) + d_t S_\text{sys}(n(t),t),
\end{align}
 is the sum of the entropy flow in the environment,
\begin{align}\label{ent_flow}
\dot \Sigma_e (t) := \sum_\rho J_\rho(t) \Sigma_\rho(n(t))-d_t S_\text{int}(n(t)),
\end{align}
plus the time derivative of the system entropy,
\begin{align}\label{sys_ent_meso}
S_\text{sys}(n,t)& := - \ln P(n,t) + S_\text{int}(n).
\end{align}
The first term on the righthand side of \eqref{sys_ent_meso} is the system's self-information or surprisal, whose mean value  is the Shannon entropy 
\begin{align}\label{Shannon_entropy}
S_\text{sh}(t):=-\sum_n P(n,t) \ln P(n,t).
\end{align}

The stochastic entropy production, i.e. the time integral of \eqref{epr}, can be obtained comparing forward and backward trajectory probabilities
\begin{align}
\Sigma[\mathcal{X}] = \int_0^\tau dt \, \dot \Sigma(t)= \ln \frac{P[ \mathcal{X}]}{\overline{P}[\mathcal{X}]},
\end{align}
where $\overline{P}[\mathcal{X}]$ is the probability of the time-reversed trajectory obtained from \eqref{prob_traj} by the change of variable $t_k \mapsto \tau - t_k$ and reversing the protocol $\vartheta(t) \mapsto \vartheta(\tau - t)$.
Therefore, the mean entropy production 
is the Kullback-Leibler divergence
\begin{align}
\mean{\Sigma[\mathcal{X}]} = \int \mathcal{D}\mathcal{X} P[\mathcal{X}] \ln \frac{P[ \mathcal{X}]}{\overline{P}[\mathcal{X}]} \geq 0.
\end{align}
Hereafter, the symbol $\mean{\dots}$ denotes the expectation value of an observable, computed with the appropriate probability, namely, $P[\mathcal{X}]$ for trajectory functionals or $P(n,t)$ for functions of the state at a single time $t$.  

The mean entropy production rate \cite{sch76} follows from averaging \eqref{epr},
\eqref{ent_flow} and \eqref{sys_ent_meso}, 
%\begin{widetext}
\begin{align}
%\begin{aligned}
\label{mean_epr}
\langle \dot \Sigma \rangle  &=\sum_{n,\rho} R_\rho(n) P(n,t) \Sigma_\rho(n) + d_t S_\text{sh}(t)\\
%\\=& \sum_{n,\rho >0} [ R_\rho(n)P(n,t)-R_{-\rho}(n+\Delta_\rho)P(n+\Delta_\rho,t)] \Sigma_\rho(n) + S_\text{sh} 
&  = \sum_{n,\rho >0}  [ R_\rho(n)P(n,t)-R_{-\rho}(n+\Delta_\rho)P(n+\Delta_\rho,t)]\nonumber \\
& \hspace{2.2cm} \times \ln \frac{R_\rho(n) P(n,t)}{R_{-\rho}(n+\Delta_\rho) P(n+\Delta_\rho,t)} \geq 0. \nonumber
%\end{aligned}
\end{align}
%\end{widetext}
The nonnegativity follows from $(x-y)\ln(x/y) \geq 0$ valid for all positive $x,y$.

The dynamics is said to be detailed balanced when $a_\rho =0$ for all $\rho$. Thus, in the absence of parametric time dependence of the rates in the long time limit, when the initial condition $P(n,0)$ is relaxed, $\Sigma(t)$ is identically zero and the stationary solution of \eqref{me}  is the equilibrium Gibbs distribution $P^{\text{eq}}(n)$,
\begin{align}\label{Gibbs}
P^{\text{ss}}(n) \underset{a_\rho =0}{=} \frac{ e^{-\Phi(n)}}{\sum_n e^{-\Phi(n)} }=:P^{\text{eq}}(n).
\end{align}
In view of \eqref{micro_ldb} and \eqref{Gibbs}, when $a_\rho =0$ the local detailed balance condition on the transition rates can also be written as the equality of forward and backward fluxes at equilibrium:
\begin{align}
P^{\text{eq}}(n) R_{\rho}(n)=P^{\text{eq}}(n+\Delta_\rho) R_{-\rho}(n+\Delta_\rho).
\end{align}

Two other useful decompositions of the entropy production rate \eqref{epr} can be introduced. First,  rearranging \eqref{epr}, we get
\begin{align}\label{nc_driv_epr}
\dot \Sigma(t) = \dot{\Sigma}_{\text{nc}} (t) +\dot \Sigma _\text{d}(t)- d_t [\Phi(n(t),t) +\ln P(n(t),t)],
\end{align}
where $ \dot \Sigma_{\text{nc}} (t) := \sum_\rho J_\rho(t) a_\rho$ is the dissipation rate due to nonconservative forces, $\dot \Sigma_\text{d}(t):=\partial_t \Phi (n,t)|_{n=n(t)} $ is the dissipation rate caused by driving parametrically the reference equilibrium, and $-[\Phi(n,t) +\ln P(n,t)]$ is the stochastic Massieu potential \cite{Rao2018a}. The decomposition \eqref{nc_driv_epr} is useful to rationalize which mechanism keeps the system out of equilibrium. In particular, we can name three limiting scenarios: in nonequilibrium stationary states only $ \dot{\Sigma}_{\text{nc}}$ is on average nonzero; in periodic detailed balance dynamics only $ \dot{\Sigma}_\text{d}$ is on average nonzero; in detailed balance dynamics relaxing from an initial nonequilibrium probability distribution only $d_t (\Phi +\ln P)$ is nonzero. It is worth mentioning that the  nonconservative dissipation rate  $\dot \Sigma_{\text{nc}} (t) $ can be cast in terms of the physical currents $\sum_{\rho>0}  \mathbb{X}_\rho I_\rho $ between the reservoirs that generate the fundamental forces $\mathcal{f}_\text{nc}$.
Note that, for systems in contact with reservoirs at the same temperature $T$, the nonconservative and driving contributions to the entropy production correspond to work contributions divided by the temperature $T$.

Second, for systems with nonautonomous dynamics, or with relaxation from a nonstationary initial condition,
it is worth recasting the entropy production rates into the adiabatic and nonadiabatic components \cite{hat01,speck2005integral,Esposito2007,Esposito2010Jul},
\begin{align}\label{ad_nonad_epr} 
\dot \Sigma(t)&= \dot \Sigma_\text{ad}(t) +\dot \Sigma_\text{na} (t),
%&= {\textstyle\sum_\rho } j_\rho(t)[ a_\rho - \Delta_\rho \cdot \partial_c  (\phi(c(t);t) - I_\text{ss}(c(t;t))) \Delta_\rho \cdot \partial_c (I(c(t),t)-I_\text{ss}(c(t);t))  ] 
\end{align}
with the adiabatic entropy production rate 
\begin{align}\label{ad_epr}
\dot \Sigma_\text{ad}(t):= {\textstyle\sum_\rho } J_\rho(t) \left( \Sigma_\rho(t) + \ln \frac{P_t^{\text{ss}}(n(t))}{P_t^{\text{ss}}(n(t)+\Delta_\rho)}  \right),
\end{align}
and the nonadiabatic entropy production rate
\begin{align}\label{na_epr}
\dot \Sigma_\text{na}(t):= -d_t \ln P(n(t),t)+ {\textstyle\sum_\rho } J_\rho(t) \ln \frac{P_t^{\text{ss}}(n(t)+\Delta_\rho)}{P_t^{\text{ss}}(n(t))}.
%\dot \Sigma_\text{na}(t):= {\textstyle\sum_\rho } J_\rho(t) \ln \frac{P(n(t),t) P_t^{\text{ss}}(n(t)+\Delta_\rho)}{P_t^{\text{ss}}(n(t)) P(n(t)+\Delta_\rho,t)}.
\end{align}
We recall that $P_t^{\text{ss}}(n)$ denotes the stationary solution of the master equation \eqref{me} with transition rates held fix at their instantaneous values $R_\rho(n,\vartheta(t))$.
Loosely speaking the splitting in \eqref{ad_nonad_epr} is in terms of the dissipation to maintain a stationary state and to drive it. This consideration becomes exact in two limiting cases: For autonomous systems at stationarity, in which $P(n,t)=P_t^{\text{ss}}(n)$, so that  \eqref{na_epr} 
 is identically zero and the adiabatic entropy production rate \eqref{ad_epr} equals $\dot \Sigma(t)$; for nonautonomous detailed balance dynamics, for which $P_t^{\text{ss}}(n)=P_t^{\text{eq}}(n)$, so that  \eqref{ad_epr} 
 is identically zero and the  nonadiabatic entropy production rate \eqref{na_epr} equals $\dot \Sigma(t)$. It can be shown that both $\dot \Sigma_\text{ad}(t)$ and $\dot \Sigma_\text{na}(t)$ are on average positive \cite{Esposito2007,Esposito2010Jul,ge2010,Rao2018Sep}. 
 Adiabatic and nonadiabatic components are also called housekeeping and excess ones, respectively, even though such names were originally assigned to the decomposition of entropy production of systems in contact with isothermal baths.

There exists another measure of the time-irreversibility of the dynamics obtained by lumping together in the master equation \eqref{me} all the transition rates $R_\rho(n)$ associated to jumps with the same size $\tilde \Delta_\rho$, as $\tilde R_\rho(n):=\sum_{\rho: \Delta_\rho=\tilde \Delta_\rho} R_\rho(n)$ \footnote{{This happens when a change in the state space can be realized via multiple pathways. See Sec. \ref{sec:ex1_SM} for an example.}}. On the resulting trajectories $\mathcal{X}'$ (which are identical to $\mathcal{X}$ when restricted to the state space) one can define the  Kullback-Leibler divergence:
\begin{align}\label{KL_state_space}
\langle \tilde \Sigma[\mathcal{X}'] \rangle = \int \mathcal{D}\mathcal{X}' P[\mathcal{X}'] \ln \frac{P[ \mathcal{X}']}{\overline{P}[\mathcal{X}']} \geq 0.
\end{align}
Such trajectories carry no detailed information about the elementary transitions, therefore \eqref{KL_state_space} is a mere information-theoretic with no general connection to thermodynamics. The rate of \eqref{KL_state_space}
is obtained as done for \eqref{mean_epr}:
\begin{align}
%\begin{aligned}
\hspace{-0.1cm}\langle \dot {\tilde \Sigma }(t) \rangle &=\sum_{\substack{n,\rho >0 \\ \rho: \Delta_\rho=\tilde \Delta_\rho}} [\tilde R_\rho(n)P(n,t)-\tilde R_{-\rho}(n+\Delta_\rho,t)P(n+\Delta_\rho)] \nonumber \\
& \hspace{1cm}\times \ln \frac{\tilde R_\rho(n)P(n,t)}{\tilde R_{-\rho}(n+\Delta_\rho)P(n+\Delta_\rho,t)} \geq 0.
\label{info_epr}
%\end{aligned}
\end{align} 
By applying the log sum inequality to \eqref{mean_epr}, it follows that $\langle \dot {\tilde \Sigma} \rangle \leq \langle\dot \Sigma \rangle$.

We conclude this short summary of the core structure of stochastic thermodynamics with three comments on the transition rates. 
First, they can be rewritten as
\begin{align}\label{R_rho}
&R_\rho(n)
=\Gamma_{\rho}(n) e^{\frac{1}{2}[-\Phi(n+\Delta_\rho)+\Phi(n)+ a_\rho]},
\end{align}
by identifying a symmetric kinetic factor $\Gamma_{\rho}(n)=\sqrt{R_\rho(n)R_{-\rho}(n+\Delta_\rho)}=\Gamma_{-\rho}(n+\Delta_\rho)$ not constrained by thermodynamics \cite{Maes2018non}.
Second, on can introduce dual (also called adjoint, or reversed) rates, 
\begin{align}\label{dualrate}
&R_\rho^{\dagger}(n) := \frac{R_{-\rho}(n+\Delta_\rho) P_t^{\text{ss}}(n+\Delta_\rho)}{P_t^{\text{ss}}(n)} \;,
\end{align}
which generate a stochastic process with the same stationary distribution as the original one, namely $P_t^{\text{ss}}={P_t^{\text{ss}}}^\dagger$, and change the sign of the average stationary currents, $\mean{I_\rho^{\dagger}}=-\mean{I_\rho}$ \cite{norris1998markov}. Third, by combining the escape rate $\sum_\rho R_\rho(n)$, which gives the mean time to leave the state $n$, with the entrance rate $\sum_\rho R_\rho(n- \Delta_\rho)$ in the same state, we introduce the inflow rate 
\begin{align}\label{inflow_rate}
\Lambda(n):=\sum_\rho [R_\rho(n- \Delta_\rho)-R_\rho(n)]
\end{align}
that measures the rate of contraction of a discrete volume centered on the state $n$ \cite{fal15c}.

\section{Macroscopic limit}\label{sec:macro_me}

\subsection{Thermodynamic and kinetic conditions}\label{sec:large_scale}

We identify a large parameter $V$ on which the rates $R_\rho$ depend. This can be for instance a mesoscopic volume which is large when measured in units of molecular volumes, or a typical particle number per state which is much larger than unity. We focus on systems such that $n \to \infty$ as $V \to \infty$ and thus define an intensive state variable  $c := n/V$. We fix the asymptotic behavior of the thermodynamic variables to match the extensivity prescribed by standard (macroscopic) thermodynamics,
\begin{align}\label{Phi_extensive}
\phi(c) :=&\lim_{V \to \infty }\frac{\Phi(n)}{V}.%, \quad A_\rho(n)= a_\rho(c) +  O(V^{-1})  ,
\end{align}
Note that $a_\rho$ is already a quantity of order $O(V^0)$ because the fundamental nonconservative forces are already expressed in terms of \emph{intensive} thermodynamic variables of the reservoirs (that are macroscopic).
We also assume that a deterministic macroscopic limit of the dynamics \eqref{me} exists, which requires that the transition rates scale asymptotically with $V$, 
\begin{align}\label{r_rho}
r_\rho(c) := &\lim_{V \to \infty} \frac{R_\rho(n)}{V} = \gamma_{ \rho}(c) e^{\frac{1}{2}[-\Delta_\rho \cdot \partial_c \phi(c) + a_\rho] },
 \end{align} 
which entails for the macroscopic kinetic coefficients
\begin{align}\label{lambda}
&  \gamma_\rho(c) := \lim_{V \to \infty} \frac{\Gamma_{\rho}(n) }{V}=\gamma_{-\rho}(c).
 \end{align}
Thus the macroscopic expression of the local detailed balance \eqref{micro_ldb} reads
\begin{align}\label{ldb_macro}
\sigma_\rho(c)&:= \lim_{V \to \infty}\Sigma_\rho(n)=\ln  \frac{r_\rho(c)}{r_{-\rho}(c)}
=-\Delta_\rho \cdot \partial_c \phi(c)\! + \!a_\rho \;.
\end{align}
Note that $\gamma_\rho(c)$ may depend in a symmetric fashion on the forces $-\Delta_\rho \cdot \partial_c \phi(c) $ and $a_\rho$. 
%If so, we assume the dependence to be analytic. Thus, since $  \gamma_\rho(c)$ is symmetric in $\rho$,  it can only depend on the product $a_\rho \Delta_\rho \cdot \partial_c \phi $ at lowest order.
 
Under these assumptions the master equation for the probability density of the intensive state variable, $p(c,t)= V^N P(n,t)$, reads
\begin{equation}
\begin{aligned}\label{MEpc}
 \partial_t p(c,t)& = V \sum_\rho\!\left[ r_\rho {\textstyle \left(c - \frac{\Delta_\rho}{V} \right)} p {\textstyle\left(c-  \frac{\Delta_\rho}{V} ,t \right)} -r_\rho(c) p(c,t)\right ] .
 %\\&= V  \underbrace{\sum_\rho\! \left[ e^{-\frac{\Delta_\rho}{V} \cdot \partial_c}-1 \right]r_\rho(c) }_{H \left(c, - \frac{\partial_c}{V} \right)} p(c,t)
 % \partial_t p(c,t)&= V  H \left(c, - \frac{\partial_c}{V} \right) p(c,t)
\end{aligned}
\end{equation}
The scaling imposed in \eqref{r_rho} leads to the concentration of probability in the macroscopic limit \cite{vanKampen,gardiner}.
Namely, the probability density $p(c,t)$ acquires the large-deviation form 
\begin{align}\label{rateI}
p(c,t) 
%=\frac{1}{\tilde Z} e^{-V \tilde I(c,t)}
%= \frac{1}{Z} e^{-V  I(c,t)+o(V)} 
\asymp e^{-V  I(c,t)},
%p(c,t) = e^{-V I(c,t)+o(V)}
\end{align}
where $ I \geq 0$, called rate function, is $V$-independent. 
%and $Z$ is a subexponential normalization.
The symbol $\asymp$ stands for the logarithm equality of the two sides of \eqref{rateI}, that is $-\lim_{V \to \infty} \frac{1}{V} \ln p(c,t) = I(c,t)$.
The scaling captures the fact that $p$ is singular in the limit $V \to \infty$, wherein the system is described by the deterministic dynamics of the minima of $I(c,t)$ as we are going to show in Sec. \ref{sec:deterministic_dyn}.
% It will be useful in the following to 
%single out in \eqref{rateI} the exponential part (in $V$) of the normalization $Z$ by introducing the rate function
%\begin{align}
% I(c,t) := - \lim_{V \to \infty} \frac 1 V \ln p(c,t) = \tilde I(c,t) - \tilde I(\mathcal{x}(t),t)
%\end{align}
%where $\mathcal{x}(t):=\mathrm{argmin}_c\, I(c,t)$ is the most likely, i.e. macroscopic state at time $t$.  

Plugging \eqref{rateI} into \eqref{MEpc} and expanding to leading order in $V$ yields the evolution equation for the rate function
\begin{align}\label{HJeq}
-\partial_t I(c,t)= \sum_\rho \left(e^{\Delta_\rho \cdot \partial_c I(c,t)} -1\right)r_\rho(c) ,
\end{align}
which is a Hamilton-Jacobi equation with momenta $\pi= \partial_c I$ and Hamiltonian function \cite{kubo1973,gang1987stationary}
\begin{align}\label{hamiltonian}
 \mathcal{H} (c, \pi)=\sum_\rho \left(e^{\Delta_\rho \cdot \pi} -1\right)r_\rho(c).
\end{align}
The stationary rate function $ I_\text{ss}(c)$, also called quasi-potential -- where it exists and is differentiable -- satisfies the time-independent version of \eqref{HJeq} \cite{mae07},
\begin{align}\label{HJ0}
\mathcal{H} (c, \partial_c I_\text{ss})=0,
\end{align}
and gives the stationary probability density function 
\begin{align}\label{rateI_ss}
p_\text{ss}(c) 
 \asymp e^{-V  I_\text{ss}(c)}.
\end{align}

For detailed balanced dynamics, the quasi-potential coincides (up to a constant) with the negative Massieu potential $\phi$, whose minima will be denoted by $\mathcal{x}^{\text{eq}}$. This can be seen by rewriting \eqref{HJ0} as
\begin{align}\label{HJ_eq}
\sum_\rho (r^{(0)}_{-\rho}-r^{(0)}_{\rho} e^{ \Delta_\rho \cdot \partial_c I_\text{ss}})=0
\end{align}
with detailed-balanced rates
\begin{align}\label{eqRates}
r_\rho^{(0)}:= \lim_{a_\rho \to 0}r_\rho=\gamma^{(0)}_\rho e^{-\frac 1 2 \Delta_\rho \cdot \partial_c \phi},
% $ \gamma^{(0)}_\rho:= \gamma_\rho|_{a_\rho=0}
\end{align}
and noting that $I_\text{ss}=\phi + \text{const}$ nullifies each summand in \eqref{HJ_eq} by virtue of \eqref{ldb_macro}. Hence, the large-scale stochastic dynamics preserves the equilibrium Gibbs distribution \eqref{Gibbs} and yields the Einstein's fluctuations formula \cite{Einstein1910}
\begin{align}\label{p_eq}
p_{\text{eq}}(c) \asymp e^{-V[\phi(c)-\phi(\mathcal{x}^{\text{eq}}) ]} ,
\end{align}
with the partition function $\sum_n e^{-\Phi(n)} 
\asymp e^{-V \phi(\mathcal{x}^{\text{eq}})}$ obtained by the Laplace approximation. Note that the equilibrium state $\mathcal{x}^{\text{eq}}$ is the one that minimizes the thermodynamic potential $\phi$.
 
For transition rates that carry an explicit time dependence, we introduce the instantaneous stationary rate function $I_\text{ss}^t(c)$  
which satisfies
\begin{align}\label{instantaneous_I}
\mathcal{H}_t (c, \partial_c I_\text{ss}^t)=0,
 \end{align}
where $\mathcal{H}_t$ is the Hamiltonian \eqref{hamiltonian} with transition rates $r_\rho(c,\vartheta(t))$ held fixed at their instantaneous value at time $t$. We end by noting that, using (\ref{dualrate}), the scaled dual rate is given by
\begin{align}\label{scaledDualRate}
r_\rho^{\dagger}(c) := \lim_{V \to \infty} \frac{R_\rho^{\dagger}(n)}{V} 
=  r_{-\rho}(c) e^{-\Delta_\rho \cdot \partial_c I_{\text{ss}}(c)} \;,
 \end{align} 
and is associated to the macroscopic log-ratio
\begin{align}
\begin{aligned}\label{ldb_macro_a}
\sigma^\text{ad}_\rho(c)&:= \lim_{V \to \infty} \ln \frac{R_\rho(n)}{R_\rho^{\dagger}(n)}
= \ln  \frac{r_\rho(c)}{r_{\rho}^{\dagger}(c)} \\
&=\Delta_\rho \cdot \partial_c [I_{\text{ss}}(c)-\phi(c)]\! + \!a_\rho.
\end{aligned}
\end{align}
{This corresponds to the macroscopic limit of the adiabatic entropy production of transition $\rho$, as can be directly seen from \eqref{ad_epr}.}
The importance of these quantities for the macroscopic dynamics and thermodynamics will be elucidated in following sections.

\subsection{Deterministic dynamics}\label{sec:deterministic_dyn}

The macroscopic dynamical equation for the intensive state variable can be obtained multiplying both sides of \eqref{MEpc} by $c$, and integrating over $c$ with the change of variable $c \mapsto c+ \Delta_\rho/V$ in the righthand side,
\begin{align}\label{dcdt}
 d_t \mean{c}= \sum_\rho \Delta_\rho \mean{r_\rho(c)}= \sum_{\rho>0} \Delta_\rho[ \mean{r_\rho(c)}- \mean{r_{-\rho}(c)}].
\end{align}
The mean value of a generic state observable $\mathcal{O}$ at time $t$ is calculated as
$\mean{\mathcal{O}(c(t))}= \int dc \, \mathcal{O}(c) p(c,t) $.
In view of the large-deviation scaling of the probability \eqref{rateI}, the probability density concentrates around its most likely state $\mathcal{x}(t)$ which corresponds to the global minimum of $I(c,t)$
%we have that $\lim_{V \to \infty}p(c,t) = \delta(c-\mathcal{x}(t))$   
\cite{Hao2011Jan, Qian2016Nov,huang2017processes}, namely,
\begin{align}\label{min_rate_function}
\mathcal{x}(t) :=\text{argmin}_c I(c,t),
\end{align}
and therefore 
\begin{align}\label{min_rate_functionBis}
I(\mathcal{x}(t),t)=\partial_c I(\mathcal{x}(t),t)=0 
\ \;, \ \ 
\partial^2_c I(\mathcal{x}(t),t)>0
\;.
\end{align}
This implies that, using Laplace method, $\lim_{V \to \infty }\mean{\mathcal{O}(c(t))}=\mathcal{O}(\mathcal{x}(t))$ and \eqref{dcdt} approaches the dynamics 
\begin{align}\label{macro_dcdt}
d_t  \mathcal{x}(t)= F(\mathcal{x}(t)),
\end{align}
with the deterministic drift field
\begin{align}\label{macro_drift}
F(c) :=\sum_\rho \Delta_\rho r_\rho(c) = \partial_\pi \mathcal{H}(c,\pi)|_{\pi=0},
\end{align}
which we recognize as the derivative of the Hamiltonian \eqref{hamiltonian} with respect to the momenta $\pi$. We will thoroughly explore the connection with the Hamiltonian dynamics in Sec. \ref{sec:hamiltonian_dyn}.
Notably, the macroscopic limit of the inflow rate \eqref{inflow_rate} is the negative divergence of the drift field \eqref{macro_drift},
\begin{align}\label{contraction_rate}
  \lambda(c):= \lim_{V \to \infty} \Lambda(n) %= -\sum_\rho \Delta_\rho \cdot \partial_c r_\rho(c)
  =-\partial_c \cdot F(c) .
\end{align}
The latter is a key quantity in the dynamical systems theory as it represents the phase space volume contraction rate \cite{dor99,gas05}, that is the negative sum of the Lyapunov exponents \cite{benettin1980lyapunov} of the deterministic dynamics \eqref{macro_dcdt}. The phase space contraction rate plays a key role in the statistical mechanics of deterministic thermostatted systems \cite{eva90} as it enters the steady state (Gallavotti-Cohen \cite{eva93, gal95,gal95b}) and the transient (Evans-Searles \cite{eva94,eva02}) fluctuation theorems, and can sometimes be identified with the entropy production rate \cite{cohen1998note}. 

For any autonomous dynamics \eqref{macro_dcdt}, the quasi-potential $I_\text{ss}$ always decreases along the solution, i.e. 
\begin{align}\label{mean_nonadiabatic_autonomous}
d_t I_\text{ss}(\mathcal{x}(t)) = F(\mathcal{x}(t)) \cdot \partial_c I_\text{ss}(\mathcal{x}(t))  \leq 0,
\end{align}
where the inequality is proved using the explicit form of \eqref{HJ0} with the fact that $e^x-1 \geq x$. 
Since \eqref{min_rate_function} implies that $I_\text{ss}$ always reaches a minimum on the long-time solutions of \eqref{macro_dcdt}, be them  {stable} fixed points $\mathcal{x}^*$ (defined by $F(\mathcal{x}^*)=0${, see Sec. \ref{sec:multi_meta}}) or time-dependent attractors $\mathcal{x}^*(t)$,
the quasi-potential is a Lyapunov function of the deterministic dynamics \eqref{macro_dcdt}.
When the dynamics is detailed balanced ($a_{\rho}=0$), \eqref{mean_nonadiabatic_autonomous} reduces to
\begin{align}\label{mean_nonadiabatic_autonomous_atEq}
 d_t \phi(\mathcal{x}(t)) \leq 0 \;,
\end{align}
and we recover a central tenet of macroscopic thermodynamics, namely that the thermodynamic potential $\phi$ is minimized by the dynamics.

We end by introducing the Hamiltonian of the dual process
\begin{align}\label{Dualhamiltonian}
\mathcal{H}^{\dagger} (c, \pi) = \sum_\rho \left(e^{\Delta_\rho \cdot \pi} -1\right)r_\rho^{\dagger}(c) \;,
\end{align}
and its deterministic drift field 
\begin{align}\label{Dual_macro_drift}
F^{\dagger}(c) :=\sum_\rho \Delta_\rho r_\rho^{\dagger}(c) 
= \partial_\pi \mathcal{H}^{\dagger}(c,\pi)|_{\pi=0} \;.
\end{align}
The dual dynamics shares not only the same fixed point as the original one, $F^{\dagger}(x^*)=F(x^*)=0$, but also the same steady state rate function, 
\begin{align}\label{DualhamiltonianAtSS}
\mathcal{H}^{\dagger}(c,\partial_c I_\text{ss})=-\mathcal{H}(c,\partial_c I_\text{ss})=0 \;.
\end{align}
The dual macroscopic trajectory $\mathcal{x}^\dagger(t)$ solution of the equation
\begin{align}\label{eq:macro_dual_dyn}
d_t {\mathcal{x}^\dagger}(t)= F^{\dagger}(\mathcal{x}^\dagger(t))
\end{align}
with initial condition $\mathcal{x}^\dagger(0)=\mathcal{x}^*$, will play a central role in the description of macroscopic fluctuations, Sec. \ref{sec:gauss_instanton}.

\subsection{Multistability vs metastability}\label{sec:multi_meta}

A fixed point is stable{, and will be denoted $\mathcal{x}^*$,} if all eigenvalues of the Jacobian matrix $\partial_c F(c)$ evaluated in $c=\mathcal{x}^*$ {have negative real part}. It is unstable when at least one eigenvalue {has positive real part}.
When $F$ is nonlinear, multiple stable fixed points {$\mathcal{x}^*_\gamma$ can be present that we indicate with the label $\gamma \in \mathbb{N}$}. More complex, time-dependent attractors such as limit cycles will only be mentioned later. 
We assume that the stable fixed points $\mathcal{x}^*_\gamma$ are separated by nondegenerate saddle points, denoted $\mathcal{x}_\nu${, i.e. fixed points with real, nonzero eigenvalues, at least one of which is positive. They} define the boundaries between the different basins of attraction of $F$.
In this case the dynamics \eqref{macro_dcdt} is nonergodic and multistable, since $\mathcal{x}(t)$, due to \eqref{mean_nonadiabatic_autonomous}, will relax within the basin of attraction selected by the initial condition $\mathcal{x}(0)$ to its fixed point. 
This has to be contrasted with the uniqueness of the stationary solution of the underlying mesoscopic master equation \eqref{me}. The discrepancy, known as Keizer's paradox \cite{keizer1978thermodynamics}, stems from the fact that the long time limit and the macroscopic limit do not commute in general. This phenomenon is understood within the spectral theory of the Markovian generator, which is summarized as follows \cite{gaveau1998,kurchan2009}. The time evolution of the probability distribution, can be expanded in the right eigenfunctions $\psi^{(R)}_{\xi}(n)$ of the matrix $\mathcal{R}$ in \eqref{me},
\begin{align}
P(n,t) = \sum_{\xi}  b_{\xi} \psi^{(R)}_{\xi}(n) e^{\omega_{\xi} t},
\end{align}
where $b_{\xi}= \sum_n \psi^{(L)}_{\xi}(n) P(n,0)$ are the overlap of the initial condition with the left eigenfunctions $ \psi^{(L)}_{\xi}(n)$.
Since $\mathcal{R}$ is a stochastic generator on a finite state space, by Perron-Frobenius theorem it has nonpositive eigenvalues  (which can be ordered by their real part $\text{Re}(\omega_{\xi}) \geq \text{Re}(\omega_{{\xi}+1})$), of which $\omega_0=0$ is nondegenerate and associated to a constant left eigenfunction. Metastability appears when there exist eigenvalues $\omega_1, \dots, \omega_{\tilde {\xi}}$ whose real part goes to $0=\omega_0$ as $V \to \infty$, while all others $\omega_{\xi}$ with ${\xi} > \tilde {\xi} $ stay finite.
The inverse of this spectral gap corresponds to a diverging time scale separating the fast dynamics within basins of attractions from the slow dynamics) between them. Ultimately, if $V \to \infty $ at finite $t$, the system probability can only converge to (linear combinations of) those $\psi^{(R)}_{\xi}(n)$ with ${\xi} \leq \tilde {\xi}$ that are selected by the initial conditions, i.e. with finite overlap $b_{\xi} \neq 0$. 

In general $\mathcal{R}$ is not similar to a symmetric matrix unless detailed balance holds, hence the eigenvalue $\omega_{\xi}$ can have a non-zero imaginary part. In this  case metastable states can be time-dependent, such as stable limit cycles at $V \to \infty$ \cite{Herpich2018Sep}.
A limit cycle is a closed trajectory in state space not surrounded by other closed trajectories. It is called stable or attracting if all neighboring trajectories approach it for large times. It appears as a periodic solution of \eqref{macro_dcdt}, $\mathcal{x}^*(t)=\mathcal{x}^*(t+t_\text{p})$, with period $t_\text{p} = 2 \pi/\text{Im}(\omega_{\xi})$ with ${\xi} \leq \tilde {\xi}$.
Hereafter, we only briefly mention such periodic attractors, while we do not explicitly consider more general quasi-periodic and chaotic attractors -- see the application in Sec.~\ref{sec:ex3} and the discussion in Sec. \ref{sec:discussion}, though.

For multistable systems a further comment is in order. %If the initial condition $I(c,0)$ is finite on the full state space, $I(c,t)$ converges to $I_\text{ss}(c)$ for $t \to \infty$. Instead, if $I(c,0)$ is finite only on a single basin of attraction $\gamma$,
%we first take the limit $V \to \infty$ corresponding to an initial probability density $p(c,0)= \delta(c-\mathcal{x}(0))$ with $\mathcal{x}(0)$ in the basin of attraction of $\mathcal{x}_\gamma^*$, 
%$I(c,t)$ converges to the \emph{local} quasi-potential $I_\text{ss}^{(\gamma)}$ for $t \to \infty$.
%if the initial condition of the dynamics is ized on a single basin of attraction $\gamma$
For general nondetailed balanced dynamics, the rate function $I_\text{ss}$ is locally nondifferentiable and \eqref{HJ0} can be solved only piecewise in each basin of attraction \cite{graham86,baek2015} obtaining the local quasi-potential $I^{(\gamma)}_\text{ss}$.
This stems from the fact that only the local stochastic dynamics can be directly determined when the macroscopic limit is taken before the long-time limit  \cite{ge2012landscapes,zhou2016construction}.
The global quasi-potential defined by first taking the long time limit before the large $V$ limit, is obtained a posteriori by fixing the normalization constants, i.e. the relative weights $\alpha_\gamma$ of the attractors, and choosing at each $c$ the minimum among the local quasi-potentials \cite{Nardini2016}:
\begin{align}\label{global_QP}
I_\text{ss}(c)= \min_\gamma (I^{(\gamma)}_\text{ss}(c)+\alpha_\gamma) - \min_\gamma \alpha_\gamma.
\end{align}
The last term ensures that $I_\text{ss}$ is zero on the most likely attractor.
The Markov jump process on attractors which is used to fix the normalization constants  \cite{freidlin,graham1987macroscopic} will be discussed in Sec. \ref{sec:emergent} as a coarse-grained description for the long-time macroscopic fluctuating thermodynamics of the systems.
If the limit $V \to \infty$ is taken before the limit $t \to \infty$ the relative weights are fixed by the initial condition (the relative probability to be on an attractor).

\subsection{Deterministic thermodynamics}\label{sec:deterministic_thermo}

The first crucial observation regards the macroscopic Shannon entropy
\begin{align}\label{Shannon_entropy_macro}
S_\text{sh}(t)=- \int dc \, p(c,t) \ln p(c,t),
\end{align}
where an irrelevant additive constant has been discarded.
Its scaled macroscopic limit is identically null, since the randomness associated to the distribution over states $c$ has vanished, see \eqref{min_rate_functionBis}:
\begin{equation}\label{shannon0}
\begin{aligned}
&-\lim_{V \to \infty} \frac{1}{V} S_\text{sh}(t)
%\int dc \, p(c,t) \ln p(c,t)%=\mean{I(c,t)}\\
%&= \int dc \, I(c,t) \frac{1}{Z} e^{-V I(c,t)} 
= I(\mathcal{x}(t),t)  =0.
\end{aligned}
\end{equation}
However, the mean of the scaled system entropy \eqref{sys_ent_meso} is in general finite and entirely given by the internal entropy evaluated on the most likely state,
\begin{align}\label{macro_s_sys}
\lim_{V \to \infty} \frac 1 V \mean{S_\text{sys}(n,t)}  = s_\text{int}(\mathcal{x}(t)) .
\end{align}
{Here we have required the internal entropy $S_\text{int}$ to be an extensive function of $V$.}

The mean variation rate of thermodynamic observables depends on the average flux of transitions, which at leading order in $V$ reads 
\begin{align}
\mean{J_\rho(t)}= V \int dc\, r_\rho(c) p(c,t).
\end{align}
The concentration of the probability \eqref{rateI} yields for the scaled macroscopic limit of the transition flux
\begin{align}\label{mean_j}
\lim_{V \to \infty}\frac 1 V \mean{J_\rho(t)}= r_\rho(\mathcal{x}(t)).
\end{align}
Hence, the function $[r_\rho(\mathcal{x}(t))-r_{-\rho}(\mathcal{x}(t))] \mathcal{o}_\rho$ represents the mean rate of variation of an intensive quantity $ \mathcal{o}_\rho=-\mathcal{o}_{-\rho}$ due to transition $\rho$.
In particular, the mean of the scaled entropy production rate $\dot \sigma:=\dot \Sigma/V$ reads in the macroscopic limit 
\begin{align} \label{epr_macro}
\lim_{V \to \infty }\!\!\mean{\dot \sigma}%&=\!{\textstyle\int_0^\tau dt  \sum_\rho} \mean{r_\rho(c(t)) \sigma_\rho(c(t))} \nonumber \\
& =  \sum_{\rho>0} [r_\rho(\mathcal{x}(t))-r_{-\rho}(\mathcal{x}(t))]\sigma_\rho(\mathcal{x}(t)) \geq 0,
\end{align}
where we have used \eqref{ldb_macro} and \eqref{shannon0} into \eqref{mean_epr} -- note the absence of the Shannon entropy with respect to \eqref{mean_epr}.
In the macroscopic limit, all average quantities at time $t$ are functions of $\mathcal{x}(t)$. However, for brevity we avoid to write explicitly this dependence.

Taking the mean value of decomposition \eqref{nc_driv_epr} and using the concentration of probability \eqref{rateI} and \eqref{mean_j}, we obtain 
\begin{align}\label{nc_driv_epr_macro_fluct}
\begin{aligned}
\lim_{V \to \infty}\mean{\dot \sigma} = \lim_{V \to \infty}\mean{\dot \sigma_\text{nc}} +\lim_{V \to \infty}\mean{\dot \sigma_\text{d}}
%& \,{\textstyle \sum_\rho} r_\rho (\mathcal{x}(t)) a_\rho \\
%&- \partial_t \phi(c,t)|_{c=\mathcal{x}(t)}
- d_t \phi (\mathcal{x}(t),t),
\end{aligned}
\end{align}
which displays the scaled mean of the nonconservative dissipation rate 
\begin{align}
\lim_{V \to \infty}\mean{\dot \sigma_\text{nc}}={\textstyle \sum_\rho} r_\rho (\mathcal{x}(t)) a_\rho \;,
\end{align}
of the driving dissipation rate 
\begin{align}
\lim_{V \to \infty}\mean{\dot \sigma_\text{d}}=\partial_t \phi (c,t)|_{c=\mathcal{x}(t)} \;, 
\end{align}
and the time derivative of the mean scaled thermodynamic potential. 
Proceeding in the same manner with \eqref{ad_nonad_epr}, we can find the alternative decomposition 
\begin{align}\label{nad_ad_epr_macro_fluct}
\lim_{V \to \infty}\mean{\dot \sigma} = \lim_{V \to \infty} \mean{\dot \sigma_\text{ad}} + \lim_{V \to \infty} \mean{\dot \sigma_\text{na}} \;,
\end{align}
in terms of mean scaled adiabatic entropy production rate 
\begin{align}\label{mean_adiabatic}
\hspace{-0.3cm}\lim_{V \to \infty} \mean{\dot \sigma_\text{ad}} 
= \sum_{\rho>0} [r_\rho(\mathcal{x}(t))-r_{-\rho}(\mathcal{x}(t))] \sigma^\text{ad}_{\rho}(\mathcal{x}(t)) \geq 0 
%&\hspace{-0.2cm}= \sum_{\rho} r_\rho(\mathcal{x}(t)) \{a_\rho + \Delta_\rho \cdot \partial_c [I_\text{ss}^t(c)-\phi(c,t)]\}|_{c=\mathcal{x}(t)} \geq 0 \;, \nonumber
\end{align}
with \eqref{ldb_macro_a}, 
and the mean scaled nonadiabatic entropy production rate
\begin{align}
\begin{aligned}\label{mean_nonadiabatic}
\lim_{V \to \infty} \mean{\dot \sigma_\text{na}} 
&= - F(\mathcal{x}(t)) \cdot \partial_c I^t_\text{ss}(\mathcal{x}(t)) \\
&=  \partial_t I_\text{ss}^t(c)|_{c=\mathcal{x}(t)}-d_t I_\text{ss}^t(\mathcal{x}(t)) \geq 0 \;,
\end{aligned}
\end{align}
where we used \eqref{ldb_macro} and \eqref{macro_drift}. 
We recall that the rate function $I_\text{ss}^t(c)$ is the solution of \eqref{instantaneous_I} with the transition rates held fixed at their instantaneous value.
Since \eqref{ad_epr} and \eqref{na_epr} are nonnegative on average, $\mean{\dot \sigma_\text{ad}} $ and $ \mean{\dot \sigma_\text{na}}$ are nonnegative as well. As a consequence, for autonomous systems ($I_\text{ss}^t=I_\text{ss}$), the positivity of the nonadiabatic entropy production \eqref{mean_nonadiabatic} provides an alternative proof that the 
quasi-potential decreases along the solution of \eqref{macro_dcdt}, as shown in \eqref{mean_nonadiabatic_autonomous}.

Finally, thanks to \eqref{r_rho} and \eqref{rateI}, the mean value of the scaled information-theoretic entropy production rate $\dot{\tilde \sigma}:=\dot{\tilde \Sigma}/V$ (cf. \eqref{info_epr}) reads
%$\dot \sigma_\text{info}(\mathcal{x}(t)) := \lim_{V \to \infty} \frac{1}{V} \langle \dot{\tilde \Sigma} (t)\rangle$
\begin{align}\label{info_epr_macro}
%\dot \sigma_\text{info}
&\lim_{V \to \infty} \langle \dot{\tilde \sigma} (t)\rangle = \\
&\hspace{1cm} \sum _{\substack{\rho >0 \\ \rho: \Delta_\rho=\tilde \Delta_\rho}} [\tilde r_\rho(\mathcal{x}(t))-\tilde r_{-\rho}(\mathcal{x}(t))] 
 \ln \frac{\tilde r_\rho(\mathcal{x}(t))}{\tilde r_{-\rho}(\mathcal{x}(t))} \geq 0, \nonumber
\end{align} 
with $\tilde r_\rho(c):=\lim_{V \to \infty}\frac 1 V \tilde R_\rho(n)$.
Using again the log sum inequality we find that $\mean{\dot{\tilde \sigma}} \leq \mean{\dot \sigma}$.

\subsection{Drift field decomposition}\label{subsec:orthogonal}

For simplicity, we focus on autonomous dynamics in this subsection. 
A useful decomposition of the macroscopic vector field $F(c)$ can be obtained
by retaining the entire Kramers-Moyal expansion of the master equation \cite{gardiner}. Namely, the Taylor expansion of the righthand side of \eqref{MEpc} yields a continuity equation with probability current $\mathfrak{j}(c,t)$:
\begin{align}\label{current}
&\qquad \qquad \qquad \partial_t p(c,t) =- \partial_c \cdot \mathfrak{j}(c,t),\\
&\mathfrak{j}(c,t):=\sum_{\rho}  \sum_{k=1}^\infty \frac{\Delta_\rho}{V^{k-1} k!} (- \Delta_\rho  \cdot \partial_c)^{k-1} [r_\rho(c) p(c,t)].\nonumber
\end{align}
Then plugging the large-deviation ansatz \eqref{rateI} into \eqref{current} and restricting to the stationary state, we identify the macroscopic limit of the probability velocity in configuration space \cite{wu2013potential},
\begin{align}
\begin{aligned}\label{velocity}
v_\text{ss}(c)&:= %\lim_{\substack{t \to \infty \\ V \to \infty}}
 \lim_{V \to \infty} \lim_{t \to \infty} \frac{\mathfrak{j}(c,t)}{p(c,t)}=\sum_\rho\Delta_\rho r_\rho(c)  \frac{e^{\Delta_\rho \cdot \partial_c I_\text{ss}}-1}{\Delta_\rho \cdot \partial_c I_\text{ss}}.
%= F(c) + \mathcal{F}(\Delta \cdot \partial_c I_\text{ss}(c),c).
%\\ \nonumber
%&= F(c) + \underbrace{\sum_\rho r_\rho(c) \sum_{k=2}^\infty \frac{\Delta_\rho}{ k!} ( \Delta_\rho  \cdot \partial_c I_\text{ss}(c))^{k-1} }_{\mathcal{F}(\partial_c I_\text{ss}(c))}.
\end{aligned}
\end{align}
%As discussed in Sec. \ref{sec:deterministic_thermo}, the macroscopic limit should follow the long-time limit, otherwise the probability mass remains confined to the attractor(s) where the initial condition has support.
As a consequence, the deterministic drift vector field
\begin{align}\label{splitting_F}
F(c)=v_\text{ss}(c) -\mathcal{F}(c),
\end{align}
splits into the asymptotic probability velocity \eqref{velocity} and a gradient-like vector field
{\begin{align}
\begin{aligned}\label{explictFcurly}
\mathcal{F}(c) &:= \sum_\rho\Delta_\rho r_\rho(c) \frac{e^{\Delta_\rho \cdot \partial_c I_\text{ss}}-\Delta_\rho \cdot \partial_c I_\text{ss}-1}{\Delta_\rho \cdot \partial_c I_\text{ss}}\\&
 =\mathcal{M}(c) \cdot \partial_c I_\text{ss},
\end{aligned}
\end{align}}
with the symmetric positive semidefinite ``mobility'' matrix
\begin{align}\label{explictM}
\mathcal{M}(c)= \sum_\rho \Delta_\rho \Delta_\rho r_\rho(c) \frac{e^{\Delta_\rho \cdot \partial_c I_\text{ss}}-\Delta_\rho \cdot \partial_c I_\text{ss}-1}{(\Delta_\rho \cdot \partial_c I_\text{ss})^2},
\end{align}
which itself depends on the rate function.
They can respectively be rewritten in terms of the Hamiltonian \eqref{hamiltonian}
\begin{align}\label{orthogonal_decomposition}
&v_\text{ss}(c)= \int_0^1 d\theta\partial_\pi \mathcal{H}(c,\pi)|_{\pi=\theta \partial_c I_\text{ss}},\\
&\mathcal{M}(c):= \int_0^1 d \theta (1-\theta) \partial^2_\pi \mathcal{H}(c,\pi)|_{\pi=\theta \partial_c I_\text{ss}} .
\end{align}
These expressions appear in \cite{gao2022revisit} for polynomial transition rates, but hold irrespective of the specific form of $r_\rho$.
%but it went unnoticed that $v_\text{ss}$ is the steady state probability velocity.

Equation \eqref{splitting_F} is an extension of the well-known ``orthogonal'' decomposition valid for {deterministic dynamical systems supplemented by weak Gaussian noise leading to diffusion processes. \cite{Bertini2015Jun,zhou2016construction}. This decomposition is here generalized to Markov jump processes, where the noise is Poissonian.} It expresses the nonlinear downhill motion of $\mathcal{x}(t)$ in the gradient of the Lyapunov function $I_\text{ss}$ superimposed to the circulation on its level sets with velocity $v_\text{ss}(\mathcal{x}(t))$. The major difference is that the mobility matrix $\mathcal{M}$ of diffusive dynamics is independent of the gradient of the rate function. 
%In Appendix \ref{app:nonlinear}, we show that $ \mathcal{F}$ can indeed be written in terms of a convex potential entailing a nonlinear gradient dynamical structure. 

To prove these properties, we first focus on the probability velocity $v_\text{ss}(c)$.
It is orthogonal to the gradient of the quasi-potential $I_\text{ss}$ since 
\begin{align}\label{orthogonal_v_I}
  v_\text{ss} \cdot \partial_c I_\text{ss}=\mathcal{H} (c, \partial_c I_\text{ss})=0,
\end{align}
thanks to the definition \eqref{velocity} and the stationary state condition \eqref{HJ0} \footnote{Stationarity of the microscopic stochastic dynamics should not be taken for stationarity of the macroscopic rate equation \eqref{macro_dcdt}. For example, a time-dependent attractor $\mathcal{x}^*(t)$, such as a stable limit cycle, corresponds to a stationary probability density $p_\text{ss}(c)$ with a rate function $I_\text{ss}$ that is  zero on the set $\{\mathcal{x}^*(t)\}_{t=0}^{t_\mathrm{p}}$.}. 
Additionally, the probability velocity is divergence-free on the fixed points
\begin{align}\label{divergenceless}
   \partial_c \cdot v_\text{ss}(\mathcal{x}^*)=0.
\end{align}
It follows from writing $v_\mathrm{ss}=\mathcal{N}(c) \cdot \partial_c I_\mathrm{ss}$, with $\mathcal{N}(c)$ an antisymmetric matrix that enforces  \eqref{orthogonal_v_I}, and using the condition \eqref{min_rate_functionBis}. Note that in general $\partial_c \cdot v_\text{ss} \neq 0$ for $c \neq \mathcal{x}^*$, as one can show by expanding the master equation \eqref{MEpc} beyond the leading order approximation \eqref{HJeq}.

Then we turn to the gradient part of the dynamics. We already showed in \eqref{mean_nonadiabatic_autonomous} that the quasi-potential $I_\text{ss}(c)$ is the Lyapunov function of \eqref{macro_dcdt}, i.e. it decreases along solutions of \eqref{macro_dcdt} and reaches a (local) minimum at a stable fixed point $\mathcal{x}^*_\gamma$ (or stable time-dependent attractor $\mathcal{x}^*(t)$). 
In fact only the gradient part of the drift vector field, $\mathcal{F}(c)$, contributes in \eqref{mean_nonadiabatic_autonomous} due to \eqref{orthogonal_v_I}, i.e.
\begin{align}\label{gradient_descent}
d_t I_\text{ss}(\mathcal{x}(t))=- \mathcal{F}(\mathcal{x}(t)) \cdot \partial_c I_\text{ss}|_{c=\mathcal{x}(t) } \leq 0.
\end{align}
%that $\mathcal{M}$ is nonnegative:
%\begin{align}\label{gradient_descent}
%d_t I_\text{ss}(\mathcal{x}(t))=- \partial_c I_\text{ss} \cdot \mathcal{M}\cdot \partial_c I_\text{ss}|_{c=\mathcal{x}(t) } \leq 0.
%\end{align}

{For detailed balance dynamics,  the velocity $v_\text{ss}$ is identically zero because $\lim_{t \to \infty }\mathfrak{j}=0$. This can also be proved starting from the explicit expression \eqref{velocity}. By using $I_\mathrm{ss}= \phi$, valid for $a_\rho=0$, and the expression \eqref{eqRates} for detailed balance transition rates, we obtain
\begin{align}
\begin{aligned}
    v_\text{ss} &\underset{a_\rho=0}{=} \sum_\rho \Delta_\rho \gamma_\rho^{(0)} e^{-\Delta_\rho \cdot \partial_c \phi/2}\frac{e^{\Delta_\rho \cdot \partial_c \phi}-1}{\Delta_\rho \cdot \partial_c \phi} \nonumber \\
    &=
    2\sum_{\rho} \Delta_\rho \gamma_\rho^{(0)} \frac{\sinh(\Delta_\rho \cdot \partial_c \phi/2)}{\Delta_\rho \cdot \partial_c \phi}
    =0,
    \end{aligned}
\end{align}
where in the last passage we split the sum over forward and backward transitions $\pm \rho$  and used the anti-symmetry of $\Delta_\rho$ and $\sinh$.}
 In this case the deterministic dynamics is a nonlinear gradient descent in the thermodynamic potential $\phi$, 
\begin{align}
\begin{aligned}
\label{F_eq}
F(c) &\underset{a_\rho=0}{=} - \mathcal{F}(c) 
\underset{a_\rho=0}{=} - D^{(0)}(c) \cdot \partial_c \phi(c) +O(( \partial_c \phi)^2) \;,
\end{aligned}
\end{align}
where we introduced 
\begin{align}\label{diffusion0}
D^{(0)}(c) := \lim_{a_\rho \to 0} D(c) 
= \frac 1 2 \textstyle  \sum_{\rho} \Delta_{\rho} \Delta_{\rho} r_\rho^{(0)}(c) \;,
\end{align}
the detailed balance limit of the positive definite symmetric diffusion matrix{, which in general reads}
\begin{align}\label{diffusion}
D(c):= \frac 1 2 \textstyle  \sum_{\rho} \Delta_{\rho} \Delta_{\rho} r_\rho(c) 
= \frac 1 2 \partial_\pi^2 \mathcal{H}(c,\pi)|_{\pi=0} \;.
\end{align}
%Using \eqref{explictM} and \eqref{diffusion} we note that $\mathcal{M}(\mathcal{x}^*)=D(\mathcal{x}^*)$.
Equation \eqref{F_eq} means that detailed balance dynamics admit only time-independent attractors, i.e., limit cycles and chaos are ruled out in equilibrium.
As indicated in the second equality of \eqref{F_eq}, neglecting higher order derivatives to get a linear gradient descent dynamics is only possible close to fixed points -- or in a continuous limit where $\Delta_\rho$ becomes infinitesimal, as will be shown in Section \ref{subsec:continuous}).
We will show in Section \ref{sec:macro_thermo} that for small but not vanishing $a_\rho$, the quasi-potential is still given by thermodynamic quantities.

The splitting \eqref{splitting_F} with the condition \eqref{orthogonal_v_I} complies with the pre-GENERIC dynamics introduced in \cite{kraaij2018deriving} -- an extension of GENERIC \cite{ottinger2005}. 
$\mathcal{F}$ entails a nonlinear gradient flow \cite{liero2013gradient,mielke2016generalization} since it can be recast as the product of the jump matrix with a gradient field in the space of transitions,
 \begin{align}
 \mathcal{F}(c) =\sum_\rho \Delta_\rho  \partial_{z_\rho} \psi(z) \vert_{z_\rho = \Delta_\rho \cdot \partial_c I_\text{ss}} ,
 \end{align}
 where the potential $\psi$ is defined for all $z_\rho \neq 0$ as 
 \begin{align}
 \begin{aligned}
 \psi(z)
 %&= \sum_\rho r_\rho(c) \left( \int^{z_\rho} \frac{e^y-1}{y} dy - z_\rho \right)\\
 &= \sum_\rho r_\rho \left[ \mathrm{Ei}(z_\rho) - \ln(|z_\rho|) -z_\rho \right],
 \end{aligned}
 \end{align}
and $\mathrm{Ei}$ is the exponential integral. One can check that $\psi(z)$ is convex and nonnegative with minimum at $z \to 0$, but it is not symmetric in $z_\rho$. These properties imply 
 \begin{align}
 \sum_\rho z_\rho  \partial_{z_\rho} \psi \geq \psi(z) - \psi(0) \geq \min_z \psi(z) - \psi(0) =0,
 \end{align}
which corresponds to \eqref{gradient_descent} with $z_\rho = \Delta_\rho \cdot \partial_c I_\text{ss}$.

We note that the drift field on any attractor reduces to the probability velocity
\begin{align}\label{vss_attractors}
   F(\mathcal{x}^*(t)) = v_\text{ss}(\mathcal{x}^*(t)) \,, \ \ \mathcal{F}(\mathcal{x}^*(t))=0 \;,
\end{align}
since $\partial_c I_\text{ss}(\mathcal{x}^*(t))=0$ in \eqref{explictFcurly}.
In the special case of time-independent attractors (i.e. fixed points),  \eqref{vss_attractors} becomes
\begin{align}\label{velocityFPzero}
F(\mathcal{x}_\gamma^*) = v_\text{ss}(\mathcal{x}_\gamma^*)=\mathcal{F}(\mathcal{x}_\gamma^*)=0,
\end{align}
i.e. the probability velocity 
nullifies on the fixed points.

Finally, it is worth considering the drift field decomposition of the dual process, $F^{\dagger}(c)=v_\text{ss}^{\dagger}(c) -\mathcal{F}^{\dagger}(c)$. We find that the dual dynamics reverts the velocity $v_\text{ss}^{\dagger}(c)=-v_\text{ss}(c)$, the velocity remains orthogonal to the gradient of the quasi-potential $v_\text{ss}^{\dagger} \cdot \partial_c I_\text{ss}=0$, and the quasi-potential remains a Lyapunov function of the dual dynamics $d_t I_\text{ss}(\mathcal{x}^\dagger(t))=- \mathcal{F}^{\dagger}(\mathcal{x}^\dagger(t)) \cdot \partial_c I_\text{ss}|_{c=\mathcal{x}^\dagger(t) } \leq 0$.

\subsection{Linearized dynamics and thermodynamics}\label{sec:macro_linear_dyn}

The deterministic dynamics \eqref{macro_dcdt} linearized around a fixed point reads
\begin{align}\label{linear_db}
d_t  \mathcal{x}(t)
= (\mathcal{x}(t)- \mathcal{x}^*) \cdot \partial_c F(\mathcal{x}^*) \;,
\end{align}
where the matrix defining the relaxation coefficients is
the Jacobian matrix of the dynamics. %which using \eqref{splitting_F}-\eqref{explictM}.
In turn, the nonadiabatic entropy production \eqref{mean_nonadiabatic} around the fixed point reads
\begin{align}\label{mean_nonadiabatic_autonomousBis}
\lim_{V \to \infty} \mean{\dot \sigma_\text{na}}
= (\mathcal{x}(t)-\mathcal{x}^*) \cdot \mathcal{S}(\mathcal{x}^*) \cdot (\mathcal{x}(t)-\mathcal{x}^*) \;,
\end{align}
where we introduced the matrix
\begin{align}\label{StabMatrix}
\begin{aligned}
&\mathcal{S}(c) := -\partial_c F(c) \cdot \partial_c^2 I_\text{ss}, 
\end{aligned}
\end{align}
{which is symmetric and positive semidefinite when evaluated in a stable fixed point:
\begin{align}
&{\mathcal{S}(\mathcal{x}^*)= \partial_c^2 I_\text{ss}(\mathcal{x}^*) \cdot D(\mathcal{x}^*) \cdot \partial_c^2 I_\text{ss}(\mathcal{x}^*) \;.}
\end{align}
We used \eqref{splitting_F}-\eqref{explictM} with \eqref{divergenceless} and \eqref{diffusion}.}
%{\bf Stability of fixed point becoming unstable (phase transitions)}

%{\bf Link to Glansdorf-Prigogine stability criterion: see equation $19$ in \cite{Nicolis1984} as well as https://arxiv.org/abs/1410.2183 as well}

%{\bf We should consider the time derivative of the EP rate (wish should be negative close to eq), as well as the ad and nonad contributions.} 
%%%%%%%%%%%%%%%%%%%%%%
\iffalse
By expanding \eqref{mean_nonadiabatic_autonomous} around the fixed point, we get
\begin{align}\label{mean_nonadiabatic_autonomousBis}
&d_t I_\text{ss}(\mathcal{x}(t)) 
= \partial_c I_\text{ss}(\mathcal{x}(t)) \cdot F(\mathcal{x}(t))  \\
&= (\mathcal{x}(t)-\mathcal{x}^*) \cdot \mathcal{S} \cdot (\mathcal{x}(t)-\mathcal{x}^*) \leq 0,\nonumber
\end{align}
where we used $F(\mathcal{x}^*)=0$ and $\partial_c I_\text{ss}(\mathcal{x}^*)=0$ and introduced the stability matrix
\begin{align}\label{StabMatrix}
\mathcal{S} := \partial_c^2 I_\text{ss}(\mathcal{x}^*) \cdot \partial_c F(\mathcal{x}^*) \;.
\end{align} 
\fi
%%%%%%%%%%%%%%%%%%%%%%%%% 

We now turn to a thermodynamically motivated linearization. 
Using \eqref{r_rho} and \eqref{ldb_macro}, we can recast the drift field \eqref{macro_drift} as
\begin{align}\label{F}
\begin{aligned}
\hspace{-0.2cm}F(c)=  2 \sum_{\rho>0}  \Delta_\rho \gamma_\rho(c) \sinh\left (\frac{1}{2}(-\Delta_\rho \cdot \partial_c \phi(c) + a_\rho) \right) 
\end{aligned}
\end{align}
which shows that $F$ is not a linear function of the thermodynamic forces, unless some limiting cases are considered.
First, when $\Delta_\rho$ and $a_\rho$ become small in an appropriate sense, as in the continuous-space limit treated in Sec. \ref{subsec:continuous}.
%---so that 
%\begin{align}
%F(c) \to   D(c) \cdot \partial_c \phi(c) + M(c) \cdot \mathcal{f}_{\text{nc}}
%\end{align}
%with $D(c)$ is the positive-definite, symmetric matrix 
%\begin{align}\label{diffusion}
%D(c):= \frac 1 2 \textstyle  \sum_{\rho} \Delta_{\rho} \Delta_{\rho} r_\rho(c)= \sum_{\rho >0} \Delta_{\rho} \Delta_{\rho} \gamma_\rho(c),
%\end{align}
%and $ M(c):=\sum_{\rho>0} \gamma_\rho(c) \Delta_\rho \mathbb{X}_\rho$ is the matrix coupling the state dynamics to the fundamental nonconservative forces $ \mathcal{f}_{\text{nc}}$.
%
%\begin{align}
%\partial_t \mathcal{x} = - \nabla \cdot \left[ \chi (\mathcal{x})\cdot \left(-   \nabla \frac{\delta \phi}{\delta \mathcal{x}}  +   f (\mathcal{x}) \right) \right]+ \sum_{\rho \in R} \Delta_\rho r_\rho(\mathcal{x})  .
%\end{align}
%Second, when the dynamics is detailed balanced and only small displacements %$\delta c:=\mathcal{x}-\mathcal{x}^* $ 
%from the equilibrium concentration $\mathcal{x}^*$ are considered. In this case Eq. \eqref{F} can be linearized in $\mathcal{x}-\mathcal{x}^*_{\gamma}$ as
%to give 
%\begin{align}
%d_t \mathcal{x}(t)= -\Gamma \cdot \mathcal{f}(\mathcal{x}(t)) 
%%(\mathcal{x}(t)-\mathcal{x}^*) 
%\end{align}
%where  $\Gamma:=  \sum_{\rho>0} \gamma^{(0)}_\rho(\mathcal{x}^*) \Delta_\rho \Delta_\rho$ is a symmetric positive-definite matrix  and $ \mathcal{f}(\mathcal{x}):=(\mathcal{x}- \mathcal{x}^*) \cdot \partial_c \partial_c \phi(\mathcal{x}^*)$ is the thermodynamic force. This thus retrieves the setting of linear irreversible thermodynamics.
Second, when the nonconservative force $a_\rho$ is small and only small displacements $\mathcal{x}(t)-\mathcal{x}^\text{eq} $ from the equilibrium state $\mathcal{x}^{\text{eq}}$ are considered \footnote{We restrict to autonomous dynamics for simplicity.}. In this latter case,  Eq. \eqref{F} can be linearized in $\mathcal{x}-\mathcal{x}^{\text{eq}}$ and $a_\rho$ 
to give 
\begin{align}
\begin{aligned}
\label{dcdt_linear}
d_t  \mathcal{x}(t) =(\mathcal{x}(t)- \mathcal{x}^{\text{eq}})  \cdot \partial_c F^{(0)}(\mathcal{x}^{\text{eq}}) 
+ M^{(0)}  \cdot \mathcal{f}_{\text{nc}}\;,
%(\mathcal{x}(t)-\mathcal{x}^*) \;. 
\end{aligned}
\end{align}
where, as can be verified using \eqref{macro_drift}, \eqref{eqRates} and \eqref{diffusion0}, 
\begin{align}
\begin{aligned}\label{linear_dbEq}
\partial_c F^{(0)}(\mathcal{x}^{\text{eq}}) 
&:= \lim_{a_\rho \to 0} \partial_c F(\mathcal{x}^{\text{eq}}) \\
&= - D^{(0)}(\mathcal{x}^{\text{eq}}) \cdot \partial_c \partial_c \phi (\mathcal{x}^{\text{eq}}) \;,
\end{aligned}
\end{align}
%\begin{align}\label{EqDiffusion}
%D^{(0)}:= \lim_{a_\rho \to 0} D(\mathcal{x}^{\text{eq}}) =  \sum_{\rho > 0} \Delta_{\rho} \Delta_{\rho} \gamma^{(0)}_\rho(\mathcal{x}^{\text{eq}}) \;,
%\end{align}
and the matrix coupling the state dynamics to the fundamental nonconservative forces $\mathcal{f}_{\text{nc}}$ reads
\begin{align}
M^{(0)}:=\sum_{\rho>0} \gamma^{(0)}_\rho(\mathcal{x}^{\text{eq}}) \Delta_\rho \mathbb{X}_\rho \;.
\end{align}
The fact that the symmetric positive-definite matrix $D^{(0)}(\mathcal{x}^{\text{eq}})$ appears in \eqref{dcdt_linear} is a statement of the Onsager reciprocal relations \cite{Onsager31, forastiere2022linear}.

For $a_\rho \neq 0$ the fixed point  $\mathcal{x}^*$ of the perturbed near-equilibrium dynamics differs from the equilibrium fixed point $\mathcal{x}^{\text{eq}}$ and is given by 
\begin{align}\label{stationary_point}
- (\mathcal{x}^*-\mathcal{x}^{\text{eq}}) \cdot \partial_c F^{(0)}(\mathcal{x}^{\text{eq}}) = M^{(0)}  \cdot \mathcal{f}_{\text{nc}},
\end{align}
which we can use to write \eqref{dcdt_linear} 
\begin{align}\label{linear_EqDyn}
d_t  \mathcal{x}(t)
= (\mathcal{x}(t)- \mathcal{x}^*) \cdot \partial_c F^{(0)}(\mathcal{x}^{\text{eq}}) \;.
\end{align}
This shows that \eqref{linear_dbEq} is a fluctuation-dissipation relation, since $\partial_c \partial_c \phi (\mathcal{x}^{\text{eq}})$ is related to the scaled correlations of the state variable (see section \ref{sec:gauss_instanton}) and $\partial_c F^{(0)}(\mathcal{x}^{\text{eq}})$ characterizes the rate of relaxation to equilibrium in an autonomous detailed balanced systems ($a_\rho=0$), as one can see using \eqref{linear_EqDyn} with $\mathcal{x}^{*}=\mathcal{x}^{\text{eq}}$. 
%\begin{align}
%\mathcal{x}(t)= e^{\Gamma^{(0)} t} \mathcal{x}(0)+\mathcal{x}^{\text{eq}}\;.
%\end{align}
For detailed balance systems we retrieve the setting of linear irreversible thermodynamics in which the only force is the (linearized) gradient of the Massieu potential  
\cite{Prigogine1961Jul,Groot1984}.

Turning now to the dynamics of the scaled nonadiabatic entropy production, using \eqref{mean_nonadiabatic_autonomousBis} to lowest order in $a_\rho$ and \eqref{linear_dbEq}, we find that \begin{align}\label{mean_nonadiabatic_autonomousEq}
\lim_{V \to \infty} \mean{\dot \sigma_\text{na}} = (\mathcal{x}-\mathcal{x}^*) \cdot \mathcal{S}^{(0)}(\mathcal{x}^{\text{eq}}) \cdot (\mathcal{x}-\mathcal{x}^*)  \geq 0 \;,
\end{align}
with the $a_\rho \to 0$ limit of the matrix \eqref{StabMatrix} 
\begin{align}
\label{StabMatrixZero}
\mathcal{S}^{(0)}(c) 
=\partial_c \partial_c \phi(c) \cdot D^{(0)}(c) \cdot \partial_c \partial_c \phi(c) \;,
\end{align}

The scaled entropy production rate corresponding to \eqref{dcdt_linear} follows by expanding \eqref{epr_macro} at second order in $\mathcal{x}-\mathcal{x}^{\text{eq}}$ and $a_\rho$,
\begin{align}\label{ep_linear}
\lim_{V \to \infty} \mean{\dot \sigma} = \sum_{\rho>0} \gamma^{(0)}_\rho (\mathcal{x}^{\text{eq}}) y_\rho(\mathcal{x}-\mathcal{x}^{\text{eq}}) y_\rho(\mathcal{x}-\mathcal{x}^{\text{eq}}),
\end{align}
with the forces (nonconservative and conservative) acting along $\rho$ 
\begin{align}
y_\rho(c):=a_\rho - \Delta_\rho \cdot \partial_c \partial_c \phi (\mathcal{x}^{\text{eq}}) \cdot c \;.
\end{align}
Rewriting $\mathcal{x}-\mathcal{x}^{\text{eq}} =(\mathcal{x}^*-\mathcal{x}^{\text{eq}})+(\mathcal{x}-\mathcal{x}^*)$ and using the stationary condition \eqref{stationary_point} to eliminate the mixed terms, \eqref{ep_linear} simplifies to
\begin{align}\label{ep_linear2}
\lim_{V \to \infty} \mean{\dot \sigma} %&=  \mathcal{f}_{\phi} (\mathcal{x}-\mathcal{x}^*) \cdot D^{(0)}\cdot \mathcal{f}_{\phi}(\mathcal{x}-\mathcal{x}^*) \\
=& (\mathcal{x}-\mathcal{x}^*) \cdot \mathcal{S}^{(0)}(\mathcal{x}^{\text{eq}}) \cdot (\mathcal{x}-\mathcal{x}^*) \\
&+ \sum_{\rho>0} \gamma^{(0)}(\mathcal{x}^{\text{eq}}) y_\rho(\mathcal{x}^*-\mathcal{x}^{\text{eq}}) y_\rho(\mathcal{x}^*-\mathcal{x}^{\text{eq}}). \nonumber
\end{align}
%where $\mathcal{f}_{\phi}(c)=\partial_c \partial_c \phi(\mathcal{x}^{\text{eq}}) \cdot c$.
In view of \eqref{mean_nonadiabatic_autonomousEq}, this is nothing but the nonadiabatic-adiabatic decomposition of the scaled entropy production \eqref{ad_nonad_epr} close to equilibrium.
%{\bf I have the suspicion that in this limit the adiabatic nonadiabatic decomposition coincides with nonconservative conservative one, because $I$ and $\phi$ become identical.}
The first (nonadiabatic) contribution in the right-hand side of \eqref{ep_linear2} describes the nonegative entropy produced as the system relaxes to the nonequilibrium steady state $\mathcal{x}^*$.
The second (adiabatic) one describes the nonnegative entropy production to sustain that steady state.
Hence, the structure of steady state thermodynamics proposed by Oono-Paniconi is recovered here 
\cite{oon98}.
{
Furthermore, we retrieve the minimum entropy production principle since \eqref{ep_linear2} states that the fixed point $\mathcal{x}^*$ minimizes the entropy production rate among all states solutions of \eqref{dcdt_linear} \cite{Nicolis1984,Prigogine1961Jul}. In general, this result holds true only for states $\mathcal{x}^*$ close to detailed balance and linear dynamics. Nevertheless, we will present in Sec.~\ref{sec:ex3} a class of model systems in which the minimum entropy production principle is valid far from equilibrium. Additionally, computing the time derivative of \eqref{ep_linear2} with the aid of \eqref{linear_EqDyn},    
\begin{align}\label{derivative_ep_linear}
d_t \lim_{V \to \infty} \mean{\dot \sigma} 
&= 2 d_t\mathcal{x} \cdot \mathcal{S}^{(0)}(\mathcal{x}^{\text{eq}}) \cdot (\mathcal{x}-\mathcal{x}^*) \\
&=- 2 z \cdot \partial_c \partial_c \phi (\mathcal{x}^{\text{eq}}) \cdot  z^T   \leq 0
\nonumber,
\end{align}
with $z=(\mathcal{x}-\mathcal{x}^*) \cdot \partial_c F^{(0)}(\mathcal{x}^{\text{eq}})$, we conclude that the entropy production rate monotonically decreases along solutions of \eqref{linear_EqDyn} \cite{maes2015revisiting}. 
}
\begin{comment}
since from 
\begin{equation}
\label{ep_gauss}
\lim_{V \to \infty} \mean{\dot \sigma} \geq 
%\mathcal{f}_{\phi} (\mathcal{x}-\mathcal{x}^*) \cdot D^{(0)}\cdot \mathcal{f}_{\phi}(\mathcal{x}-\mathcal{x}^*) \geq 0 ,
(\mathcal{x}-\mathcal{x}^*) \cdot \mathcal{S}^{(0)}(\mathcal{x}^{\text{eq}}) \cdot (\mathcal{x}-\mathcal{x}^*) \geq 0 ,
\end{equation}
\end{comment}

\section{Macroscopic fluctuations}\label{sec:gauss_instanton}

In this section we describe the asymptotic stochastic dynamics and the associated thermodynamics. 

\subsection{Dynamics of the state variable}\label{sec:hamiltonian_dyn}

The macroscopic master equation \eqref{MEpc} can be recast as
\begin{align}\label{MEexp}
 \partial_t p(c,t)&= V  \mathcal{H} \left(c, - V^{-1} \partial_c \right) p(c,t),
 \end{align}
by introducing the (scaled) generator of the stochastic dynamics 
\begin{align}\label{H_operator}
\mathcal{H} (c, -V^{-1}\partial_c) =\sum_\rho\! \left[ e^{-V^{-1}\Delta_\rho \cdot \partial_c}-1 \right]r_\rho(c),
\end{align}
 which contains the operator $e^{-V^{-1} \Delta_\rho \cdot \partial_c}$ that shifts by $-V^{-1}\Delta_\rho $ the argument of the function it is applied to. 
Its identification allows us to switch to an equivalent representation of the stochastic dynamics, consisting in the probability density of stochastic trajectories conditioned on the initial value $c(0)$, namely,  the ordered set of states $c(t)$ in some time interval $[0,\tau] \ni t$ \cite{doi1976second,dedominicis78,grassberger1980fock,peliti1985path,WeberFrey}:
\begin{align}\nonumber
P[\{c(t)\}| c(0)] &=  \int \mathcal{D}\pi \, e^{V\! \int_0^\tau dt \left[- \pi(t) \cdot d_t c(t) + \mathcal{H} (c(t),\pi(t)) \right]}\\
&= \int \mathcal{D}\pi  \,e^{V \mathcal{A}[\{c(t)\},\{\pi(t)\}]} .
\label{Ptraj}
\end{align}
Here $\pi$ is an auxiliary variable to integrate out  in order to pass from the (Poissonian) generating function of the transitions, $e^{V \! \int_0^\tau dt \mathcal{H} (c,\pi)}$, to the probability distribution of trajectories \cite{gaveau1999variational,lefevre2007dynamics}. Equation \eqref{Ptraj} can be formally obtained by applying a time-slicing to the solution of \eqref{MEexp} -- that is $p(c,t)= e^{V \! \int_0^\tau dt \mathcal{H} (c,-V^{-1}\partial_c)} p(c,0)$, with a time-ordered exponential -- very analogously to the derivation of the quantum path integral representation of the Schr\"{o}dinger equation \footnote{More rigorously, one can see \eqref{action} as the action associated to the Hamilton-Jacobi equation defined by \eqref{HJeq}.}.
Because of the appearance of $V$ in the exponential, the functional integral in \eqref{Ptraj} is dominated for large $V$ by the trajectories that maximize the action functional  \cite{dykman1994large, smith2011large,Lazarescu2019Aug}
\begin{align}\label{action}
\mathcal{A}=   \int_0^\tau dt  \bigg[- \pi(t) \cdot d_t c(t) + \sum_\rho r_\rho(c(t))  (e^{\Delta_\rho \cdot \pi(t) } -1 )\bigg]   ,
\end{align} 
supplemented by the appropriate boundary conditions \cite{Lazarescu2019Aug}.
Namely, the solutions of the Hamiltonian equations
\begin{align}\label{hamilton_eq}
\begin{aligned}
 d_tc&= \partial_\pi  \mathcal{H}(c,\pi)= \sum_\rho \Delta_\rho r_\rho(c) e^{\Delta_\rho \cdot \pi } \\
 d_t\pi&=-\partial_c \mathcal{H}(c,\pi) =-\sum_\rho \partial_c r_\rho(c) \left(e^{\Delta_\rho \cdot \pi} -1\right), 
\end{aligned}
\end{align}
are the most likely paths, which when inserted into \eqref{action}, give their (exponentially small) probability 
\begin{align}\label{P_action_initialI}
%P[\{c(t)\}]=P[\{c(t)\}|c(0)]p(c(0),0)\asymp e^{V\{ \mathcal{A}[\{c(t),\pi(t)\}]- I(c(0),0)\}}.
P[\{c(t)\}|c(0)]p(c(0),0)\asymp e^{V\big( \mathcal{A}[\{c(t),\pi(t)\}]- I(c(0),0)\big)}.
\end{align} 

While small fluctuations can be described by Langevin dynamics (whose range of validity is discussed in \ref{sec:Onsager_Machlup}), large deviations from the average dynamics \eqref{macro_dcdt} are correctly described only by \eqref{hamilton_eq}. In particular, an important class of solutions is that of fluctuating trajectories, called instantons \cite{coleman1988}, that connect in a long (formally infinite) time the attractor $\mathcal{x}^*$ to an arbitrary state $c$ in the same basin. These trajectories, starting from a stable fixed point\footnote{{In infinite time, the most likely trajectory starting from any point}{of the attractor displays an initial relaxation to the local stable fixed point in which the velocity is null. This deterministic relaxation nullifies the action, thus can be neglected and the initial condition can be directly placed in the fixed point.}}, $c(0)=\mathcal{x}^*_\gamma$ , are characterized by $\pi(0)=0$ and thus $\mathcal{H} (c(t),\pi(t))=0$ for all $t$, since $\mathcal{H}$ is a constant of motion in absence of explicit time dependence of the transition rates $r_\rho$. 
It follows from \eqref{DualhamiltonianAtSS} that $\pi= \partial_c I_\text{ss}$ on the manifold $\mathcal{H}=0$. 
Hence, instantons are solutions of 
\begin{align}\label{instanton_dynamics}
d_t c(t) =\sum_\rho \Delta_\rho r_\rho(c(t))e^{\Delta_\rho \cdot \partial_c I_\text{ss}(c(t)) }=-F^\dagger(c(t))
\end{align}
namely, they are the time reversal of the dual-dynamics trajectories $\mathcal{x}^\dagger(t)$ introduced in \eqref{eq:macro_dual_dyn}.

As a consequence, the long-time transition probability from $\mathcal{x}^*_{\gamma}$ to $c$ -- that is the local stationary probability in the basin of attraction $\gamma$, $p^{(\gamma)}_\text{ss}(c)$  -- is related at the leading order to the difference of the local  quasi-potential \cite{bouchet2016generalisation},
\begin{align}\label{trans_rate}
&\lim_{t \to \infty}P(c(t)=c|c(0)=\mathcal{x}^*)= p^{(\gamma)}_\text{ss}(c) \\ &\hspace{1cm}
\asymp e^{- V \int_{c(0)=\mathcal{x}^*}^{c(\infty)=c} dt \pi(t) \cdot d_t c(t)}
= e^{- V[I^{(\gamma)}_\text{ss}(c)-I^{(\gamma)}_\text{ss}(\mathcal{x}^*_\gamma)] }. \nonumber
\end{align}

We note that the integral in \eqref{trans_rate} can be multivalued. This means that a given state $c$ can correspond to different values of $\pi$ on the instanton. Such Hamiltonian trajectory generates folds, called caustics, when projected onto the state space \cite{graham1985weak,Maier1993effect,luchinsky1998analogue,Nardini2016}. In this case, $I_\text{ss}^{\gamma}(c)$ is obtained by taking the minimum value over $\pi$ of the integral in \eqref{trans_rate}\footnote{A caustic, in analogy with geometrical optics, is then the set of states $c$ with multiple minimizers.}. Since such minimizer can jump from one branch of the instanton to another when we move away from $c$ in different directions of the state space, the local quasi-potential is generally not differentiable in such points (unless the dynamics is detailed balance), a phenomenon called Lagrangian phase transition \cite{bertini2010lagrangian}. 

\subsection{Fluctuating Thermodynamics}\label{sec:macro_thermo}

In this subsection, we formulate the second law along single fluctuating trajectories and use it to derive a bound on the variation of the rate function. 
To do so, we first present the asymptotic dynamics that describe the evolution of the states \emph{and} the single transitions jointly. 
Because of \eqref{lambda}, the transition flux is expected to asymptotically scale linearly with $V$, i.e., $ J_\rho(t)= V j_\rho(t) +o(V)$, with $ j_\rho(t)$ independent of $V$. So, dividing both sides of \eqref{eq_ndot} by $V$, we obtain the asymptotic stochastic evolution of the state vector 
\begin{align}\label{eq_cdot}
d_t c(t) = \sum_\rho \Delta_\rho j_\rho(t).
\end{align}
Since the flux $ j_\rho(t)$ can be described by a Poissonian distribution conditioned on the present state $c(t)$ with average $r_\rho(c(t))$, the average value of \eqref{eq_cdot} coincides with the equation \eqref{dcdt} for the mean concentration obtained from the asymptotic form of the master equation \eqref{MEexp}.

%Since the mesoscopic dynamics enjoys the condition of detailed balance \eqref{micro_ldb}, the entropy production in the time interval $[0,\tau] \ni t$ is defined as the functional $\Sigma[\mathcal{X}]= \sum_0^\tau dt \dot \Sigma(t)$ with entropy production rate
%\begin{align}\label{EP}
%\dot \Sigma(t)=\!  {\textstyle \sum_\rho} J_\rho(t) \Sigma_\rho(n(t)) - d_t \ln P(n(t),t)
%- \ln\frac{P(n(\tau),\tau)}{P(n(0),0)}.
%\end{align}
%Here $\Sigma_\rho(n(t)) $ is defined in \eqref{micro_ldb}, and the second term in \eqref{EP} amounts to the variation of the stochastic entropy (or self-information) of the probability distribution, $-\ln P(n,t)$. 

Using the scaling of the flux $J_\rho$ and of the probability distribution \eqref{rateI}, the scaled entropy production rate \eqref{epr} takes the asymptotic form
\begin{equation}\label{ep}
\begin{aligned}
\dot \sigma(t) & %\underset{V \to \infty}{=} 
=\!{\textstyle \sum_\rho} j_\rho(t)\sigma_\rho(c(t))  + d_t I(c(t),t) .
\end{aligned}
\end{equation}
The decomposition \eqref{nc_driv_epr} becomes
\begin{equation}\label{nc_driv_epr_macro}
\begin{aligned}
 \dot \sigma(t)%&= \textstyle{  \sum_\rho} j_\rho (t)[a_\rho+\Delta_\rho \cdot \partial_c \phi(c(t)) ] \\
 &= \dot \sigma_{\text{nc}} (t) +\dot \sigma_\text{d}(t)+d_t (I-\phi) (c(t),t),
\end{aligned}
\end{equation}
namely, it displays the scaled dissipation rate due to nonconservative forces $ \dot \sigma_{\text{nc}} (t) := \sum_\rho j_\rho(t) a_\rho$ and the scaled dissipation rate needed to parametrically drive the reference equilibrium, $\dot \sigma_\text{d}(t):=-\partial_t \phi (c,t)|_{c=c(t)} $, plus the total time derivative of the scaled stochastic Massieu potential $ I(c,t)-\phi(c,t) $.
Note the presence of the scaled variation of self-information on the righthand side of \eqref{ep} and \eqref{nc_driv_epr_macro}. While it vanishes on average, it is nonzero along macroscopic fluctuating trajectories so that
\begin{equation}
\begin{aligned}
\lim_{V \to \infty} \frac 1 V S_\text{sys}(n(t),t)& =  I(c(t),t)+s_\text{int}(c(t)) .
\end{aligned}
\end{equation}

%For systems with nonautonomous dynamics, i.e. with rates $r_\rho(c;t)$ that depends parametrically on time, or with relaxation dynamics, i.e. time-independent $r_\rho(c)$ but initial rate function $I(c,t) \neq I_\text{ss}(c)$, it is convenient to split the entropy production into adiabatic and nonadiabatic components. 
%For the macroscopic systems considered here, the splitting can be implemented directly on the scaled entropy production rate, i.e. the time derivative of the functional \eqref{ep}. 
The decomposition \eqref{ad_nonad_epr} becomes 
\begin{align}\label{ad_nonad_epr_macro}
\dot \sigma(t)&={\textstyle\sum_\rho } j_\rho(t) \sigma^\text{ad}_\rho+\dot \sigma_\text{na},
\end{align}
%where it is useful to identify the adiabatic entropy production in each transition $\rho$,
%\begin{align}\label{ad_rho_macro}
%\sigma^\text{ad}_\rho(t)&:= \ln  \frac{r_\rho(c(t))}%{r_{\rho}^{\dagger}(c(t))} \\
%&=\Delta_\rho \cdot \partial_c[I_\text{ss}^{t}(c)-%\phi(c,t)]|_{c=c(t)} +a_\rho \;,\nonumber
%\end{align}
using the scaled adiabatic entropy production of each transition, \eqref{ldb_macro_a}, and the scaled nonadiabatic entropy production rate
\begin{align}\label{na_rho_macro}
\dot \sigma_\text{na}(t):=
d_t[I(c(t),t)-I_\text{ss}^{t}(c(t))] +\partial_t I_\text{ss}^{t}(c)|_{c=c(t)}.
\end{align}
Again, for autonomous systems at stationarity $I(c,t)=I_\text{ss}(c)=I_\text{ss}^t(c)$, so that \eqref{na_rho_macro} 
 is identically zero and the scaled adiabatic entropy production rate 
 \begin{align}\label{ad_rho_macro}
\dot \sigma_\text{ad}(t)= {\textstyle\sum_\rho } j_\rho(t)\sigma^\text{ad}_\rho(t) 
\end{align} 
equals $\dot \sigma(t)$. And for nonautonomous detailed balance dynamics $I_\text{ss}^t(c)=\phi(c,t)+\mathrm{const}$, so that \eqref{ad_rho_macro} 
is identically zero for all $\rho$  and the scaled nonadiabatic entropy production rate $\dot \sigma_\text{na}(t)$ equals $\dot \sigma(t)$.
The adiabatic-nonadiabatic decomposition \eqref{ad_nonad_epr_macro} can be used to provide a strengthening of the second law of thermodynamics. 
This follows from averaging \eqref{na_rho_macro} and taking the limit $V \to \infty$
%, $\mean{\dot \sigma_\text{na}(t)}=-d_t I_\text{ss}^t(\mathcal{x}(t)) + \partial_t I_\text{ss}^t(\mathcal{x}(t))$, which 
combined with $\mean{\dot \sigma_\text{na}} \geq 0$ and $\mean{\dot \sigma_\text{ad}} \geq 0$, %(see Appendix \ref{app:ad_ep})
which yield
\begin{align}\label{time_emerging_2nd}
\lim_{V \to \infty} \mean{\dot \sigma(\mathcal{x}(t))} \geq -d_t I_\text{ss}^t(\mathcal{x}(t)) + \partial_t I_\text{ss}^t(c)|_{c=\mathcal{x}(t)} \geq 0. 
\end{align}
In particular, for autonomous relaxation dynamics the equality $I_\text{ss}^t=I_\text{ss}$ simplifies \eqref{time_emerging_2nd} to \cite{Freitas2021NatCom}
\begin{align}\label{emerging_2nd}
\lim_{V \to \infty} \mean{\dot \sigma(\mathcal{x}(t))} \geq -d_t I_\text{ss}(\mathcal{x}(t))  \geq 0,
\end{align}
where the last inequality is the Lyapunov property \eqref{mean_nonadiabatic_autonomous}. Equation \eqref{emerging_2nd} is the nonlinear extension of \eqref{ep_gauss} and first appeared in \cite{gaveau1998dissipation} for reaction-diffusion systems.
Note that $I_\text{ss}$ is replaced by the local quasi-potential $I_\text{ss}^{(\gamma)}$ if the initial condition of the dynamics is localized on a single basin of attraction $\gamma$, a fact that will be used in Sec. \ref{sec:bounds_kappa}. 
The first inequality in \eqref{emerging_2nd} becomes tight when $a_\rho$ is infinitesimal \cite{falasco21local,Freitas2021NatCom}. In that limit the  quasi-potential is still determined by thermodynamics. Indeed, a first order expansion in $a_\rho$ of \eqref{HJ0} gives within each basin $\gamma$
\begin{align}\label{linear_resp}
I^{(\gamma)}_\text{ss}(c) = \phi(c) +\sigma_\text{nc}^{(0)}(c,\mathcal{x}^\text{eq}_{\gamma}) ,
\end{align}
where 
\begin{align}
\sigma_\text{nc}^{(0)}(c,\mathcal{x}^\text{eq}_{\gamma}):= \int_{\mathcal{x}(0)=c}^{\mathcal{x}(\infty)=\mathcal{x}_\gamma^\text{eq}} dt  \sum_\rho r^{(0)}_\rho (\mathcal{x}(t))a_\rho 
\end{align}
is the dissipation of nonconservative forces along the solution of detailed-balanced deterministic dynamics
\begin{align}\label{equilibrium_relaxation}
d_t \mathcal{x}(t)= \sum_\rho \Delta_\rho r^{(0)}_\rho(\mathcal{x}(t)) ,
\end{align}
connecting $\mathcal{x}(0)=c$ to the equilibrium fixed point $\mathcal{x}(t \to \infty)=\mathcal{x}_{\gamma}^\text{eq}$ \cite{falasco21local,freitas2021linear}.
Note that \eqref{linear_resp} is valid for any $c$ in the basin of attraction $\gamma$, not only close to $\mathcal{x}^*_\gamma$.
The linear response formula \eqref{linear_resp} yields the large-deviation form of McLennan-Zubarev probability distribution function \cite{mclennan1959,zubarev1994,Maes10mclennan,Colangeli11expansion} by replacing the average of the near-equilibrium dissipation over all trajectories with its most likely value.
 
By relating the macroscopic entropy production during relaxation to steady state macroscopic fluctuations, the inequalities \eqref{time_emerging_2nd} and \eqref{emerging_2nd} can thus be seen as generalized fluctuation-dissipation relations.
Standard fluctuation-dissipation relations are recovered using a parabolic approximation of $I^{(\gamma)}_\text{ss}(c)$ around an equilibrium fixed point \cite{freitas2021linear}, which is equivalent to the alternative derivation based on the Langevin approximation given in Sec. \ref{sec:Onsager_Machlup}. 
Finally, the nonadiabatic entropy production provides a second (upper) bound valid for rare fluctuations (instanton), that is derived in Appendix \ref{app:bound} and will be used in Section \ref{sec:bounds_kappa}.

\subsection{Gaussian fluctuations}\label{sec:Onsager_Machlup}

When comparing \eqref{hamilton_eq} with \eqref{macro_dcdt}, we recognize that the solution $\pi(t)=0$, giving $\mathcal{A}=0$, corresponds to the deterministic trajectories. 
This suggests that small fluctuations around the deterministic behavior are characterized by small values of $\pi(t)$. In particular, Gaussian fluctuations around each deterministic solution can be obtained by expanding the action \eqref{action} around $\pi(t)=0$ and $c(t)=\mathcal{x}(t)$ to quadratic order in $\pi(t)$ and $\varrho(t) := c(t)-\mathcal{x}(t)$, respectively:
\begin{align}\label{Action_Gaussian}
\mathcal{A}_\text{G}= 
 \int_0^\tau \! dt \bigg[- \pi \cdot d_t \varrho + \varrho \cdot \partial_c F(\mathcal{x}) \cdot \pi +\pi \cdot D(\mathcal{x}) \cdot \pi  \bigg].
\end{align}
This approximation holds only for times much shorter than the escape time from a basin of attraction, which will be discussed in Sec. \ref{sec:bounds_kappa}.
It is equivalent to the first order in van Kampen's system size expansion {of \eqref{MEpc}}  \cite{vanKampen}, known as linear noise approximation \cite{thomas2015}. The action \eqref{Action_Gaussian} {contains the scaled generator 
\begin{align}
    \mathcal{H}_\mathrm{FPE}(\varrho,\pi)= \varrho \cdot \partial_c F(\mathcal{x}) \cdot \pi +\pi \cdot D(\mathcal{x}) \cdot \pi,
\end{align}
that depends parametrically on time via $\mathcal{x}(t)$ and
 corresponds to the Fokker-Planck equation
\begin{align}\label{FPElinear}
 \partial_t  p(\varrho,t) = -\partial_\varrho \cdot \left[\varrho \cdot \partial_c F(\mathcal{x})p(\varrho,t) - \frac{D(\mathcal{x})}{V}\partial_\varrho p(\varrho,t) \right].
\end{align}
The associated linear Langevin equation with Gaussian additive noise $\eta$ reads}
\begin{align}\label{linear_langevin}
d_t \varrho(t)=  \varrho(t) \cdot \partial_c F(\mathcal{x}(t)) +\frac{1}{ \sqrt{V}} \eta(t).
\end{align}
%\begin{align}\label{linear_langevin}
%d_t\varrho(t)= \varrho(t) \cdot \partial_c F(\mathcal{x}(t)) +\frac{1}{ \sqrt{V}} \eta(t).
%\end{align}
Here $\eta$ is delta-correlated, with mean zero and covariance matrix $2 D(\mathcal{x}(t))$ defined in \eqref{diffusion} and $\partial_c F$ is the Jacobian matrix of the deterministic drift $F$ given in \eqref{macro_drift}. Equation \eqref{linear_langevin} also amounts to a quadratic approximation of the rate function as 
\begin{align}
I(\varrho,t) = \frac 1 2  \varrho \cdot \partial_c^2 I (\mathcal{x}(t)) \cdot \varrho + O(\varrho^3)\;,
\end{align}
where the scaled covariance matrix $\mathcal{C}=(\partial_c^2 I)^{-1}$ is obtained multiplying \eqref{linear_langevin} by $\varrho(t)$ and averaging over $\eta$,
\begin{align}
\frac{d\mathcal{C}}{dt}(t)=& \partial_c F^T(\mathcal{x}(t))  \cdot \mathcal{C}(t)+\mathcal{C}(t)  \cdot \partial_c F (\mathcal{x}(t)) \nonumber\\
 &  + 2 D(\mathcal{x}(t)), 
 \label{dC/dt}
\end{align}
with $\partial_c F^T$ the transpose of $\partial_c F$ \cite{tomita1974irreversible}. When the expansion is carried out around a limit cycle $\mathcal{x}(t)=\mathcal{x}^*(t)$, \eqref{linear_langevin} allows one to study transversal and longitudinal Gaussian fluctuations \cite{Dykman1993Stationary,vance1996fluctuations,boland2008limit,Sheth2018}. The latter are unsuppressed, i.e., free diffusion takes place along the limit cycle with the effective diffusion coefficient proportional to $ \tau_\text{p}'/2V $, $\tau_\text{p}'$ being the period variation of the Hamiltonian orbit upon a small perturbation of {$\mathcal{H}_\mathrm{FPE}$} \cite{gaspard02,gaspard2002trace}. These fluctuations are stochastic Goldstone modes since they cause the decay in time of correlation functions and ultimately restore the time-translation invariance of the microscopic dynamics, which is spontaneously broken in the limit $V \to \infty$ \cite{SmithMorowitz2016}. 
Recently, the relation between the number of coherent oscillations and the thermodynamic force \cite{Remlein2022coherence} -- or the entropy production \cite{marsland2019} -- has been studied.

When the expansion is performed around a stable fixed point $\mathcal{x}^*_{{\gamma}}$, \eqref{dC/dt} simplifies to the steady state fluctuation-dissipation theorem \cite{keizer1978thermodynamics,dykman1994large,pro09},
\begin{align}\label{fluctuation_dissipation_ss}
\partial_c F^T(\mathcal{x}^*_\gamma)  \cdot \mathcal{C}_\text{ss}+\mathcal{C}_\text{ss}  \cdot \partial_c F (\mathcal{x}^*_\gamma)   =- 2 D(\mathcal{x}^*_\gamma) \;,
\end{align}
also known as Lyapunov equation, with $\mathcal{C}_\text{ss}=(\partial_c^2 I_\text{ss})^{-1}{(\mathcal{x}^*_\gamma)}$.  
Multiplying \eqref{fluctuation_dissipation_ss} by $\mathcal{C}_\text{ss}^{-1}$ and taking the trace, we obtain the relation
\begin{align}\label{contraction_fluctuations}
\lambda_\gamma =  D(\mathcal{x}^*_\gamma) : \mathcal{C}_\text{ss}^{-1}
\end{align}
connecting the volume contraction rate \eqref{contraction_rate} on the attractor, $ \lambda_\gamma:=-\partial_c \cdot F(\mathcal{x}^*_\gamma)$,  to the state and noise covariance.

Close to detailed balance dynamics, \eqref{linear_langevin} can be recast using \eqref{stationary_point} and \eqref{linear_dbEq} as 
\begin{align}\label{linear_langevin_OM}
d_t \varrho(t)=  \varrho(t) \cdot \partial_c F^{(0)}(\mathcal{x}^{\text{eq}})+\frac{1}{ \sqrt{V}} \eta(t),
\end{align}
which adds to \eqref{linear_EqDyn} small fluctuations driven by the Gaussian noise $\eta$ with covariance equal to $2D^{(0)}$.
For $a_\rho=0$ (i.e. $\mathcal{x}^*=\mathcal{x}^{\text{eq}}$), \eqref{linear_langevin_OM} corresponds to the linear theory of fluctuations introduced by Onsager and Machlup \cite{onsagermachlup53}, and \eqref{fluctuation_dissipation_ss}, using \eqref{linear_dbEq}, reduces to 
$\mathcal{C}^{-1}_\text{ss} = \partial_c^2 \phi(\mathcal{x}^{\text{eq}})$.
%\eqref{fluctuation_dissipation_eq}.
%Finally, in Sec. \ref{sec:path_langevin} we derive that the righthand side of \eqref{ep_gauss} is the information-theoretic entropy production  associated to $c(t)=\mathcal{x}(t)+\varrho(t)$, with $\mathcal{x}(t)$ and $\varrho(t)$ solutions of \eqref{dcdt_linear} and \eqref{linear_langevin_OM}, respectively. It is always smaller than the thermodynamic entropy production unless all $a_\rho$ are identically zero.

\subsection{Truncation of the Kramers-Moyal expansion}\label{sec:KM_truncation}

It is worth stressing that an expansion of \eqref{action} which is quadratic in $\pi(t)$ but retains nonlinearities in $c(t)$ is equivalent to an uncontrolled truncation of the full Kramers-Moyal expansion of the master equation \cite{kampen1961power}. 
Note that the derivation of \eqref{HJeq} hinges on the Taylor expansion of the rate function as
\begin{align}
p {\textstyle \left(c - V^{-1}\Delta_\rho,t \right)} = p(c,t) e^{\Delta_\rho \cdot \partial_c I(c,t) + O(V^{-1})}    ,
\end{align}
which is in general different from a direct expansion of the probability density
\begin{align}\label{wrong_exp}
p {\textstyle \left(c - V^{-1}\Delta_\rho,t \right)} \approx \left[1 - \frac{\Delta_\rho}{V} \cdot \partial_c + \frac{1}{2 V^2} ( \Delta_\rho  \cdot \partial_c)^2 \right] p(c,t).
\end{align}
Equation \eqref{wrong_exp} wrongly assumes that $p(c,t)$ remains a smooth function as $V$ approaches infinity and reduces \eqref{MEexp} to the Fokker-Planck equation
\begin{align}\label{chemical_FPE}
 \partial_t p(c,t)& \approx - \partial_c \cdot \left \{F(c) p(c,t) -\frac 1 V \partial_c \cdot [D(c) p(c,t)] \right\}.
\end{align}
Albeit generally incorrect, this approximation is often used. Equation \eqref{chemical_FPE} becomes accurate if $\Delta_\rho$ or $\partial_c I(c,t)$ are infinitesimal, namely, when a certain continuous limit exists (as in Sec.\ref{subsec:continuous}) or when only Gaussian fluctuations are considered, as in Sec. \ref{sec:Onsager_Machlup}. {The latter case corresponds to the linearization  around a deterministic solution $\mathcal{x}$ of the drift field and the diffusion matrix in \eqref{chemical_FPE}, which yields \eqref{FPElinear}.}

Equation \eqref{chemical_FPE} corresponds to the Ito nonlinear Langevin equation \cite{gillespie2000chemical,Vastola_2020}
\begin{align}\label{nonlinear_langevin}
d_tc(t)=  F(c(t)) +  \frac{1}{\sqrt{V}}\eta(t)
\end{align}
with multiplicative Gaussian noise $ \eta(t)$ having covariance $2D(c(t)) $. This equation in general does not correctly captures the fluctuations of $c(t)$ beyond the Gaussian level. Indeed, as already discussed for some specific models \cite{hanggi1988,gaveau1997master,vellela2009stochastic,Gopal2022}, the approximation \eqref{nonlinear_langevin} mistakes the rate function $I_\text{ss}$ away from $\mathcal{x}^*_\gamma$ and $\mathcal{x}_\nu$.
The discrepancy can be made explicit for one-step processes in one dimension, i.e. $\Delta_{\pm \rho}= \pm 1$ and $N=1$, whose exact- and diffusion-approximated rate function can be obtained analytically. For such systems \eqref{HJ0} reads
\begin{align}\label{one_step}
\sum_{\rho>0} r_{\rho} (c)(e^{d_c I_\text{ss}(c)}-1) = \sum_{\rho>0} r_{-\rho}(c) (1-e^{-d_c I_\text{ss}(c)}) ,
\end{align}
which is solved by 
\begin{align}\label{one_step_Iss}
d_c I_\text{ss}(c)= \ln \frac{\sum_{\rho>0} r_{-\rho} (c)}{\sum_{\rho>0} r_{\rho} (c)},
\end{align}
while the expansion of \eqref{one_step} to second order in $d_c I_\text{ss}$ (corresponding to \eqref{nonlinear_langevin}) has solution
\begin{align}\label{one_step_Iss_NLE}
 d_c I_\text{NLE}(c)= \frac{\sum_{\rho>0} [r_{-\rho} (c)-r_{\rho} (c)]}{\frac 1 2\sum_{\rho>0} [r_{\rho} (c)+r_{-\rho} (c)]}= -\frac{F(c)}{D(c)},
\end{align}
as can be directly verified by substitution. {With the subscript NLE we denote hereafter quantities corresponding to the nonlinear Langevin equation \eqref{nonlinear_langevin}.}
Hence, for this simple model it is easy to see that $I_\text{NLE}(c) \neq I_\text{ss}(c)$ unless $c$ is infinitesimally close to $\mathcal{x}^*_\gamma$ or $\mathcal{x}_\nu$. Indeed, in a neighborhood of the fixed points where $F=\varepsilon \to 0$, it holds that $\sum_{\rho>0}  r_{-\rho}=\sum_{\rho>0} r_\rho +\varepsilon$ and thus $d_c I_\text{ss} =\varepsilon/ (\sum_{\rho>0} r_\rho)  = d_c I_\text{NLE}$ at first oder in $\varepsilon$. The use of equation \eqref{one_step_Iss} will be exemplified in Secs. \ref{sec:ex1} and \ref{sec:ex2_CMOS}.

\subsection{Informational entropy production of Langevin equations}\label{sec:path_langevin}

The nonlinear Langevin approximation of Sec. \ref{sec:KM_truncation} is also thermodynamically inconsistent in addition to being incorrect to describe fluctuations beyond the Gaussian level.
 %the nonlinear drift field $F(c)$ and the noise covariance $D(c)$ do not satisfy the Einstein relation.
Here we derive the apparent entropy production associated to \eqref{nonlinear_langevin} and show that its mean value differs from \eqref{epr_macro}. In particular, it vanishes in any stationary state. In Sec. \ref{sec:macro_ft} we will also explicitly show that \eqref{nonlinear_langevin} breaks the fluctuation theorem for currents.
As discussed in Sec. \ref{subsec:continuous}, the nonlinear Langevin approximation of an underlying jump process is accurate only in a continuous limit that restores the validity of the Einstein relation.

Given a nonlinear Langevin equation of the type \eqref{nonlinear_langevin}, we can write the associated path probability 
$P[\{c(t)\}| c(0)]$ starting from the  conditional Gaussian weight of the noise $P[\{\eta(t)\}|c(t)] \propto e^{-\frac 1 4 \int_0^\tau dt \eta(t) \cdot D^{-1}(c(t)) \cdot \eta(t)}$ and implementing a change of variables $\eta(t) \mapsto c(t)$ \cite{Wiegel,zinn02}:
\begin{align}\label{path_langevin}
P[\{c(t)\}| c(0)] \asymp e^{- \frac V 4 \int_0^\tau dt [d_t c-F(c)] \cdot D^{-1}(c) \cdot[d_t c-F(c)]}.
\end{align} 
Here we disregarded the functional Jacobian $|\frac{\delta \eta}{ \delta c}|$ that is sub-exponential in $V$. This is equivalent to the statement that the choice of the stochastic calculus to handle \eqref{nonlinear_langevin} is irrelevant at leading order in $V$. Also the dependence on time are not explicitly written to avoid clutter. 
Note that the exponent in \eqref{path_langevin} can be made linear in $d_t c -F$ by means of a Hubbard-Stratonovich transformation \cite{negele2018quantum}, i.e. by introducing the auxiliary Gaussian field $\pi(t)$ such that
\begin{align}\label{path_HS_langevin}
P[\{c(t)\}| c(0)] \asymp \int \mathcal{D} \pi \, e^{-V\int_0^\tau dt[\pi \cdot D \cdot \pi - \text{i} \pi \cdot (d_t c- F) ] },
\end{align} 
with $\mathcal{D} \pi $ the appropriately normalized measure.

The path probability associated to a time reversed trajectory satisfying \eqref{nonlinear_langevin} is obtained by swopping the sign of the time derivative in \eqref{path_langevin},
\begin{align}\label{path_langevin_timerev}
\overline{P}[\{c(t)\}| c(0)] \asymp e^{-V \frac 1 4 \int_0^\tau dt [d_t c+F] \cdot D^{-1} \cdot[d_t c+F]}.
\end{align} 
The scaled  entropy production estimated by \eqref{nonlinear_langevin} is thus given by the log ratio between forward and time-reversed path probabilities, \eqref{path_langevin} and \eqref{path_langevin_timerev}. It reads, using \eqref{rateI} for the initial and final probability densities (with the truncation of \ref{sec:KM_truncation})
\begin{equation}
\begin{aligned}\label{ep_nle}
 \sigma_\text{NLE} := &\frac{1}{V}  \ln \frac{P[\{c(t)\}| c(0)]}{\overline{P}[\{c(t)\}| c(t)]}
\\&
+  I_\text{NLE}(c(\tau),\tau) -  I_\text{NLE}(c(0),0).
\end{aligned}
\end{equation}
This is only an apparent entropy production inferred solely from the dynamics in configuration space. We already explained in Sec. \ref{sec:gauss_instanton} the shortcoming of the boundary term in \eqref{ep_nle}. We now show that the informational entropy flow, the first line of \eqref{ep_nle}, is also flawed. Using \eqref{path_langevin}, we obtain for the rate of \eqref{ep_nle}
\begin{align}
\dot \sigma_\text{NLE} = d_t c(t)  \cdot D^{-1}(c(t)) \cdot F(c(t))+d_t I_\text{NLE}(c(t),t) .
\end{align}
Taking its average and replacing $d_t c$ by means of \eqref{nonlinear_langevin}, we arrive at the expression
 \begin{align}\label{ep_nle_mean}
\lim_{V \to \infty} \mean{\dot \sigma_\text{NLE} }= F(\mathcal{x}(t))  \cdot D^{-1}(\mathcal{x}(t)) \cdot F(\mathcal{x}(t)),
\end{align}
recalling that $\mean{\eta(t)   \cdot D^{-1}(c(t)) \cdot F(c(t))}=O(V^{-1})$ .
Equation \eqref{ep_nle_mean} is the Kullback-Leibler divergence between forward and backward probability densities of paths solutions of \eqref{nonlinear_langevin}, i.e. it is the diffusive approximation of 
\eqref{info_epr}.
Correlations between fluctuations are at least of order $O(V^{-1})$ and thus do not appear in the macroscopic limit. Hence, \eqref{ep_nle_mean} is also the mean entropy production rate of the linear Langevin equation desribed in \ref{sec:Onsager_Machlup}, once $F$ is linearized around a stable state.

In any stationary state the apparent entropy production rate vanishes identically,
 \begin{align}\label{ep_nle_mean_stationary}
\lim_{V \to \infty} \mean{\dot \sigma_\text{NLE} }= F(\mathcal{x}^*)  \cdot D^{-1}(\mathcal{x}^*) \cdot F(\mathcal{x}^*)=0,
\end{align}
since $F(\mathcal{x}^*) =0$ by definition of fixed point. Namely, the macroscopic entropy production predicted by a Langevin dynamics misses altogether the contribution 
needed to sustain a time-independent attractor, i.e. the quantity $\mean{\dot \sigma_\gamma}$ defined in Sec. \ref{sec:emergent}.
In particular, if we consider small deviations from a stable fixed point, \eqref{ep_nle_mean} becomes
 \begin{align}\label{ep_Gaussian_mean}
&\lim_{V \to \infty} \mean{\dot \sigma_\text{NLE} }\\
&= (\mathcal{x} -\mathcal{x}_\gamma^*)\cdot \partial_c F(\mathcal{x}_\gamma^*)  \cdot D^{-1}(\mathcal{x}_\gamma^*) \cdot \partial_c  F(\mathcal{x}_\gamma^*) \cdot (\mathcal{x} -\mathcal{x}_\gamma^*),
\nonumber
\end{align}
that reduces to the first term of \eqref{ep_linear2} close to detailed balance dynamics. Therefore, $\mean{\dot \sigma_\text{NLE} }$ misses the adiabatic component and thus underestimates the thermodynamic entropy production rate.
We emphasize that the entropy production has the same formal structure for linear and nonlinear Langevin equations, in the sense that it is a quadratic form of the forces -- the Langevin equation linearizes the fluxes in the forces. For the linear Langevin equation, the force is further linearized in the displacement of the state $c$ from its fixed point.

{\subsection{Nonreciprocity}\label{sec:nonrec}
In conservative mechanical systems that follow Newton's third law of action and reaction, interactions between different degrees of freedom are said to be reciprocal. This holds true when they are brought in contact with thermal reservoirs in such a way that their stochastic dynamics remain detailed balanced. 
Nonreciprocity arises in presence of dissipative processes, be them nongradient forces or effective time-delayed interactions \cite{Ivlev2015third,ste16,loos2020irreversibility,fruchart2021nonreciprocal,dinelli2023nonreciprocal}.
In stochastic systems characterized by the deterministic asymptotic dynamics \eqref{macro_dcdt}, nonreciprocity can be defined by means of the Jacobian matrix,
\begin{align}\label{state_response}
\lim_{V \to \infty} \frac{\partial \dot c}{\partial c} (t)=  \partial_c F(\mathcal{x}(t)),
\end{align}
which measures the variation rate of a degree of freedom to perturbations in a different one. The trace of \eqref{state_response} gives the variation rate of phase space volumes introduced in \eqref{contraction_rate}. To evaluate \eqref{state_response} in a stable fixed point, we use the decomposition \eqref{orthogonal_decomposition}. We simplify the gradient part of the drift as
\begin{align}
\partial_c \mathcal{F}(\mathcal{x}^*)&= \nonumber
 \mathcal{M}(\mathcal{x}^*) \cdot \partial^2_c I_\mathrm{ss} (\mathcal{x}^*) +\partial_c \mathcal{M}(\mathcal{x}^*) \cdot \partial_c I_\text{ss}(\mathcal{x}^*) \\
 &= D(\mathcal{x}^*) \cdot \partial^2_c I_\mathrm{ss} (\mathcal{x}^*).
\end{align}
while for the dissipative part we employ the property \eqref{divergenceless} to write $\partial_{c} v_\mathrm{ss}  (\mathcal{x}^*)=:\varpi=\mathcal{N}(\mathcal{x}^*)\cdot \mathcal{C}^{-1}_\mathrm{ss}$.
 The matrix $\varpi$ has purely imaginary eigenvalues, referred to as cycling frequencies and proposed as empirical quantities to measure irreversibility \cite{bat16,Mura2018scaling}. 
We thus obtain
\begin{align}\label{state_response_linear}
 \partial_c F(\mathcal{x}^*)= -D(\mathcal{x}^*) \cdot  \mathcal{C}_\mathrm{ss}^{-1}  + \varpi.
\end{align}
Taking the trace, we regain the relation for the phase-space contraction rate \eqref{contraction_fluctuations}.
Factoring out the kinetic matrix that sets the time scale of the interactions, we can define the macroscopic response matrix
\begin{align}\label{response}
 \mathcal{R} \equiv    D^{-1}(\mathcal{x}^*) \cdot   \partial_c F(\mathcal{x}^*) =  -  \mathcal{C}_\mathrm{ss}^{-1}  +  D^{-1}(\mathcal{x}^*) \cdot \varpi.
 \end{align}
 For detailed balanced dynamics, $\varpi=0$, which implies that $ \mathcal{R}$ is symmetric:
 \begin{align}
     \mathcal{R} \underset{a_\rho=0}{=} -\partial_c^2 \phi(\mathcal{x}^\mathrm{eq}).
 \end{align}
Hence, we can extract a measure of nonreciprocity of the coupling between two degrees of freedom from the antisymmetric part of \eqref{response},
\begin{align}\label{vorticity}
\begin{aligned}
\mathcal{R}- \mathcal{R}^T =D^{-1}(\mathcal{x}^*)\cdot \varpi -\varpi ^T \cdot D^{-1}(\mathcal{x}^*).
   \end{aligned}
\end{align}
If multiple fixed points exist, such relation is valid for each attractor for $t$ much smaller than the escape time from its basin. 
}

\subsection{Irreversibility of relaxation and instanton}\label{sec:large_dev}

Under the condition of detailed balance, $a_\rho=0$, the instantons $c(t)$ that solve \eqref{hamilton_eq} are the time reversal of the deterministic trajectories solutions of \eqref{macro_dcdt}:
%%%%%%%%%%
%\begin{equation}\label{reversibility}
%\begin{aligned}
%d_t c(t)& \underset{a_\rho=0}{=} \sum_\rho \Delta_\rho r^{(0)}_\rho(c(t)) e^{\Delta_\rho \cdot \pi(t) }\\
% & = -\sum_\rho \Delta_{-\rho} r^{(0)}_{-\rho}(c(t)) e^{\Delta_\rho \cdot [\pi(t)-\partial_c \phi(c(t)) ]}\\
% &= - d_t \mathcal{x} (t).
%\end{aligned}
% \end{equation}
%To prove that, we used in order: the detailed balance %\eqref{ldb_macro}, the property $\Delta_{\rho}=-\Delta_{-\rho}$,  %the equality $\pi(t)=\partial_c \phi(c(t))$ valid for $a_\rho=0$ on %the manifold $\mathcal{H} =0$, and a relabeling of the transitions %$\rho \mapsto -\rho$. 
%%%%%%%%%%
\begin{equation}\label{reversibility}
d_t c(t)\underset{a_\rho=0}{=} - d_t \mathcal{x} (t).
 \end{equation}
This relation readily follows from the equality $r_\rho^\dagger = r_\rho$ valid for detailed balance systems. This result is ultimately a consequence of the symmetry
\begin{align}\label{symmetry_H}
\mathcal{H} (c,\pi)\underset{a_\rho=0}{=}\mathcal{H} (c,-\pi+\partial_c \phi),
\end{align}
that holds in presence of detailed balance \cite{dykman1994large}, and is responsible for the validity of \eqref{linear_resp} \cite{falasco21local}. Its generalization will lead to the fluctuation theorems discussed in Sec. \ref{sec:macro_ft}.

Instead, when the dynamics is not detailed balanced, $a_\rho \neq 0$, the adiabatic entropy production quantifies the time-asymmetry between relaxational and instantonic trajectories. Equation \eqref{reversibility} is replaced by
%\begin{equation}\label{no_reversibility}
%\begin{aligned}
%d_t c(t)&= \sum_\rho \Delta_\rho r_\rho(c(t)) e^{\Delta_\rho \cdot \pi(t) }\\
% & = -\sum_\rho \Delta_{\rho} r_{\rho}(c(t)) e^{-\sigma^\text{ad}_{\rho}(t)} = - d_t \mathcal{x} (t) +O(\sigma^\text{ad}_{\rho})
%\end{aligned}
% \end{equation}
\begin{align}\label{no_reversibility}
d_t c(t)&= \sum_\rho \Delta_\rho r_\rho(c(t)) e^{\Delta_\rho \cdot \pi(t) }=-F^{\dagger}(c(t))\\
 & = -\sum_\rho \Delta_{\rho} r_{\rho}(c(t)) e^{-\sigma^\text{ad}_{\rho}(t)}= - d_t \mathcal{x} (t) +O(\sigma^\text{ad}_{\rho}) \nonumber 
 \end{align}
where we used $\pi(t)=\partial_c I_\text{ss}(c(t))$ on the manifold $\mathcal{H} =0$, the dual drift \eqref{Dual_macro_drift}, the relation 
\begin{align}\label{dual_rate_thermo}
r^\dagger_{\rho}(c)=r_{\rho}(c)e^{-\sigma^\text{ad}_{\rho}(c)}
\end{align}
%$r^\dagger_{-\rho}=r_{\rho}(c)e^{\Delta_\rho \cdot \partial_c I_\text{ss}(c)}=r_{-\rho}(c)e^{\sigma^\text{ad}_{\rho}}$
valid for autonomous dynamics, and the relabelling of the transitions $\rho \mapsto -\rho$. 
We point out two important aspects of the instanton dynamics. First, in the case of Gaussian noise %relaxation and instantonic trajectories are known to be related by the simple transformation
the time-reversed dual dynamics that defines the instanton is a simple transformation of the drift field which flips the sign of the gradient part \cite{chernyak2006path,Bertini2015Jun}, i.e. $- D(c) \cdot \partial_c I_\text{ss}(c)  \mapsto D(c) \cdot \partial_c I_\text{ss}(c)$, without changing the probability velocity $v_\text{ss}(c)$. Away from the diffusive limit, the dual dynamics has to be implemented at the level of single transition rates, i.e., $r_\rho \mapsto r_\rho^\dagger$, rather than directly on quantities defined in configuration space. Second, the dual transition rates $r^\dagger_{\rho}$ do not have in general the same functional form of the transition rate $r_{\rho}$ unless detailed balance holds, cf. \eqref{dual_rate_thermo}. This means that trajectories akin to the most likely fluctuations cannot be generated in the deterministic system only by tuning the nonconservative forces, rather new transitions belonging to different physical classes are required. See Sec. \ref{sec:ex1_SM} for an example.

\subsection{Thermodynamic constraints on transition rates between attractors}\label{sec:bounds_kappa}

The distribution of exit locations from an attractor asymptotically peaks on the saddle points \footnote{Or close to them, in the case of saddles with flat stable directions.} dividing two basins \cite{day1990large,maier1997limiting,Luchinsky1999observation}.
Therefore, when the end point $c$ of an instanton coincides with a saddle point $\mathcal{x}_\nu$ on the separatrix dividing the basin of attraction of $\mathcal{x}^*_\gamma$ and $\mathcal{x}^*_{\gamma'}$, \eqref{trans_rate} is the macroscopic transition rate $\kappa_\nu$ from the attractor $\gamma$ to $\gamma'$ through the saddle point $\nu$ (see Appendix \ref{app:transition_rate}):
 \begin{align}\label{trans_saddle}
 \kappa_\nu\asymp  e^{- V[I^{(\gamma)}_\text{ss}(\mathcal{x}_\nu)-I^{(\gamma)}_\text{ss}(\mathcal{x}^*_\gamma)] }.
\end{align}
Since exit times from each attractor are exponential distributed in the large $V$ limit \cite{bovier2002,day83,freidlin}, the inverse of the transition rate coincides with the mean first passage time $1/ \kappa_\nu:= \inf\{t \geq\! 0:\! c(0)= \mathcal{x}^*_\gamma ,c(t)=\mathcal{x}_\nu\}$.
 Hereafter $\kappa_{-\nu}$ will be used to indicate the rate of the opposite transition, namely from $\mathcal{x}^*_{\gamma'}$ to $\mathcal{x}^*_{\gamma}$ through the same saddle $\nu$. Crucially, if one used the quasi-potential $I_\text{NLE}$ obtained by the nonlinear Langevin equation in \eqref{trans_rate} and \eqref{trans_saddle}, the transition probability and the escape rate from the attractor would be misestimated with an exponential error \cite{Bressloff2014,assaf2017wkb}.
 
For detailed balance dynamics $I_\text{ss}$ equals $\phi$ up to a constant, as shown in Sec. \ref{sec:macro_thermo}. Hence \eqref{trans_saddle} takes the form of an Arrhenius rate \cite{arrhenius1889,hanggi90kramers} with the large parameter $V$ playing the role of the (small) inverse temperature:
\begin{align}\label{trans_saddle_eq}
%\lim_{a_\rho \to 0}\kappa_\nu=:
\kappa_\nu^{\text{eq}} \asymp  e^{ -V[\phi(\mathcal{x}_\nu)-\phi(\mathcal{x}^*_\gamma)] }.
\end{align}
For later convenience, we have defined an Arrhenius rate
$\kappa_\nu^{\text{eq}}$ with respect to the fixed and saddle points of the nonequilibrium dynamics. In fact, $\kappa_\nu^{\text{eq}} \neq \lim_{a_\rho \to 0}\kappa_\nu $ since $\mathcal{x}^*_\gamma \neq \mathcal{x}^\text{eq}$ and $\mathcal{x}_\nu$ may even be absent for $a_\rho=0$. In Secs. \ref{sec:ex1} and \ref{sec:ex2_CMOS} we give examples of such metastables states created by continuous dissipation in chemical reaction networks and electronic circuits. Note that Eyring-Kramers formula \cite{Eyring1935activated,kramers1940,landauer1961frequency,langer1969statistical} for the transition rate is the extension of \eqref{trans_saddle_eq} that includes subexponential prefactors.

Close to detailed balance, i.e. at linear order in $a_\rho$, the  quasi-potential is given by \eqref{linear_resp} and thus the jump dynamics between attractors inherits the local detailed balance property \cite{falasco21local}, namely,
\begin{align}\label{ldb_macro_jump}
\lim_{V \to \infty} \frac{1}{V}\ln \frac{\kappa_\nu}{\kappa_{-\nu}} =  \phi(\mathcal{x}^\text{eq}_{\gamma})-\phi(\mathcal{x}^\text{eq}_{\gamma'}) + \sigma_\nu^{(0)},
\end{align}
where %$\sigma_\nu^{(0)} := -\sigma_\text{nc}^{(0)}(\mathcal{x}_\nu,\mathcal{x}^\text{eq}_{\gamma}) + \sigma_\text{nc}^{(0)}(\mathcal{x}_\nu,\mathcal{x}^\text{eq}_{\gamma'}) $
$\sigma_\nu^{(0)} := -\sigma_\text{nc}^{(0)}(\mathcal{x}_\nu,\mathcal{x}^\text{eq}_{\gamma}) + \sigma_\text{nc}^{(0)}(\mathcal{x}_\nu,\mathcal{x}^\text{eq}_{\gamma'}) $ becomes the dissipation of nonconservative forces along the equilibrium instanton from $\mathcal{x}^\text{eq}_{\gamma}$ to $\mathcal{x}_\nu$ -- the time-reversed solution of \eqref{equilibrium_relaxation}, see \eqref{reversibility} -- plus the dissipation in the deterministic relaxation from $\mathcal{x}_\nu$ to $\mathcal{x}^\text{eq}_{\gamma'}$. 
We will show in Sec. \ref{sec:emergent} that the property \eqref{ldb_macro_jump} allows us to retrieve standard stochastic thermodynamics when the latter is formulated for the jump dynamics on macroscopic attractors.

Remarkably, thermodynamics constrains the macroscopic transition rate $\kappa_\nu$ from the attractor $\gamma$ to $\gamma'$ through the saddle point $\nu$ in terms of the entropy $\sigma_{\gamma \to \nu}$ (resp. $\sigma_{\nu \to \gamma}$) produced along the instanton (rep. relaxation): 
\begin{align}\label{upper_lower_bounds}
 - \sigma_{\nu \to \gamma} \leq \lim_{V \to \infty} \frac 1 V  \ln \kappa_\nu  \leq  \sigma_{\gamma \to \nu}.
\end{align}
We first derive the lower bound  in \eqref{upper_lower_bounds}.
Integrating \eqref{emerging_2nd} along a relaxation trajectory from a neighborhood of the saddle $\mathcal{x}_\nu$ to a neighborhood of the fix point $\mathcal{x}^*_\gamma$  \footnote{See appendix \ref{app:bound} for a discussion about the exact choice of the trajectory endpoints.}, 
%The trajectory duration $\tau$ must be finite, i.e. the endpoint of the trajectory should not coincide with $c^*_\gamma$, otherwise $ \sigma_{\nu \to \gamma}$ would diverge due to the entropy production of the fixed point (see Sec. \ref{sec:emergent}), which is zero only close to detailed balance.} 
and using \eqref{trans_saddle}, we immediately obtain
\begin{align}\label{lower_bound_kappa}
%\kappa_\nu  \geq e^{-V \sigma_{\nu \to \gamma}}.
\lim_{V \to \infty} \frac 1 V  \ln \kappa_\nu \geq - \sigma_{\nu \to \gamma}.
\end{align}
The bound \eqref{lower_bound_kappa} is the conceptual analog of \cite{falasco2020dissipationtime,Neri2022} for state observables, i.e. a recently discovered speed limit which is an upper bound on the rate of processes defined for current observables.
In general, splitting the entropy produced in the relaxation 
\begin{align}\label{sigma_rel_splitting}
 \sigma_{\nu \to \gamma} =  \phi(\mathcal{x}_\nu)-\phi(\mathcal{x}_\gamma^*) +\sigma_\text{nc}^{\nu \to \gamma}
 %+{\textstyle \int_0^\infty dt \sum_\rho}  a_\rho r_{\rho}(\mathcal{x}(t))
\end{align}
into the equilibrium part, i.e. the variation of the Massieu potential, and the dissipation of the nonconservative forces along the relaxation trajectory, i.e.  
\begin{align}
\sigma_\text{nc}^{\nu \to \gamma}:= \int_{\mathcal{x}(0)=\mathcal{x}_\nu}^{\mathcal{x}(\infty)=\mathcal{x}^*_\gamma} dt \sum_\rho a_\rho r_\rho(\mathcal{x}(t)) \;,
\end{align}
we can rephrase the bound \eqref{lower_bound_kappa} in terms of the Arrhenius  transition rate \eqref{trans_saddle_eq},
\begin{align}
\frac{\kappa_\nu }{\kappa_\nu^{\text{eq}}} \geq e^{-V \sigma_\text{nc}^{\nu \to \gamma}}.
\end{align}

The additional upper bound in \eqref{upper_lower_bounds} can be obtained in a similar fashion. The derivation, given in appendix \ref{app:bound}, is based on averaging the entropy production decomposition \eqref{ad_nonad_epr_macro} by conditioning the initial and final points of trajectories of infinite duration, and on showing that the adiabatic entropy production remains nonnegative along these paths.

For detailed balance dynamics, $\sigma_{\nu \to \gamma}=-\sigma_{\gamma \to \nu} > 0$ (the sign comes from the positivity of the mean entropy production \eqref{emerging_2nd}), as these entropy productions are just the difference of the Massieu potential between the saddle and the stable fixed point. Therefore, the upper and lower bounds in \eqref{upper_lower_bounds} converge to each other in this limit.
Moreover, at linear order in $a_\rho$, the quasi-potential is given by \eqref{linear_resp}, and thus \eqref{upper_lower_bounds} holds again in the form of equality. In this case we can read \eqref{upper_lower_bounds} as a maximum entropy production principle: the attractor with the largest life-time is the one with the largest relaxation (or smallest instantonic) entropy production. This statement remains true for all systems in which relaxation and instantonic entropy productions are close to each other. However, note that $\sigma_{\gamma \to \nu}$ need not be negative far from equilibrium, which means that the righthand side of \eqref{upper_lower_bounds} can be a loose bound.

We can also isolate the equilibrium contribution to the entropy production of the instanton as done in \eqref{sigma_rel_splitting}, 
 \begin{align}\label{sigma_inst_splitting}
 \sigma_{\gamma \to \nu} =  \phi(\mathcal{x}_\gamma^*)-\phi(\mathcal{x}_\nu) +\sigma_\text{nc}^{\gamma \to \nu},
 %+{\textstyle \int_0^\infty dt \sum_\rho}  a_\rho r_{\rho}(\mathcal{x}(t))
 \end{align}
to recast the upper bound in \eqref{upper_lower_bounds} as
 \begin{align}\label{upper_bound}
 \frac{\kappa_\nu }{\kappa_\nu^{\text{eq}}} \leq e^{V \sigma_\text{nc}^{\gamma \to \nu}}.
 \end{align}
This inequality is an exact result -- asymptotically in $V$ -- and thus improves a similar, but approximated result obtained by neglecting changes in the system's dynamical activity \cite{Kuznets-Speck2021}. The main differences are that $\sigma_\text{nc}^{\gamma \to \nu}$  is replaced in \cite{Kuznets-Speck2021} by one half of the dissipation caused by nonconservative forces along the entire trajectory connecting the two basins of attractions and the Arrhenius rate in \eqref{upper_bound} is calculated with respect to nonequilibrium fixed points. 

Finally, we note that these bounds can be tightened by considering the macroscopic limit of the information-theoretic entropy production \eqref{info_epr_macro}.
If the jump vectors $\tilde \Delta_\rho$ are linearly independent, the condition $F(\mathcal{x}^*)=0$ implies that  $\tilde r_\rho(\mathcal{x}^*)-\tilde r_{-\rho}(\mathcal{x}^*)=0$ for all $\rho$ and the function \eqref{info_epr_macro} is zero on the fixed point $\mathcal{x}^*$ \cite{Freitas2021NatCom}. Therefore, replacing $\sigma$ with $\tilde \sigma$ in \eqref{upper_lower_bounds} we obtain tighter bounds -- no longer connected to thermodynamics, though. See Sec. \ref{sec:ex2_CMOS} for an application of such bounds to a model of electronic circuit.
 
 \begin{figure}
\includegraphics[width=0.5\textwidth]{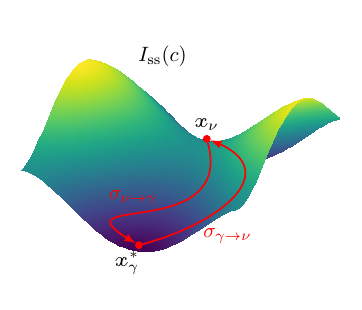}
\caption{Schematic representation, for a system without detailed balance, of relaxation and instanton trajectories connecting the local minimum $\mathcal{x}^*_\gamma$ of the rate function $I_\text{ss}$ and a saddle point $\mathcal{x}_\nu$. \label{fig:rel_inst_ep}}
\end{figure}

The ability to obtain exact solutions of \eqref{macro_dcdt} and \eqref{hamilton_eq} underlies the practical applicability of the bounds \eqref{upper_lower_bounds}. While relaxation trajectories are relatively easy to find by a direct numerical integration of an initial value problem, instantons require more advanced approaches in view of the boundary value problem which defines them. One typically employs the shooting method \cite{press2007numerical,keller2018numerical} for low-dimensional systems. Alternatively, one resorts to the minimum action method \cite{weinan2004minimum,grafke2017long,kikuchi2020ritz,zakine2023minimum} that is a fictitious-time gradient descent on the (negative of the) action \eqref{action} in the trajectory space satisfying the appropriate boundary conditions.

\subsection{Macroscopic time-integrated observables and fluctuation theorems}\label{sec:macro_ft}
%\section{Macroscopic generating function of state and current observables}

The trajectory description of the previous section can be complemented by an analysis of the full statistics of thermodynamic observables
\cite{garrahan2009first,speck2012large,hurtado2014thermodynamics,Esposito2007,chetrite2008fluctuation}.
For observables that are time integrated functionals of the trajectory $\mathcal{X}$ of the form,
\begin{align}\label{time_int_obs} 
\mathcal{O}[\mathcal{X}]= \int_0^\tau dt \underbrace{{\textstyle\sum_\rho }J_\rho(t) \mathcal{O}_\rho(n(t))}_{\dot{\mathcal{O}}(t)},
\end{align}
 we are interested in the generating function $G_\mathcal{O}(z,t):=\mean{e^{z\mathcal{O}[\mathcal{X}]}}$, which gives all the moments upon differentiation with respect to $z$, i.e. $\partial^m_z G_\mathcal{O}(z,\tau)|_{z=0} =\mean{\mathcal{O}^m}$. 
To this aim, it is customary to start from the time evolution equation of the joint distribution  $P(n,\mathcal{O},t)$ 
 %:= \mean{\delta(n(t)-n) \delta( B(t)-B)} 
 of the occupation number $n$ and the observable $\mathcal{O}$,
\begin{align}\label{mater_POn}
&\partial_t P(n, \mathcal{O}, t) =\\
& \sum_\rho \big[
R_\rho(n - \Delta_\rho)
P(n - \Delta_\rho,\mathcal{O}-\mathcal{O}_\rho,  t) -R_\rho(n) P(n, \mathcal{O}, t) \big].\nonumber
\end{align}
For extensive observables, i.e. such that $\lim_{V\to \infty} \mathcal{O}_\rho(n) = \mathcal{o}_\rho(c)$, we can perform a large scale expansion analogous  to that worked out for the master equation in Sec. \ref{sec:large_scale}. The resulting evolution equation for the joint probability of the intensive variables $c$ and the scaled observable $\lim_{V\to \infty}  \mathcal{O}[\mathcal{X}]/V =\mathcal{o} [\mathcal{X}] $ reads
\begin{align}\label{ME_gen}
&\partial_t p(c,\mathcal{o},t)=\\
& V \sum_\rho\!\bigg[ r_\rho {\textstyle \left(c - \frac{\Delta_\rho}{V} \right)} p {\textstyle\left(c-  \frac{\Delta_\rho}{V} , \mathcal{o} - \frac{\mathcal{o}_\rho}{V},t \right)} \nonumber -r_\rho(c) p(c,\mathcal{o},t)\bigg].
\end{align}
Equation \eqref{ME_gen} admits a solution of the large deviations form 
\begin{align}\label{rate_fluxes}
p(c,\mathcal{o},t)\asymp e^{-VY(c,\mathcal{o},t)}
\end{align}
 and suggests to define the scaled cumulants  generating function 
\begin{align}\label{scgf}
K_\mathcal{o}(z,t) := \lim_{V \to \infty} \frac{1}{V} \ln G_\mathcal{O}(z,t) ,
\end{align}
 which encodes all the relevant statistics in the macroscopic limit. It can be obtained in the following way.
 Multiplying \eqref{ME_gen} by $e^{Vz \mathcal{o}}$ and  integrating over $\mathcal{o}$ yields a master equation for $G_\mathcal{O}(c,z,t):=\mean{\delta(c(t)-c)e^{Vz \mathcal{o}[\mathcal{X}]}}$, which is the generating function of $\mathcal{O}$ conditioned on the macroscopic state $c$ at time $t$: 
{
\begin{equation}
\begin{aligned}
 \partial_t G_\mathcal{O}(c,z,t)%& V \sum_\rho\!\bigg[  e^{q \mathcal{o}_\rho} r_\rho {\textstyle \left(c - \frac{\Delta_\rho}{V} \right)} G {\textstyle\left(c-  \frac{\Delta_\rho}{V} , q ,t \right)} \\
%& \qquad\qquad -r_\rho(c) G_\mathcal{O}(c,q,t)\bigg] \\=
%&V  \underbrace{\sum_\rho\! \left[   e^{q \mathcal{o}_\rho -V^{-1}\Delta_\rho \cdot \partial_c}-1 \right] r_\rho(c)}_{\mathcal{H}_z \left(c, - V^{-1}\partial_c \right)} G_\mathcal{O}(c,q,t) .
&=V \mathcal{H}_z \left(c, - V^{-1}\partial_c \right) G_\mathcal{O}(c,z,t) .
\end{aligned}
\end{equation}
The latter is expressed in terms of  the `tilted' operator 
\begin{align}
\mathcal{H}_z \left(c, - V^{-1}\partial_c \right):=\sum_\rho\! (   e^{z \mathcal{o}_\rho -V^{-1}\Delta_\rho \cdot \partial_c}-1 ) r_\rho(c)\;, 
\end{align}
which reduces to the generator of the stochastic dynamics \eqref{H_operator} for $z=0$, since $G_\mathcal{O}(c,z,t)|_{z=0}=p(c,t)$. Similarly to Sec. \ref{sec:gauss_instanton}, we can obtain the (unconditioned) generating function as the functional integral 
\begin{align}\label{gen_func}
G_\mathcal{O}(z,\tau) = \int \mathcal{D} c \int \mathcal{D} \pi \, e^{V \{\mathcal{A}_z[\{c(t),\pi(t)\}]-I(c(0),0)\}} ,
\end{align}
that at leading order in $V$ can be evaluated by Laplace method:
 \begin{align}\label{gen_func_saddle}
G_\mathcal{O}(z,\tau) \asymp  e^{V \max_{c(t),\pi(t)} \{\mathcal{A}_z[\{c(t),\pi(t)\}]-I(c(0),0)\}} .
\end{align}
Namely, we are left with maximizing the tilted (or biased) action 
 \begin{align}\label{tilted_action}
\mathcal{A}_z=   \int_0^\tau dt  \bigg[- \pi(t) \cdot d_t c(t) + \mathcal{H}_z(c(t),\pi(t))\bigg]   
\end{align} 
with respect to $c(t)$ and $\pi(t)$, that is to find the solutions of the Hamiltonian equations
\begin{align}\label{hamiltonian_eqs_z}
&d_t c= \partial_\pi  \mathcal{H}_{z}(c,\pi) && d_t \pi= -\partial_c  \mathcal{H}_{z}(c,\pi),
\end{align}
with the appropriate boundary conditions. These read $\pi(0)= \partial_c I(c(0),0)$ and $\pi(\tau)=0$ \cite{Lazarescu2019Aug} for trajectories drawn from the unconstrained ensemble with initial distribution $p(c(0),t) \asymp e^{-V I(c(0),0)}$, or $c(0)= \mathcal{x}_\gamma^*$.
%and $c(\tau)=\mathcal{x}_\nu$ for a transition path.
Finally, the (contracted) rate function of the observable $\mathcal{O}$ is obtained by the Legendre-Fenchel transform
\begin{align}\label{rate_function_obs}
Y(\mathcal{o},\tau):=\sup_{z} \{ K_\mathcal{o}(z,\tau) -z \mathcal{o} \}.
\end{align}
Note that if $K_\mathcal{o}(z,\tau)$ has nondifferentiable points, the rate function has a nonconvex part, but \eqref{rate_function_obs} returns only its convex envelope. 

From \eqref{tilted_action}, one can directly extract the mean value of the observable along trajectories conditioned on the boundaries, $ \mean{\dot{\mathcal{o}}}= \partial_z \mathcal{H}_{z}|_{z=0}(c(t),\pi(t))$, with $c(t)$ and $\pi(t)$ the solution of \eqref{hamiltonian_eqs_z} at $z=0$, which is \eqref{hamilton_eq}. In particular, following a derivation very analogous to that of  \ref{sec:large_dev} we find 
\begin{align}\label{currents_inst_rel}
\mathcal{o}_{\gamma \to \nu}&=\int dt \sum_{\rho} \mathcal{o}_\rho r_\rho(c(t))e^{\Delta_\rho \cdot \partial_c I_\text{ss}(c(t))} \\
&=\pm \int dt \sum_{\rho} \mathcal{o}_\rho r_\rho^{\dagger}(c(t)) = \pm \mathcal{o}_{\nu \to \gamma} +O(\sigma_\rho^\text{ad}), \nonumber
\end{align}
where the positive (resp. negative) sign is for observables $\mathcal{o}_\rho=\mathcal{o}_{-\rho}$ (resp. $\mathcal{o}_\rho=-\mathcal{o}_{-\rho}$).

\subsubsection{Entropy production at steady state}\label{sec:epr_ft}
Choosing $\mathcal{o}_\rho(c)=\sigma_\rho(c)$ defined in \eqref{ldb_macro} and adding the boundary terms $q[I(c(\tau),\tau) - I(c(0),0)]$ to the action in \eqref{gen_func}, we obtain the scaled cumulant generating function of the entropy production \eqref{ep}:
\begin{equation}
\begin{aligned}\label{scgf}
K_\sigma(z,\tau)&= \max_{c(t),\pi(t)} \big\{\mathcal{A}_z [\{c(t),\pi(t)\}] \\
&\quad+  z I(c(\tau),\tau) - (z+1)I(c(0),0)] \big\}.
\end{aligned}
\end{equation}
If we restrict to a stationary state, i.e. $\mathcal{H}_z$ has no explicit time dependence and $I(c,\tau) = I(c,0)= I_\text{ss}(c)$, the scaled cumulant generating function satisfies the symmetry
\begin{align}\label{ft_K}
K_\sigma(z,\tau)=K_\sigma(-z-1,\tau).
\end{align}
 This follows from the local detailed balance property \eqref{ldb_macro}, which ensures the Lebowitz-Spohn symmetry of the tilted operator \cite{leb99,kur98}
\begin{equation}
\begin{aligned}\label{symmetry_LS}
\mathcal{H}_z(c, \pi) &=
\sum_\rho r_{-\rho}(c) \left( e^{ -\Delta_\rho \cdot \pi  - z \sigma_\rho(c) }-1\right)\\
&=\sum_\rho r_{\rho}(c) \left( e^{- \Delta_\rho \cdot \pi - (z+1) \sigma_\rho(c) }-1\right)\\
&=\mathcal{H}_{-z-1}(c, -\pi) .
\end{aligned}
\end{equation}
Stationarity implies that the generating function can as well be obtained by integrating over the time-reversed trajectories of $c(t)$ and $\pi(t)$, defined by the change of variable $\pi \mapsto -\pi$ and  $t  \mapsto \tau-t$, which maps the tilted action as $\mathcal{A}_{z} [\{c(t),\pi(t)\}]  \mapsto \mathcal{A}_{-z-1} [\{c(t),\pi(t)\}] $ thanks to \eqref{symmetry_LS}. Hence, by the Laplace approximation the scaled cumulant generating function is recast as
\begin{equation}
\begin{aligned}\label{ft_time_rev}
K_\sigma(z,\tau)&= \max_{c(t),\pi(t)} \big\{\mathcal{A}_{-z-1}[\{c(t),\pi(t)\}] \\
&\qquad+ z I_\text{ss}(c(0)) - (z+1)I_\text{ss}(c(\tau))] \big\}.
\end{aligned}
\end{equation}
Comparing with \eqref{scgf} evaluated at stationarity, i.e. setting $I=I_\text{ss}$, we arrive at the announced symmetry \eqref{ft_K}.

By a Legendre-Fenchel transform of $K_\sigma(z,\tau)$ that gives the {(convex envelop of the)} rate function of the entropy production  \eqref{rate_function_obs}, the symmetry \eqref{ft_K} results in finite time the fluctuation theorem
\begin{equation}\label{ft}
Y(\sigma,\tau)-Y(-\sigma,\tau)= \sigma.
\end{equation}

Finally, one can directly check that the diffusion-type approximation of the generating function, obtained by expanding the tilted action \eqref{tilted_action} to second order in $\pi(t)$, does preserve the symmetry \eqref{symmetry_LS} and the fluctuation theorem \eqref{ft}, provided that the functional dependence on $q$ is left untouched. This means that preserving the fluctuation relation requires to identify the entropy production at the mesoscopic level, i.e. by resolving the entropy production associated to each transition.  

\subsubsection{Transition currents at steady state}\label{sec:currents_ss}

Choosing $\mathcal{o}_{\pm \rho}= \pm 1$  and the counting field $z_\rho=-z_{-\rho}$,
we obtain the scaled cumulant generating function $K_\iota(z,\tau)$ of all the time-integrated currents with the reservoirs, each of which reads  $\iota_\rho[\mathcal{X}]:= \int_0^\tau dt I_\rho(t)$. We assume that the initial state $c(0)$ is sampled from the stationary probability density function $p_\text{ss}(c)$.
Thanks to the symmetry
\begin{equation}
\begin{aligned}\label{symmetry_LS_currents}
\mathcal{H}_z(c, \pi) &=
\sum_\rho r_{-\rho}(c) \left( e^{ -\Delta_\rho \cdot \pi  - z_\rho }-1\right)\\
&=\sum_\rho r_{\rho}(c) \left( e^{- \Delta_\rho \cdot \pi - z_\rho- \sigma_\rho(c,t) }-1\right)\\
&=\mathcal{H}_{-z-a}(c, -\pi +\partial_c\phi), 
\end{aligned}
\end{equation}
obtained using \eqref{ldb_macro}, 
the tilted action is mapped as $\mathcal{A}_{z} [\{c(t),\pi(t)\}]  \mapsto \mathcal{A}_{-z-a} [\{c(t),\pi(t)\}] + o(\tau)$ 
 by the change of variables $\pi \mapsto -\pi +\partial_c \phi$ and the time-reversal $ t  \mapsto \tau-t$. Therefore, the time- and size scaled cumulant generating function $K_j(z):= \lim_{\tau \to \infty} \frac 1 \tau K_\iota(z,\tau)$ satisfies the symmetry 
 \begin{align}\label{ft_K_infty}
 K_\iota(z)= K_\iota(-z-a),
 \end{align}
 where $a$ is the vector of transition forces $a_\rho$.
It is easy to check that the symmetry \eqref{symmetry_LS_currents}, and thus \eqref{ft_K_infty}, are in general broken when $\mathcal{H}_z(c, \pi)$ is expanded at second order in $\pi$, i.e. when the Langevin approximation \eqref{nonlinear_langevin} is employed. Since the symmetry of the full action holds only asymptotically, the fluctuation theorem for the currents is valid for $\tau \to \infty$. Namely,
\begin{equation}\label{ft_currents}
Y(\iota)-Y(-\iota)= \sum_{\rho>0} a_\rho \iota_\rho,
\end{equation}
with the long-time large-$V$ rate function of the currents corresponding to probability $p(\iota,t) \asymp e^{-VtY(\iota)}$.

For systems with multiple deterministic fixed points $\mathcal{x}^*_\gamma$, in general there exist multiple $q$-dependent  vectors $(c(t),\pi(t))$ that maximize \eqref{symmetry_LS_currents} as $z \to 0$. When $z$ departs from zero, the \eqref{symmetry_LS_currents} attains a different value on each of these optimal solutions. Hence, the function $\partial_z K_j(z)$ has a jump discontinuity at $z=0$, reflecting the coexistence of macroscopic states with different mean currents. 
Accordingly, Legendre-Fenchel transforming $ K_j(z)$ gives only the convex envelope of the rate function. Such phenomenon, is called dynamical phase transition \cite{GarrahanLecomtePrl2007,garrahan2009first,hurtado11,espigares2013,gingrich14,nyawo2017}, given the formal analogy with equilibrium phase transitions in which the free energy develops singularities. Its generality was shown in \cite{Lazarescu2019Aug} for chemical reaction networks, but extends to all systems displaying multiple stable fixed points in the macroscopic limit.

Without spelling out the derivation, very analogous to that of \cite{Herpich2020}, we note that \eqref{symmetry_LS_currents} implies a finite-time symmetry for the scaled cumulant generating function of nonautonomous systems initially in thermal equilibrium with only one reservoir. 

Mind that the validity of such detailed fluctuation theorems hinges on the initial and final probability distribution having support on the same attractors, otherwise absolute irreversibility should be taken into account \cite{Murashita14,buffoni2022spontaneous}.

\subsubsection{Fluctuation theorems from dual dynamics }

Other detailed fluctuations theorems can be obtained invoking additional symmetries of a suitably tilted Hamiltonian. In particular, one can leverage the equality 
\begin{align}\label{symmetry_dual_tilted}
\mathcal{H}_z(c,\pi)&=\sum_\rho r_\rho \left (e^{\Delta_\rho\cdot \pi +z \sigma^\text{ad}_\rho}-1 \right) \nonumber \\
&=\sum_\rho r^\dagger_\rho \left (e^{\Delta_\rho\cdot \pi +(z+1) \sigma^\text{ad}_\rho}-1 \right) \nonumber \\
&=\mathcal{H}^\dagger_{-z-1}(c,\pi),
\end{align}
that follows from the fact that the  escape rate of the dual dynamics is the same of the original dynamics, $\sum_\rho r_\rho =\sum_\rho r^\dagger_\rho $, and that $(\sigma^\text{ad}_\rho)^\dagger=-\sigma^\text{ad}_\rho$ changes sign under the dual transformation. 
From  \eqref{symmetry_dual_tilted}, we obtain the fluctuation relation that links the rate functions of the adiabatic entropy production in the  original dynamics, $Y$, and in the dual one, $Y^\dagger$:
\begin{equation}\label{ft_sigma_ad}
Y(\sigma_\text{ad},\tau)-Y^\dagger(-\sigma_\text{ad},\tau)= \sigma_\text{ad}.
\end{equation}
Equation \eqref{ft_sigma_ad} also follows, in the large deviations formulation, from the log ratio of path probabilities of states and jumps \cite{sun06}, that is reported in \eqref{ft_ad_epr}. Analogously, a fluctuation theorem  for the nonadiabatic entropy production \eqref{na_rho_macro} can be derived by combining the dual dynamics with the time reversal \cite{Rao2018Sep}. The positivity of the mean adiabatic and nonadiabatic entropy production readily follows from such symmetry relations.

\section{Macroscopic field theory}

\subsection{Continuous-space limit: dynamics}\label{subsec:continuous}

We now specify more the structure of the system dividing the transitions $\rho$ into two sets, named $R$ and $C$. {The label $R$ denotes transition that keep a discrete character (akin to chemical reactions), while $C$ indicates jumps that become infinitesimal in size (such as continuous drift or diffusion is real space).}
Correspondingly, we enforce a bipartite structure by splitting the entries of the vector $c$, denoted $c_{\alpha,x}$, so that transitions belonging to $R$ (resp. $C$) only act on the label $\alpha$ (resp. $x$). For concreteness, we can think of $\alpha$ as a label for different particle species and $x$ as a $d$-dimensional vector for the (discrete) space coordinate. 
Within this picture, we assume that to each pair of first neighbors $(x,x')$ we associate pairs of transitions $\pm \rho^{\alpha}_{(x,x')} \in C$ (each pair acting on a single label $\alpha$, for simplicity) whose rates are conveniently written as $r^{\alpha}_{(x,x')}$ and $r^{\alpha}_{(x',x)}$, respectively.
%
% transitions $\rho \in C$ are associated to transitions between $x$ and $x+\epsilon \delta_\rho^{x}$, where $\epsilon$ is a small parameter and $ \delta_\rho=  -\delta_{-\rho}$ is a unit vector pointing from $x$ to its nearest neighbors.
%

We look for the continuous-space limit of \eqref{action} when a macroscopic scale $\Omega$, e.g. the volume of the whole system, is large with respect to the mesoscopic scale $V$ in which the variable $c_{\alpha,x}$ is defined.
To this aim we introduce a continuous variable  $r= x \epsilon$ with $\epsilon^d=V/\Omega$, such that $c_{\alpha,x} \to c_\alpha(r)$ and  $\pi_{\alpha,x} \to \pi_\alpha(r)$ become smooth fields as $\epsilon \to 0$.
%\begin{align*}
%&\{\alpha, x\} \underset{\rho \in R }{\longrightarrow} \{\alpha', x\}   \qquad \quad \{\alpha, x\} \underset{\rho \in C }{\longrightarrow}\{\alpha, x'=x+\epsilon\} 
% \end{align*}
 With the rewriting $V \sum_x = \Omega \sum_x \frac{V}{\Omega}\to \Omega \int dr$, the action in \eqref{action} takes the form
\begin{align}\label{Amacro}
 V \mathcal{A} &=   \Omega \int_0^\tau \! dt \! \int \!dr \bigg\{-  \pi(r,t) \cdot d_t c(r,t)\\
 &\quad \quad + \mathcal{H}_R(c(r,t),\pi(r,t))
+  \mathcal{H}_C(c(r,t),\pi(r,t)) ]  \bigg  \}. \nonumber
\end{align}
Here we have isolated the generator of transitions $\rho \in R$ 
 \begin{equation}\label{HRdef}
 \begin{aligned}
\mathcal{H}_R(c,\pi):={\textstyle \sum_{\rho \in R} }r_\rho(c)  (e^{\Delta_\rho \cdot \pi } -1 ) ,
\end{aligned}
\end{equation}
which maintains the form of the moment generating function of a Poisson noise, and the generator of transitions $\rho \in C$ ,
\begin{align}\label{HCdef}
 &\mathcal{H}_C(c,\pi):=  \\
 &\hspace{0.6cm}\lim_{\epsilon \to 0}  \sum_{\alpha,x'} \bigg( r^{\alpha}_{+\rho}(c)  (e^{\Delta_{+\rho} \cdot \pi } -1 )+ r^{\alpha}_{-\rho}(c)  (e^{\Delta_{-\rho} \cdot \pi } -1 ) \bigg), \nonumber
\end{align}
with $+\rho=(x,x')$ and $-\rho=(x',x)$, and $\Delta_{\pm\rho}$ is a discrete gradient acting as, e.g., $\Delta_{\pm\rho} \cdot \pi =\pm( \pi_{\alpha,x} -\pi_{\alpha,x'})$.
To avoid clutter, we employ the same symbol ``$\cdot$'' to denote the scalar product in the reduced space of $\alpha$, e.g., $\pi(r) \cdot d_t c(r)= \sum_\alpha \pi_\alpha(r) d_t c_\alpha(r)$, and also in the physical space coordinatized by $r$.

To evaluate the limit in \eqref{HCdef} we use the following expressions for the discrete gradient of functions
\begin{equation}\label{cont_limit}
\begin{aligned}
& \Delta_{\pm \rho} \cdot \pi =\pm \left(\epsilon \hat{\nabla} \pi(r) + \frac{1}{2} \epsilon^2 \hat{\nabla}^2 \pi(r)\right ) + O(\epsilon^3),\\
&  \Delta_{\pm \rho} \cdot \partial_c \phi(c)  = \pm \epsilon \hat{\nabla}  \frac{\delta}{\delta c(r)}\phi[c(r)] + O(\epsilon^2), \\
\end{aligned}
\end{equation}
which yield the expansion of the transition rates \footnote{We omit to write down the explicit form of terms of order $\epsilon$ and $\epsilon^2$ which are symmetric, as they cancel in the sum of \eqref{HCdef}.}
\begin{align}\label{r_rho_cont}
&  r^\alpha_{\pm \rho}(c) = \epsilon^{-2}[ \chi_\alpha(c(r)) + O(\epsilon)] \\
& \qquad \qquad \times \left[1 \mp \frac \epsilon 2  \hat{\nabla} \frac{\delta  \phi}{\delta c_\alpha(r)} \pm  \frac \epsilon 2 e_{x,x'} f_\alpha (r)  + O(\epsilon^2)  \right] 
\nonumber
\end{align}
where we write $\hat \nabla := e_{x,x'} \cdot \nabla$, with $e_{x,x'}$ the unit vector pointing from $x$ to $x'$. 
Here the mobility matrix $\chi(c)$ (diagonal with entries $\chi_\alpha$) and the nonconservative force vector $f(r)$, as well as the thermodynamic functional $\phi(\{c_{x,\alpha}\}) \to \phi[c(r)]$ are independent of $\epsilon$. The scaling $\gamma_\rho(c_{x,\alpha}) = \epsilon^{-2} \chi_\alpha(c(r)) +O(\epsilon^{-1})$ of the kinetic part of the rates is imposed to ensure a proper diffusive limit on the same time-scales of transitions $\rho \in R$. In addition, we required that the nonconservative force $a_\rho=  \epsilon e_{x,x'} f_\alpha (r) +O(\epsilon^2)$ is small on the macroscopic scale, a condition called `weak asymmetry' in the lattice gas literature.

Therefore, neglecting contributions of $O(\epsilon^2)$ and higher, \eqref{cont_limit} and \eqref{r_rho_cont} reduce the generator $\mathcal{H}_C$ to
\begin{align}\label{HC}
\mathcal{H}_C(c,\pi) = \nabla \pi  \cdot \chi(c)\cdot  \left[\left( - \nabla \frac{\delta \phi}{\delta c}  +   f  \right)  +  \nabla \pi  \right ] ,
\end{align}
{which has the structure (quadratic in $\pi$) of the cumulant generating function of a Gaussian noise field.}
If the set $R$ is empty, \eqref{HC} is the generator of the full dynamics, corresponding to the functional Langevin equation
\begin{align}\label{mft}
\begin{aligned}
&\partial_t c(r,t) = - \nabla \cdot \left( j[c] +\frac{1}{\sqrt{\Omega}} \xi (r,t)\right), \\
& j[c]:= \chi \cdot \left(- \nabla \frac{\delta \phi}{\delta c}  +   f  \right),
\end{aligned}
\end{align}
where $\xi(r,t)$ is zero-mean, Gaussian with correlations $\mean{\xi(r,t)\xi(r',t')}= 2  \chi(c(r,t))\delta(r-r')\delta(t-t')$.
%delta-correlated both in space and in time with correlation matrix $ 2 \chi(c)$. 
This can be easily seen by performing the Gaussian integral over $\pi$ and recognizing that the resulting action is the standard one corresponding to the Langevin equation \eqref{mft} (cf. \eqref{path_HS_langevin} for the finite dimensional case).
 A few comments are in order. First, the stochastic partial differential equation \eqref{mft} is ill-defined \cite{konarovskyi2020dean}, and has to be interpreted as a formal restatement of the large deviations principle given by the action \eqref{Amacro}. Alternatively, it can be viewed as an effective low-wave-number field theory, which is equivalent to restoring a minimal distance -- the underlying lattice constant $\epsilon$. Also, \eqref{mft} is a continuous-space limit of the underlying lattice model, not a hydrodynamic equation for a macroscopic field obtained by locally integrating (i.e. coarse-graining) the variables $c_{\alpha, x}$ \cite{spohn2012large}.   

Equation \eqref{mft} is at the basis of the macroscopic fluctuation theory \cite{Bertini2015Jun}, in which the diffusion flux is recast as the Fick law 
\begin{equation}
\chi_\alpha(c) \nabla \frac{\delta \phi}{\delta c_\alpha} = D_\alpha(c) \nabla c ,
\end{equation}
with the diffusivity obeying the Einstein relation 
\begin{equation}
D_\alpha = \chi_\alpha \partial_{c_\alpha}^2\phi .
\end{equation}
This is a form of fluctuation-dissipation relation that replaces in continuous space the detailed balance relation. In fact, the continuous-space limit reduces the initial local detailed balance assumption to a condition of local equilibrium: since the transitions $\rho \in C$ are associated to points in physical space, the reservoirs responsible for them must be accordingly distributed in space, thus creating a continuum background of local  equilibria. 

A particularly known instance of \eqref{mft} is the Dean-Kawasaki equation, which describes the density of identical particles interacting through a two-body potential $U(r-r')$ \cite{dean1996langevin,kawasaki1998}. It is obtained by setting 
\begin{align}\label{isothermal_phi}
\phi[c]=\frac{1}{T}u[c]-s_\text{int}[c],
\end{align}
with the scaled internal energy and entropy
\begin{align}\label{interacting_BM}
\begin{aligned}
&u[c]=\frac{1}{2}  \int dr \int dr' c(r) U(r-r') c(r') \\
&s_\text{int}[c]= -\int dr c(r) (\ln c(r) -1) ,
\end{aligned}
\end{align}
as well as $\chi(c) \propto c$. 
While it was originally derived starting from the overdamped Langevin equations for the coordinates of an ensemble of identical particles, here it follows from the continuous limit of a Markov jump process on the lattice with transition rates satisfying the above mentioned scaling \cite{lefevre2007dynamics}. For $f=0$, the average value of \eqref{mft} reduces to the central equation of dynamic density functional theory \cite{marconi1999dynamic}.

Another important class is that of simple exclusion processes obtained using \eqref{isothermal_phi} with the free energy of an ideal binary mixture,
\begin{align}
&u[c]=0  \\
&s_\text{int}[c]= -\int dr [c(r) \ln c(r)+(1-c(r))\ln(1-c(r))] ,\nonumber
\end{align}
and the mobility $\chi(c) \propto c (1-c)$\cite{Bertini2015Jun,lefevre2007dynamics,tailleur2007mapping}.}
% This fluctuating field theory is derived by considering transition rates of the form
%$R_\rho(c_{\alpha,x}) \propto c_{\alpha,x} (1-c_{\alpha,x'})$ for the underlying lattice gas dynamics 

%When $R$ is not empty, the full action takes the form
%\begin{align*}
%  \mathcal{A}&=  \Omega \!\int \! dt \! \int \! dr  \bigg\{\!-\pi  \cdot d_t c + \sum_\rho r_\rho(c)  (e^{\Delta_\rho \cdot \pi } -1 )  \\
%&\quad+   \chi(c) \cdot\left[ \left(-   \nabla \frac{\delta \phi}{\delta c}  +   f (c) \right)\cdot \nabla \pi  + \frac 1 2 (\nabla \pi)^2  \right]   \bigg\} ,
%\end{align*}
The full action \eqref{Amacro} shows that for large $\Omega$ the most likely trajectories dominate the statistical averages. 
In particular, the probability density function of the stochastic field $c(r,t)$ acquires the large deviations form 
\begin{align}\label{rateI_continuous}
p[c(r),t] \asymp e^{-\Omega I[c(r),t] },
\end{align}
with rate functional $I$. Analogous considerations on the macroscopic dynamics discussed previously for the finite-dimensional case apply here to the field theoretical description \cite{Bertini2015Jun}. In particular, when the set $R$ is empty, the functional analog of the Hamilton-Jacobi equation \eqref{HJeq} can be obtained from 
the Fokker-Planck equation associated to \eqref{mft},
\cite{zinn02}
\begin{align}\label{FPE_continuous}
\begin{aligned}
&\partial_t p =\! \int dr \frac{\delta}{\delta c(r)} \left[-p \nabla \cdot j  + \frac{1}{\Omega}\frac{\delta( \chi p)}{\delta c(r)}  \right] =: - \int dr \frac{\delta \mathfrak{j}}{\delta c(r)} . 
%%&\mathfrak{j} :=  - p \nabla \cdot j - \frac{1}{\Omega}\frac{\delta}{\delta c} \chi p.
\end{aligned}
\end{align}
We have introduced the probability current $\mathfrak{j}[c]$ as done in Section \ref{subsec:orthogonal} for the finite dimensional case. It should not be confused with $j[c]$ that is the current of the physical field $c(r,t)$.

Plugging the large-deviation ansatz for $p[c]$, Eq. \eqref{rateI_continuous}, into \eqref{FPE_continuous}, we obtain the Hamilton-Jacobi equation
\begin{align}\label{HJ_continuous_time}
-\partial_t I[c(r),t]=\int dr \, \mathcal{H}_C  \! \left (c(r), \frac{\delta I}{\delta c(r)} \right),
\end{align}
whose stationary limit (when it exists) gives
\begin{align}\label{HJ_continuous}
\int dr \, \mathcal{H}_C  \! \left (c(r), \frac{\delta I_\text{ss}}{\delta c(r)} \right) =0.
\end{align}
{The instanton dynamics outlined in Sec. \ref{sec:hamiltonian_dyn} remains formally unaltered \cite{Elgart2004rare}}.

Moreover, one recognizes that the stationary probability velocity in the macroscopic limit (corresponding to the definition \eqref{velocity} for the finite dimensional case), 
\begin{align}
\begin{aligned}
 \lim_{V \to \infty}\lim_{t \to \infty} \frac{\mathfrak{j}[c]}{p[c]} &=-\nabla \cdot \left[ j(c) +  \chi(c) \cdot \nabla \frac{\delta I_\text{ss}}{\delta c}\right] \\
&= -\nabla \cdot  j_A(c),
\end{aligned}
\end{align}
is the negative divergence of the nonconservative part of the macroscopic physical current $j$. This generalizes \eqref{velocity} to field theories and gives the orthogonal decomposition (cf. Eq. \eqref{splitting_F}) for the physical current
\begin{align}\label{decomposition_continuous}
 j=  - \chi \cdot  \nabla \frac{\delta I_\text{ss}}{\delta c}  + j_{A},
\end{align}
with $\int dr \,   j_{A} \cdot \nabla \frac{\delta I_\text{ss}}{\delta c}  =0$.
A similar decomposition holds for nonautonomous dynamics, just replacing $I_\text{ss}$ with $I_\text{ss}^t$, the solution of \eqref{HJ_continuous} with driving frozen at its instantaneous value at time $t$.

 In the most general case, i.e. when the set $R$ is not empty, the macroscopic noiseless equation read
\begin{align}\label{reaction_diffusion_type}
\partial_t \mathcal{x}(r,t)= - \nabla \cdot  j + \sum_{\rho \in R} \Delta_\rho r_\rho(\mathcal{x}(r,t)),
\end{align}
which is the general form of a reaction-diffusion-advection equation of interacting \cite{aslyamov2023nonideal} and driven (or active) particles (see Sec. \ref{sec:ex3} for an example).
The last term in \eqref{reaction_diffusion_type} makes the stochastic field $c$ not conserved, i.e. in general $d_t \int dr c(r,t) = \int dr \sum_{\rho \in R} \Delta_\rho r_\rho(c(r,t)) \neq 0$. However, there might exist globally conserved quantities $c \cdot \ell$, with $\ell$ the left null vector(s) of the matrix $\Delta_\rho$, namely, $\ell \cdot \Delta_\rho =0$ for all $\rho \in R$ \cite{Rao2018a,Falasco2018a,Avanzini2019a}.
 Note that the decompositions \eqref{splitting_F} and \eqref{decomposition_continuous} are still valid, but the orthogonal conditions are replaced by the single property $\int dr [v_\text{ss} -\nabla \cdot j_{A} ] \frac{\delta I_\text{ss}}{\delta c} =0$. 

Extending the Gaussian theory of Sec. \ref{sec:Onsager_Machlup} to the infinite-dimensional case, we can linearize  the stochastic dynamics around the solutions $\mathcal{x}(r,t)$ of the deterministic equation \eqref{reaction_diffusion_type} setting $\varrho(r,t)=c(r,t)-\mathcal{x}(r,t) $. This amounts to a parabolic approximation of the rate functional $I[\varrho,t] = - \frac 1 2 \int dr \int dr' \varrho(r) \cdot \mathcal{C}^{-1}(r,r',t) \cdot \varrho(r') +O(\varrho^3)$, and gives an equation for the covariance matrix $\mathcal{C}$ analogous to \eqref{dC/dt}, and to the Lyapunov equation \eqref{fluctuation_dissipation_ss} under stationary conditions.
This approach was employed to derive analytic expressions for the correlation length of noisy Turing patterns \cite{vance1999spatial}, but a connection to thermodynamics has not been deeply explored -- see \cite{rana2020}, though.

\subsection{Continuous-space limit: entropy production}\label{sec:entropy_continuous}

The continuous-space limit can be implemented on the deterministic limit of the microscopic expression of the entropy production \eqref{epr_macro}. For transitions $\rho \in C$, the difference of transition rates and the nonconservative forces read at leading order in $\epsilon$, respectively,
\begin{equation}
\begin{aligned}\label{continuous_limit}
&r_\rho(c)-r_{-\rho}(c) \to \epsilon^{-2} \chi(c) \cdot \epsilon \left[- \nabla \frac{\delta \phi}{\delta c}+ f \right] =\frac{1}{\epsilon} j[c] \\
& \Delta_\rho \partial_c \phi+ a_\rho \to \epsilon \left[-\nabla \frac{\delta \phi}{\delta c}+ f \right]= \epsilon  \chi(c)^{-1}  \cdot j[c] ,
\end{aligned}
\end{equation}
where $j[c]$ is the hydrodynamic current defined in \eqref{mft}.
Plugging \eqref{continuous_limit} into \eqref{epr_macro} we get 
\begin{equation}\label{ep_mft}
\begin{aligned}
 \lim_{\Omega \to \infty } \frac 1 \Omega \langle \dot \Sigma \rangle &={ \int dr } j[\mathcal{x}] \cdot  \chi(\mathcal{x}(r,t))^{-1} \cdot j[\mathcal{x}]  \\
 &  \;+{  \int dr \sum_{\rho \in R}} r_\rho(\mathcal{x}(r,t))\sigma_\rho(\mathcal{x}(r,t)),
\end{aligned}
\end{equation}
where the first term corresponds to the entropy production associated to the Langevin equation \eqref{mft}. The second term in \eqref{ep_mft}, due to transitions unaffected by this limit, is different from the informational entropy production associated to the Langevin equation (see Sec. \ref{sec:path_langevin}). The thermodynamic inconsistencies arising from an incorrect application of the Langevin approximation have been observed for various specific model systems \cite{brillouin1950, Horowitz2015, Ceccato2018}. Here, we have shown that such inconsistency extends to field theories that try to infer the entropy production solely on the basis of the dynamics in state space. This means, for example, that in field theories of active matter all internal processes must be resolved in general, such as the chemical reactions necessary to self-propulsion. Also, we remark that the second line in \eqref{ep_mft} differs from the entropy production obtained by linearizing the dynamics of transitions $\rho \in R$ \cite{markovich2021}, unless they are close to detailed balance (i.e. $a_{\rho \in R} \to 0$).

The first term in \eqref{ep_mft} can be decomposed as done in \eqref{nc_driv_epr_macro} using \eqref{mft} and integrating by parts:
\begin{align}\label{decomposition_epr_continuous}
\begin{aligned} 
&\lim_{\Omega \to \infty } \frac 1 \Omega \langle \dot \Sigma \rangle =  -d_t   \phi [\mathcal{x}(r,t),t] + \partial_t \phi[c,t]|_{c=\mathcal{x}(r,t)} + \dot \sigma_{\text{nc}}.
\end{aligned}
\end{align}
On the righthand side, the first two terms are the total variation in time of the thermodynamic potential plus the dissipation of driving (zero only for autonomous dynamics), while the third is the total dissipation due to nonconservative forces acting in the discrete and continuous space,
 \begin{align}\label{wnc_continuous}
  \dot \sigma_{\text{nc}}=\int dr \left[ \sum_{\rho \in R} r_\rho(\mathcal{x}(r,t))a_\rho(\mathcal{x}(r,t)) + j[\mathcal{x}]\cdot f(r) \right].
 \end{align}
Here we have assumed that the flow $j \cdot \frac{\delta \phi}{\delta c}$ at the boundaries is zero. However, different boundary conditions can be imposed by a specific scaling of a set of transitions $\rho \in R$ located at the boundaries \cite{Bertini2015Jun}.

When the set $R$ is empty, one can show that the orthogonal decomposition \eqref{decomposition_continuous} naturally induces a splitting of the entropy production into the nonadiabiatic component (corresponding to the ``excess work'' of \cite{bertini2013clausius} for autonomous relaxation),
\begin{align}\label{ep_na_continuous}
\begin{aligned}
\lim_{\Omega \to \infty } \frac 1 \Omega \langle \dot \Sigma_\text{na} \rangle  &= -d_t I_{t}[\mathcal{x}(r,t)]+ \partial_t I_\text{ss}^t[c]|_{c=\mathcal{x}(r,t)}\\
\quad &={ \int dr } \nabla \frac{\delta I_{t}}{\delta \mathcal{x}}  \cdot \chi(\mathcal{x}) \cdot \nabla \frac{\delta I_{t}}{\delta \mathcal{x}} \geq 0,
\end{aligned}
\end{align}
and the adiabatic component,
\begin{align}\label{ep_ad_continuous}
\begin{aligned}
\lim_{\Omega \to \infty } \frac 1 \Omega \langle \dot \Sigma_\text{ad} \rangle &= \lim_{\Omega \to \infty } \frac 1 \Omega ( \langle \dot \Sigma \rangle - \langle \dot \Sigma_\text{na} \rangle   )  \\
\quad &={ \int dr } j_A[\mathcal{x}] \cdot  \chi(\mathcal{x})^{-1} \cdot j_A[\mathcal{x}] \geq 0.
\end{aligned}
\end{align}
Both \eqref{ep_na_continuous} and \eqref{ep_ad_continuous}, being quadratic forms, are nonnegative. 
We can combine \eqref{decomposition_epr_continuous} and \eqref{ep_ad_continuous} to eliminate the entropy production rate,
\begin{equation}
\begin{aligned}\label{renormalized_excess_work}
&  \dot \sigma_{\text{nc}} + \partial_t \phi[c,t]|_{c=\mathcal{x}(r,t)} - \lim_{\Omega \to \infty } \frac 1 \Omega \langle \dot \Sigma_\text{ad} \rangle 
  \\&=  d_t   \phi [\mathcal{x}] +\lim_{\Omega \to \infty } \frac 1 \Omega \langle \dot \Sigma_\text{na} \rangle  \\
  &\geq   d_t   \phi [\mathcal{x}].
 \end{aligned}
\end{equation}
Here, for systems with reservoirs at the same temperature $T$, the first line is the ``renormalized work'' rate (times the factor $T$) and the last line is the nonequilibrium Clausius inequality as defined by macroscopic fluctuation theory \cite{bertini2013clausius}. The ``renormalized work'' counts only the energy needed in a transformation, discarding the energy which keeps the system away from equilibrium.  In the context of stochastic thermodynamics, different definitions have been proposed for such a quantity \cite{hat01,mae14}, inspired by the early phenomenological approaches to nonequilibrium thermodynamics \cite{oon98,sas06}. Notice that \eqref{renormalized_excess_work} is not peculiar to the continuous-space limit, but can be equally written for the intensive quantities appearing in \eqref{nc_driv_epr_macro}, \eqref{mean_nonadiabatic} and \eqref{mean_adiabatic}.

\section{Emergent stochastic thermodynamics}\label{sec:emergent}

The picture appearing in Sec. \ref{sec:gauss_instanton} is that of a stochastic dynamics asymptotically characterized by two distinct regimes: relatively fast relaxation in a basin  (Eq. \eqref{macro_dcdt}) followed by small Gaussian fluctuations on the attractor (Eq. \eqref{linear_langevin}); rare large fluctuations corresponding to jumps between nearby attractors along transition trajectories, i.e. the sum of instanton and noiseless relaxation paths. Hence, it is natural to asymptotically describe the stochastic dynamics as a Markov jump process on the space of the various attractors (see Fig. \ref{fig:scales}) with transition rates given by \eqref{trans_saddle} at leading order in $V$ \cite{Qian2016Nov,smith2020intrinsic} . 
Such description is valid when the macroscopic dynamics \eqref{macro_dcdt} have isolated fixed points and sufficiently regular time-dependent attractors and is far from a bifurcation (see  Sec.~\ref{sec:ex3} for an example in which this condition is not met) \cite{freidlin,graham1987macroscopic}. 
For simplicity we restrict to autonomous dynamics in the following.

For large $V$ the probability density of states can be approximated as 
\begin{align}\label{emerging_p}
p(c,\tau) \asymp e^{-V I(c,\tau)} \to \sum_\gamma \mathcal{p}_\gamma(\tau) \delta_V(c-\mathcal{x}^*_\gamma)
\end{align}
where 
  $\delta_V(c-\mathcal{x}^*_\gamma)$ is a strongly peaked function with support on the ball $\mathcal{B}_\gamma$ of radius $O(V^{-1/2})$\footnote{The fact that Gaussian fluctuations are of order $O(V^{-1/2})$ can be seen from the Sec. \ref{sec:Onsager_Machlup}.} centered on $\mathcal{x}^*_\gamma$ (whose boundary is denoted $\partial \mathcal{B}_\gamma$).
  It is normalized such that  $\mathcal{p}_\gamma(\tau) := \int_{\mathcal{B}_\gamma} dc \, p(c,\tau)  $ and tends to a Dirac delta as $V \to \infty$. 
  
  The probability of finding the system in the neighborhood of the attractor $\gamma$ at time $\tau$, i.e. $\mathcal{p}_\gamma(\tau)$, is determined by the master equation
\begin{align}\label{me_emerging}
d_\tau \mathcal{p}_\gamma(\tau) = \sum_\nu \delta_{\mathtt{o}(\nu) \gamma}[    \kappa_{-\nu} \mathcal{p}_{\mathtt{o}(-\nu)}(\tau)-\kappa_\nu \mathcal{p}_\gamma(\tau)],
\end{align}
where $\mathtt{o}(\nu)$ denotes the origin of the transition $\nu$. 
 
Time-integrated observables of the form \eqref{time_int_obs} are split by summing the intra- and inter-basins contributions as
\begin{equation}
\begin{aligned}\label{B_split}
\mathcal{O}[\mathcal{X}] &=  \sum_\gamma \int_0^\tau dt \, \dot{\mathcal{O}}(t) \Theta_\gamma(t) \\
& \quad + \sum_\nu  \int_0^\tau dt \, \dot{\mathcal{O}}(t) \Theta_\nu(t) + \varepsilon[\mathcal{X}].
\end{aligned}
\end{equation}
Here, $ \Theta_\gamma(t)=1$ if $n(t) \in \mathcal{B}_\gamma$ and 0 otherwise, while $\Theta_{+\nu}(t)=1$ (resp. $\Theta_{-\nu}(t)=1$) if $n(t)$ belongs to a {tube of width $V^{-1/2}$ centered on the transition trajectory going from $\partial\mathcal{B}_\gamma$ to $\partial\mathcal{B}_{\gamma'}$ (resp. from $\partial\mathcal{B}_{\gamma'}$ to $\partial\mathcal{B}_{\gamma}$), and 0 otherwise. The functional $\varepsilon[\mathcal{X}]$ accounts for trajectories exiting and reentering the same ball, incomplete transition attempts, and transitions between different balls happening far from the transition path.} Since they become exponentially unlikely in the macroscopic limit, we neglect $\varepsilon[\mathcal{X}]$ in the following. 
Therefore, for large $V$ and for $\tau$ much larger than the typical relaxation times within each basin, we can recast the macroscopic rate of the scaled observable \eqref{B_split} as
%\begin{align}\label{b_split}
%\mathcal{o}[\mathcal{X}] & =  \sum_\gamma \dot{\mathcal{o}}_\gamma  {\textstyle \int_0^\tau dt} \Theta_\gamma(t) + \sum_\nu  \mathcal{o}_\nu \mathcal{N}_\nu(\tau) ,
%\end{align}
\begin{align}\label{b_split}
\frac{\mathcal{O}[\mathcal{X}]}{V \tau} =: \dot{ \mathcal{o}} & =  \sum_\gamma \dot{\mathcal{o}}_\gamma \mathcal{P}_\gamma(\tau) + \sum_\nu  \mathcal{o}_\nu \mathcal{I}_\nu(\tau) ,
\end{align}
where $\dot{\mathcal{o}}_\gamma, \mathcal{o}_\nu$, $\mathcal{P}_\gamma$ and $\mathcal{I}_\nu$ are independent stochastic variables defined in the following.

The observable $\mathcal{P}_\gamma(\tau):= \frac 1 \tau {\textstyle \int_0^\tau dt} \Theta_\gamma(t)$ is the empirical density on the attractor $\gamma$ satisfying $\mean{\mathcal{P}_\gamma}=\mathcal{p}_\gamma$.
The number of transitions between attractors $\gamma$ and $\gamma'$ through $\mathcal{x}_\nu$ divided by the time-span $\tau$ is given by $\mathcal{I}_\nu(\tau)$, which asymptotically approaches a conditional Poisson distribution with intensity $  \kappa_\nu$ given by \eqref{trans_saddle} \cite{freidlin,graham1987macroscopic,day83}. In other words, it is the empirical flux through the saddle $\nu$ satisfying $\mean{\mathcal{I}_\nu}=\kappa_\nu \mathcal{p}_{\mathtt{o}(\nu)}$. 
Note that the sum over $\nu$, in \eqref{b_split} and hereafter, counts transitions in the two directions.

The observable
\begin{align}
  \dot{\mathcal{o}}_\gamma := \frac{1}{V} \frac{\int_0^\tau dt \, \dot{\mathcal{O}}(t) \Theta_\gamma(t)}{\int_0^\tau dt \, \Theta_\gamma(t)}
\end{align}
 is the time-independent rate of $\mathcal{o}[\mathcal{X}]$ inside the attractor $\gamma$, i.e. in an infinitesimal neighborhood of $\mathcal{x}^*_\gamma$. Its scaled cumulant generating function can be obtained by the scaled tilted action \eqref{tilted_action} evaluated on the constant optimal trajectories $\bar c, \bar \pi$, i.e. the solutions of the time-independent version of \eqref{hamiltonian_eqs_z} such that $\bar c=\mathcal{x}^*_\gamma$ and $\bar \pi=0$ as $z \to 0$:
  \begin{align}\label{Kbgamma}
K_{\dot{\mathcal{o}}_\gamma}(z)= \mathcal{H}_z(\bar c, \bar \pi).
 \end{align}
In practice one may want to approximate the statistics of \eqref{Kbgamma} at Gaussian level by retaining only terms of order $O(z^2)$ in $\mathcal{A}_z$ \cite{nguyen2020exp}. 

The observable $\mathcal{o}_\nu$ is the contribution of $\mathcal{o}[\mathcal{X}]$ along the transition path through $\mathcal{x}_\nu$,
\begin{align}\label{obs_transition}
 \mathcal{o}_\nu := \frac{1}{V} \int_0^{\tau^\mathrm{tr}_\nu} dt \, \dot{\mathcal{O}}(t) \Theta_\nu(t)  .
\end{align}
Its scaled cumulant generating function can be approximated by the tilted action \eqref{tilted_action} evaluated on the optimal trajectories $\bar c(t), \bar \pi(t)$, i.e. the solutions of \eqref{hamiltonian_eqs_z} such that $\bar c(0) \in \partial \mathcal{B}_\gamma$ and $\bar c(\tau^\mathrm{tr}_\nu)\in \partial \mathcal{B}_{\gamma'}$ as $z \to 0$:
   \begin{align}\label{Kbnu}
K_{\mathcal{o}_\nu}(z,\tau^\mathrm{tr}_\nu)=  \mathcal{A}_z[\{\bar c(t), \bar \pi(t)\}].
 \end{align}
{The duration $\tau^\mathrm{tr}_\nu$ of such trajectory is a stochastic variable itself, whose mean differs from the mean escape time $1/\kappa_\nu$ from a basin of attraction: $\tau^\mathrm{tr}_\nu$ refers to the transition time of those (rare) trajectories that have just exited $\mathcal{B}_\gamma$, thus it does not account for the (exponentially long) dwelling time in the neighborhood of a fixed point -- which on the contrary enters $1/\kappa_\nu$. To the best of our knowledge, its probability distribution has been evaluated only for detailed balance systems with Gaussian noise. In this case, the most probable value of $\tau^\mathrm{tr}_\nu$ and its mean $\mean{\tau^\mathrm{tr}_\nu}$ grow as the logarithm of the inverse noise strength and coincide in the limit of vanishing noise \cite{caroli1979diffusion,caroli1980wkb,caroli1981diffusion,malinin2010transition}. 
Even though we are not aware of similar results for systems with nonconservative forces and nonGaussian noise, one can speculate to replace $\tau^\mathrm{tr}_\nu$  in \eqref{Kbnu} with its mean value, as measured in computer simulations or experiments.
We remark that not only the statistics of transition times but also those of (thermo)dynamic observables conditioned on transition paths remains a subject to be explored in systems lacking detailed balance.}

%Exact calculations can only be performed assuming a quadratic energy barrier -- and shows a sub-linear scaling in the latter. 
%In the large $V$ limit its cumulative distribution is obtained by maximizing the action over trajectories that start on the boundary of $\mathcal{B}_\gamma$ and end on the boundary of $\mathcal{B}_{\gamma'}$ in a large 
%at fixed $\tau^\mathrm{tr}_\nu$, and is expected to concentrate on $\mean{\tau^\mathrm{tr}_\nu}$.
 %For this reason one can approximate $\mathcal{o}_\nu(\tau^\mathrm{tr}_\nu)$ as the mean value of $\dot{\mathcal{O}}$ integrated along an instanton plus relaxation trajectory passing through $\mathcal{x}_\nu$ but excluding a neighborhood of the saddle -- in which the idealized trajectory would spend a diverging time and thus can contribute an unphysical large value to $\mathcal{o}_\nu$.

%Finally, dividing by $\tau$ we obtain the macroscopic rate along coarse-grained  stochastic trajectories
%\begin{align}
%\frac{\mathcal{o}[\mathcal{X}] }{\tau}&\asymp  \sum_\gamma \dot{\mathcal{o}}_\gamma \mathcal{P}_\gamma(\tau) + \sum_\nu  \mathcal{o}_\nu \mathcal{I}_\nu(\tau) ,
%%\end{align}
%where $\mathcal{P}_\gamma(\tau):= \frac 1 \tau {\textstyle \int_0^\tau dt} \Theta_\gamma(t)$ is the empirical density on the attractor $\gamma$ satisfying $\mean{\mathcal{P}_\gamma}=\mathcal{p}_\gamma$, and $\mathcal{I}_\nu(\tau) :=\frac 1 \tau \mathcal{N}_\nu(\tau)$ 

In particular, we can write the mean entropy production rate on the network of attractors,
\begin{align}
 \mean{\dot \sigma} &= \sum_\gamma \mean{\dot \sigma_\gamma} \mathcal{p}_\gamma + \sum_{\nu>0}  (\mean{\sigma_\nu }\kappa_\nu  \mathcal{p}_{\mathtt{o}(\nu)} + \mean{\sigma_{-\nu} }  \kappa_{-\nu}  \mathcal{p}_{\mathtt{o}(-\nu)}) \nonumber \\
  & \quad \quad - d_\tau \sum_\gamma \mathcal{p}_\gamma \frac{\ln \mathcal{p}_\gamma}{V},
\label{macro_jump_epr}
\end{align}
where the last term is obtained by using the expression for the probability density \eqref{emerging_p} in the Shannon entropy \eqref{Shannon_entropy_macro}.
Equation \eqref{macro_jump_epr} has three important differences with respect to its definition on the mesoscopic dynamics.
(i) The existence of a mean state entropy production rate,
\begin{align}\label{ep_metastable_states}
\mean{\dot \sigma_\gamma} =  \sum_{\rho >0}[r_\rho(\mathcal{x}^*_\gamma) -r_{-\rho}(\mathcal{x}^*_\gamma)] \ln \frac{r_\rho(\mathcal{x}^*_\gamma)}{r_{-\rho}(\mathcal{x}^*_\gamma)}\geq 0,
\end{align}
 independent of macroscopic transitions, stemming from the continuous dissipation needed to sustain the attractors. (ii) The mean entropy production associated to a macroscopic transition, $\mean{\sigma_\nu}$,
 not given by the log ratio of macroscopic transition rates $\kappa_{\pm \nu}$. Indeed, as shown in Sec. \ref{sec:gauss_instanton}, the transition rates do not in general obey the local detailed balance condition, unless close to equilibrium, see \eqref{ldb_macro_jump}.
(iii) The presence of all fluxes in \eqref{macro_jump_epr}, which cannot be written in terms of currents since in general $\sigma_{-\nu }\neq -\sigma_{\nu}$.
This is because the breaking of time reversibility in the underling mesoscopic dynamics implies that the transition path between $\gamma $ and $\gamma'$ is not equal to the time-reversed transition path between $\gamma'$ and $\gamma$. Even when these paths are the same, such as for the low dimensional systems of Secs. \ref{sec:ex1_SM} and \ref{sec:ex2_CMOS}, $\sigma_{-\nu }$  need not equal $-\sigma_{\nu}$ because of \eqref{currents_inst_rel}.

At linear order in $a_\rho$ around detailed balance, \eqref{macro_jump_epr} reduces to the standard expression of traditional stochastic thermodynamics \eqref{epr}. In this limit, we find that $ \mean{\dot \sigma_\gamma} =  \sum_\rho r^{(0)}_\rho(\mathcal{x}_{\text{eq}}) a_\rho  =0 $, the transition entropy production $\sigma_\nu$ changes sign under path reversal because of \eqref{reversibility}, and the transition rates satisfy local detailed balance \eqref{ldb_macro_jump}. %Therefore, $\mean{\dot \ttsigma} = \sum_{\nu>0}(\mean{\sigma_\nu }\kappa_\nu  \mathcal{p}_{\mathtt{o}(\nu)} )  + O(a_\rho^2)$.

%Using the master equation \eqref{me_emerging} and the probability conservation $\sum_\gamma \mathcal{p}_\gamma=1$, \eqref{macro_jump_epr} can be written as
%\begin{align}
%\mean{\dot \sigma} &= \sum_\gamma \mean{\dot \sigma_\gamma} \mathcal{p}_\gamma + \sum_{\nu}  (\mean{\sigma_\nu }+\ln \frac{\mathcal{p}_{\mathfrak{o}(\nu)}}{\mathcal{p}_{\mathfrak{o}(-\nu)}})\kappa_\nu  \mathcal{p}_{\mathtt{o}(\nu)} 
%\end{align}

Finally, we observe that the constant $\alpha_\gamma$ introduced in \eqref{global_QP} as the relative weight of the attractor $\gamma$ is related to the stationary probability $\mathcal{p}^\text{ss}_\gamma$ solution of \eqref{me_emerging} as
$\alpha_\gamma=-\frac 1 V \ln \mathcal{p}^\text{ss}_\gamma$. In the Freidlin-Wentzell theory \cite{freidlin}, $\mathcal{p}^\text{ss}_\gamma$ is calculated by the martrix-tree theorem for Markov chains \cite{zia07,polettini2015cycle,Gunawardena2022reformulating,dal2023geometry}, retaining only the leading spanning tree monomial in the limit $V \to \infty$ \cite{maes2013heat}.

\section{Applications}\label{sec:applications}

We now exemplify the general theory of the previous sections in three prototypical classes of model systems, namely, networks of chemical reactions, nonlinear electronic circuits and driven interacting units.

\subsection{Chemical reaction networks}\label{sec:ex1}

We consider chemical species in a volume $V$ of solvent in equilibrium at temperature $T$. Mass transport (e.g. by diffusion or advection) is taken to be fast enough to homogenize the local distribution of chemicals on the typical time scales of the reactions. All the internal (e.g. vibrational, rotational) degrees of the chemicals are assumed to be at equilibrium so that the only relevant dynamical variables are the number of particles $n_i$ of the dynamical species labelled by $i $. In a closed system, chemical reactions exclusively involve the $N$ chemical species. In open systems, they also involve $N_Y$ chemostatted species, whose concentrations $c_y$ are externally prescribed \cite{Rao2016}. 

We first consider closed systems and assume that chemical reactions 
are well described by a Markov jump process with transitions $\rho \in R$. The associated transition rates read \cite{gil92,Schmiedl2007}
\begin{align}\label{R_CRN}
R_\rho(n)= V k_\rho \prod_{i=1}^N  \frac{1}{V^{\nu^i_\rho}} \frac{n_i!}{(n_i-\nu^i_\rho)!},
\end{align}
where the stoichiometric coefficient $\nu^i_\rho$ (resp. $\nu^i_{-\rho}$) is the number of reactants (resp. products) of species $i$ involved in the reaction $\rho$. The entries of the jump matrix are given by $\Delta^i_{\rho}=\nu^i_{-\rho}-\nu^i_{\rho}$. 
Equation \eqref{R_CRN} is known as mass-action kinetics and holds for elementary reactions between diluted chemicals \cite{Avanzini2021a}.
The reaction constants $k_\rho$ are independent of $V$ and satisfy
\begin{align}
T \log \frac{k_\rho}{k_{-\rho}}= -\sum_{i=1}^N \Delta^i_\rho  \mu_i^\circ,
\end{align}
where $ \mu_i^\circ$ is the standard chemical potential of the dynamical chemical species $i$, including the contribution of the solvent. The mesoscopic local detailed balance \eqref{micro_ldb} holds in the form 
\begin{align}\label{bare_ldb}
T \Sigma_\rho(n)= - \Phi_\text{can} (n+\Delta_\rho)+ \Phi_\text{can} (n)
\end{align}
with the `canonical' Gibbs free energy of a mixture of $N$ ideal gasses,
\begin{align}\label{free_ene}
\Phi_\text{can} (n)= \mu^\circ \cdot n + T \ln n! \,,
\end{align}
The system is closed, namely, no matter is exchanged with the environment. The system dynamics is detailed balanced since \eqref{bare_ldb} does not display nonconservative forces, $a_{\rho}=0$. Moreover, the quantities $L(n):=\ell \cdot  n$ are conserved, with the vectors $\ell$ defined by $\ell \cdot   \Delta_\rho=0$ for all $\rho$ . The conservation follows from multiplying \eqref{eq_ndot} by $\ell$. For instance, the number of atoms of each chemical element and the charges are conserved. The system relaxes to the Gibbs distribution 
 \begin{align}
 P_\text{eq}(n) \propto e^{-\Phi_\text{can}(n)/T} \delta(L(n)-L)
 \end{align}
 constrained by the value of the conserved quantities fixed by the initial conditions. 
 
We now turn to open systems where chemostats are present, i.e. $N_Y \geq 1$ chemical species (labelled by $y$) have a large copy number controlled by external reservoirs while $N$ species continue to evolve stochastically. The transition rates \eqref{R_CRN} become
\begin{align}\label{R_CRN_open}
R_\rho(n)= V k_\rho \prod_{y=1}^{N_Y} c_y ^{\nu^y_\rho} \prod_{i=1}^{N}  \frac{1}{V^{\nu^i_\rho}} \frac{n_i!}{(n_i-\nu^i_\rho)!},
\end{align}
where $c_y=n_y/V$ is the externally controlled concentration of chemostatted species $y$. The local detailed balance of the open system reads
\begin{align}\label{bare_ldb_open}
T \Sigma_\rho(n)= -\Phi_\text{can} (n+\Delta_\rho)+ \Phi_\text{can} (n) - \sum_{y=1}^{N_Y} \Delta_\rho^y   \mu_y ,
\end{align}
where the last sum represents the free energy exchanged with the reservoirs with $\mu_y=\mu_y^\circ+RT \ln c_y$, the chemical potential of the chemostats. 

We obtain the Massieu potential $\Phi(n)$ that appears in \eqref{micro_ldb} by subtracting the free-energy contribution of the exchanged matter. The `semigrandcanonical'  Gibbs free energy reads
\begin{align}\label{free_ene_semigrand}
 \Phi (n)=\Phi_\text{can} (n)  -  \tilde \mu \cdot n,
\end{align}
where $\tilde \mu_i$ is the chemical potential of the part of species $i$ that is exchanged with the chemostats.
This is obtained by means of the quantities $L(n)$ that are no longer conserved when the system is opened.
The nonconservative forces $a_\rho$ can be given in terms of the projection of cycle affinities $\mathcal{f}^k_\text{nc}=\sum_\rho C^k_\rho \sum_{y=1}^{N_Y} \Delta_\rho^y \mu_y$ on the transition $\rho$,
\begin{align}\label{a_rho_CRN}
a_\rho %=\sum_{y=1}^{N_Y} \Delta_\rho^y \mu_y  +  \Delta_\rho \cdot  \tilde \mu 
= \sum_{k} \mathcal{f}^k_\text{nc}  \mathbb{X}^k_\rho ,
\end{align}
where $\mathbb{X}^k_\rho$ counts how many times the reaction $\rho$ takes place in the cycle $k$.
A cycle \footnote{Only \emph{emergent} cycles enter \eqref{a_rho_CRN}, i.e. those that appear when the system is chemostatted.} is a sequence of transitions $C_\rho$, defined by $\sum_\rho \Delta_\rho^i C_\rho =0 $ for all $i=\{1,\dots,N\}$, that overall leave the dynamical species unchanged but results in exchange of matter between reservoirs. 
 
In the macroscopic limit $V \to \infty$ the concentration $c_i=n_i/V$ of the species $i$ stay finite.  From \eqref{R_CRN} it is straightforward to 
%see that the kinetic coefficient $\Gamma(n) $ is of order $O(V)$ thus conforming to \eqref{lambda}, 
derive that the macroscopic transition rates \eqref{r_rho} are the polynomials 
\begin{align}\label{r_CRN}
%r_\rho(c)= k_\rho  \prod_{y=1}^{N_{y}} c_y ^{\nu^y_\rho}  \prod_{i=1}^N c_i^{\nu^i_\rho}.
r_\rho(c)= k_\rho   \prod_{j=1}^{N} c_j^{\nu^j_\rho}.
\end{align}
Moreover the Massieu potential \eqref{free_ene_semigrand_density} is extensive as required by \eqref{Phi_extensive}, and its density is obtained by applying Stirling's approximation:
\begin{align}\label{free_ene_semigrand_density}
 \phi (c)= \mu_\circ \cdot c + T \sum_{i=1}^N(c_i \log c_i-c_i) - \tilde \mu \cdot c.
\end{align}
At infinite volume $V$, the deterministic dynamics \eqref{macro_dcdt} is the chemical rate equation \cite{Kurtz1972,Kurtz2011}
\begin{align}\label{rate_eq_chem}
%d_t c=  \textstyle \sum_{\rho>0}\Delta_\rho \left(k_\rho \prod_{i=1}^N c_i^{\nu^i_\rho}-k_{-\rho} \prod_{i=1}^N c_i^{\nu^i_{-\rho}}
d_t c=  \textstyle \sum_{\rho}\Delta_\rho k_\rho \prod_{j=1}^{N} c_j^{\nu^j_\rho}.
\end{align}
Since \eqref{free_ene_semigrand_density} has a single minimum for a given value of conserved quantities, a macroscopic chemical reaction network at detailed balance can only have a single stable fixed point \cite{snarski2021hamilton}. To create multiple attractors it is necessary either to consider interactions (i.e. to go beyond the mass-action kinetics \cite{Avanzini2021a}) that make \eqref{free_ene_semigrand_density} non-convex or to introduce nonconservative forces that induce a quasi-potential $I_\text{ss}$ with multiple minima. Nonconservative forces can also create more complex time-dependent attractors, such as limit cycles \cite{boland2008limit} or chaos \cite{epstein1998,gaspard2020}. In the following we provide examples for dissipative multistability and periodic attractors.

%We show next that for limit cycles the general approach presented in the previous sections remain valid.

\subsubsection{Dissipative metastability: Schl\"{o}gel model}\label{sec:ex1_SM}

We consider the Schl\"{o}gel model \cite{schlogl1972chemical,vellela2009,Lazarescu2019Aug}, an autocatalytic chemical system displaying bistability far from equilibrium.
It consists in the following set of chemical reactions,
\begin{equation}
\ce{2X + A <=>[$k_{+1}$][$k_{-1}$] 3X}\:\: ; \: \quad \ce{ B  <=>[$k_{+2}$][$k_{-2}$] X \:.}
\end{equation}
 Here, species A and B are chemostated reservoirs with constant concentration $a$ and $b$, respectively. The only degree of freedom is the number $n$ of molecules of species X. The mesoscopic reaction rate \eqref{R_CRN} read
\begin{align}
\begin{aligned}
&R_1(n)=  \frac{k_{+1}}{V}  a\, n(n-1),\qquad  \qquad R_2= k_{+2} b \\
&  R_{-1}(n)=  \frac{k_{-1} }{V^2} n(n-1)(n-2), \quad R_{-2}(n)= k_{-2} n.
\end{aligned}
\end{align}
and in the macroscopic limit $V \to \infty$ they give for \eqref{r_CRN}
\begin{align}
\begin{aligned}
&r_1(c)=  k_{+1}  a\, c^2 , &&r_{-1}(c)=  k_{-1} c^3 ,\\
   &r_2= k_{+2} b,   && r_{-2}(c)= k_{-2} c.
\end{aligned}
\end{align}
The chemical rate equation reads \eqref{rate_eq_chem}
\begin{align}\label{rate_eq}
d_t \mathcal{x} = r_1(\mathcal{x})-r_{-1}(\mathcal{x}) + r_2-r_{-2}(\mathcal{x}).
\end{align}
 The only cycle affinity $\mathcal{f}_\text{nc}=\mu_A-\mu_B=: \Delta \mu$ is the chemical potential difference of the two reservoirs and corresponds to the log ratio of rates along a cycle 
 \begin{align}\label{f_nc_schlgel}
  \Delta \mu = \ln \frac{r_1(c)r_{-2}(c)}{r_{-1}(c)r_{2}}= \ln \frac{a k_1  k_{-2}}{b k_{-1}k_{2} }.
 \end{align}
If we set $k_{+1}a =1=k_{+2}b$ by choosing appropriate units for time and concentrations, \eqref{f_nc_schlgel} gives the parametrization of the rate constant $k_{-1}=k_{-2}e^{-\Delta \mu}$ in terms of $k_{-2}$ and $\Delta \mu$.
Detailed balance is realized for $\Delta \mu=0$, such that both currents vanish $r_1(\mathcal{x}^\text{eq})-r_{-1}(\mathcal{x}^\text{eq})=0=r_2-r_{-2}(\mathcal{x}^\text{eq}) $. This condition yields the equilibrium fixed point $\mathcal{x}^\text{eq}=1/k_{-2}$. 

\begin{figure}
\center
\includegraphics[width=0.5\textwidth]{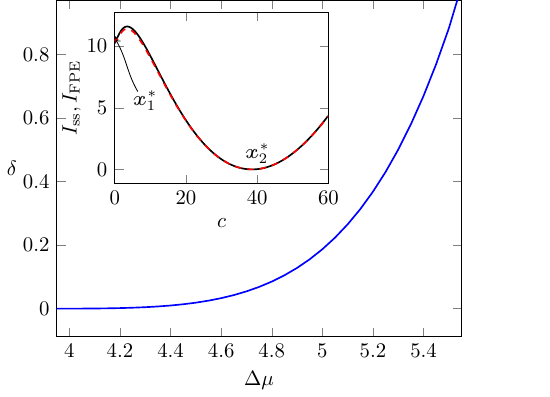}
\caption{Difference in the barrier height \eqref{delta_barrier} for the fixed point $\mathcal{x}_2^* > \mathcal{x}_1^*$ as a function of the chemical potential difference $\Delta \mu$ for $k_{-2}=3.5$. Bistability is present above the critical value $\Delta \tilde \mu \simeq 3.85$. \emph{Inset:} The quasi-potential $I_\mathrm{ss}$  given by \eqref{eq:Iss_schlogl} (solid) and its approximation ${I_\mathrm{NLE}}$  (dashed)  for $k_{-2}=3.5$ and $\Delta \mu=5$. 
\label{fig:Schlogl_delta}}
\end{figure}

Since the model is a one-step jump process in one dimension, the exact quasi-potential $I_\text{ss}$ is obtained by integrating  \eqref{one_step_Iss},
 \begin{align}
 \begin{aligned}
  I_\text{ss}(c)%&=\int^c \log \left(\frac{y (1+e^{-\Delta \mu} y^2)}{\mathcal{x}^\text{eq}(1+y^2)}\right) \, dy  \\
  &=c \log \left(\frac{c (1+e^{-\Delta \mu } c^2)}{\mathcal{x}^\text{eq}(1+c^2)}\right)-c + \text{const}\\
  &+2 e^{\frac{\Delta\mu}{2}} \arctan \left(e^{-\frac{\Delta \mu}{2}} c\right)-2 \arctan(c),
 % &=\underbrace{x\log(k_{-2}x)-x}_{\beta \phi(c)}
%  \underbrace{+ \Delta \mu(\arctan(c)-x) + \mathcal{O}((\Delta \mu)^2)}_{g(c)}
  \label{eq:Iss_schlogl}
  \end{aligned}
\end{align}
where the constant is the normalization ensuring that $\min_\gamma I_\text{ss}(x^*_\gamma)=0$. For $\Delta \mu =0 $ the first line in \eqref{eq:Iss_schlogl} is the free energy \eqref{free_ene_semigrand_density}.
The rate function is non-convex for $\Delta \mu \geq \ln 9$ and can display two local minima, $\mathcal{x}^*_1<\mathcal{x}^*_2$, only for $k_{-2} \geq \sqrt{3}$, corresponding to two stable fixed points of \eqref{rate_eq}. The appearance of metastability at the stochastic level corresponds to {a saddle-node or a super-critical pitchfork bifurcation at the deterministic level, depending on how the parameters $k_{-2}$ and $\Delta \mu$ are varied \cite{remlein2024nonequilibrium}}. At large $V$ the transition times between the two metastable states are determined by the quasi-potential barrier height $\Delta I_\mathrm{ss}:=I_\mathrm{ss}(\mathcal{x}_\nu)-I_\mathrm{ss}(\mathcal{x}^*)$, according to \eqref{trans_saddle}.
We compare the exact expression \eqref{eq:Iss_schlogl} with the approximated quasi-potential \eqref{one_step_Iss_NLE} obtained by truncating the master equation into a chemical Fokker-Planck equation \eqref{chemical_FPE}. Fig. \ref{fig:Schlogl_delta} displays the difference in barrier height between the two expressions, 
\begin{align}
\label{delta_barrier}
\delta:= \Delta I_\mathrm{ss}- \Delta I_\mathrm{FPE},
\end{align}
varying $\Delta \mu$ at fixed $k_{-2}$. The chemical Fokker-Planck equation is accurate close to the bifurcation while it systematically underestimates the barrier height away from it \cite{gaveau}. Note that a difference $\delta \simeq 0.2$, for $V \simeq 30$, would result in a transition rate around 400 times larger than $\kappa_\nu$.

Since the system is one-dimensional, the stationary probability current \eqref{current} must be zero everywhere in order to vanish at infinity and to have zero derivative.
Therefore, $v_\text{ss}(c)=0$ for all $c$ and the deterministic drift $F(c)=-\mathcal{M}(c) \partial_c I_\text{ss}$ is only composed of the gradient part with mobility \eqref{explictM}, $\mathcal{M}= (\tilde r_1-\tilde r_{-1})/\ln \frac{\tilde r_1}{\tilde r_{-1}}$, a  monotonically growing function of $c$. Here, $\tilde r_1= r_1+r_2$ and $\tilde r_{-1}= r_{-1}+r_{-2}$ are the lumped rates that enter the information-theoretic entropy production rate \eqref{info_epr_macro},
\begin{align}\label{ep_info_schlogl}
\dot{\sigma}_\text{info} (c)=\underbrace{(\tilde r_1(c)-\tilde r_{-1}(c))}_{F(c)}\underbrace{\ln \frac{\tilde r_1(c)}{\tilde r_{-1}(c)}}_{-\partial_c I_\text{ss}(c)}.
\end{align}
%which results in $\dot{\sigma}_\text{info} (\mathcal{x}(t))= - d_t I_\text{ss} (\mathcal{x}(t)) $.
Hence, the dynamics in configuration space appears reversible \footnote{One can also write the drift field as the derivative of a function $F(c)=-d_c h(c)$ such that $d_t h(\mathcal{x}(t))=-F^2(\mathcal{x}(t)) \leq 0$, which has no connection with the quasi-potential, though.} and one can check by direct substitution of \eqref{one_step_Iss} into \eqref{instanton_dynamics} that relaxation and instanton dynamics are one the time reversal of the other:
\begin{align}
\begin{aligned}\label{schlogl_reversibility}
d_t c(t)
%&=\sum_{\rho>0} \Delta_{\rho} \left(r_\rho e^{\Delta_\rho \partial_c I_\text{ss}}-r_{-\rho }e^{-\Delta_\rho \partial_c I_\text{ss}} \right)\\
&=\tilde r_1 e^{ \partial_c I_\text{ss}}-\tilde r_{-1} e^{- \partial_c I_\text{ss}} =-(r_1-r_{-1}) \\
&=-d_t \mathcal{x}(t).
\end{aligned}
\end{align}
However, the underlying chemical reactions are not in equilibrium as witnessed by the fact that the dual rates 
\begin{align}
\begin{aligned}
&r^\dagger_1=\frac{c^4+c^2}{e^{\Delta \mu }+c^2}, \quad \quad r^\dagger_{-1}=\frac{e^{-\Delta \mu }  c^5+ c^3}{\mathcal{x}^\mathrm{eq}(x^2+1)},\\
& r^\dagger_{2}=\frac{e^{\Delta \mu } \left(c^2+1\right)}{e^{\Delta \mu }+c^2}, \quad r^\dagger_{-2}=\frac{e^{-\Delta \mu } c^3+ c}{\mathcal{x}^\mathrm{eq}(c^2+1)},
\end{aligned}
\end{align}
equal the physical rates $r_\rho$ only at $\Delta \mu=0$. Interestingly, for $\Delta \mu \neq 0$ they do not follow the mass action law \eqref{r_CRN}.

\begin{figure}
\center
\includegraphics[width=0.55\textwidth]{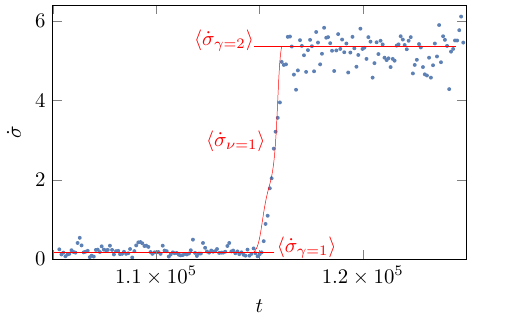}
\caption{Instantaneous entropy production rate \eqref{ep} along a stochastic trajectory (obtained via Gillespie algorithm \cite{gil77} with $\Delta \mu =2.84$, $k_{-2}=2.07$ and $V=10^2$) starting in the basin of attraction of $\mathcal{x}^*_1$ and ending in the one of $\mathcal{x}^*_2$. The solid lines are the constant values of the macroscopic mean entropy production in the two fixed points \eqref{ep_metastable_states} and the mean entropy production rate along the transition trajectory \eqref{ep_transition}. }
\label{fig:Schlogl_epr}
\end{figure}

Since bistability is created only far from equilibrium, $\sigma_{\gamma \to \nu}$ and $|\sigma_{\nu \to \gamma}|$ are very large, thus making the bound \eqref{upper_lower_bounds} very loose. Nevertheless, the bound turns into an equality if the thermodynamic entropy production is replaced by \eqref{ep_info_schlogl}, which equals the time derivative (resp. negative time derivative) of the rate function \eqref{eq:Iss_schlogl} along the instanton (resp. the relaxation) thanks to \eqref{schlogl_reversibility}:
\begin{align}
\dot{\sigma}_\text{info}  =d_t I(c(t)) =-d_t I(\mathcal{x}(t)).
\end{align}
For long times the system can be described by a 2-state Markov jump processes with transition rates $\kappa_{\pm 1}$ \cite{vellela2009}. The entropy production \eqref{ep} of the underlying dynamics is compared in Fig. \ref{fig:Schlogl_epr} with the internal entropy production of the metastable states \eqref{ep_metastable_states} and the entropy production rate calculated along the transition path 
from a neighborhood of $\mathcal{x}_1^*$ to a neighborhood of $\mathcal{x}_2^*$ in time $\mean{\tau^\mathrm{tr}}$,
 \begin{align}\label{ep_transition}
\mean{\dot \sigma_\nu(t)} =  \sum_\rho  r_\rho(c(t))e^{\Delta_\rho \cdot \pi(t)} \ln \frac{r_\rho(c(t))}{r_{-\rho}(c(t))}.
\end{align}
Here, $c(t)$ and $\pi(t)$ are solution of \eqref{hamilton_eq} on the manifold $\mathcal{H}(c(t),\pi(t)) =E $ implicitly determined by the relation 
\begin{align}
\mean{\tau^\mathrm{tr}}=\int_0^\mean{\tau^\mathrm{tr}} dt= \int_\mathcal{D} \frac{dc}{\dot c(c,\pi(c,E))},
\end{align}
with integration domain $\mathcal{D}=[\mathcal{x}_1^* + 1/\sqrt{\Omega},\mathcal{x}_2^* - 1/\sqrt{\Omega}]$.
We note that, since there exist no exact formula for $\mean{\tau_{tr}}$, its value is estimated using the expression valid for detailed balanced dynamics \cite{malinin2010transition}, replacing the height of the energy barrier with the one of the quasi-potential.

\subsubsection{Entropy production in a limit cycle: Brusselator model}\label{sec:ex1_bru}

We consider the Brusselator model, a prototypical model displaying a limit cycle far from equilibrium \cite{prigogine1968symmetry,lefever1988brusselator,AndrieuxGaspardJCP2008,nguyen2020exp}.
It is an autocatalytic chemical reaction network,
 \begin{align}
 &\ce{Y_1 <=>[$k_{+1}$][$k_{-1}$] X_1}\:\: ; \: \quad \ce{ X_1 +Y_2  <=>[$k_{+2}$][$k_{-2}$] X_2 +Y_3 }\\
&\ce{2X_1 + X_2 <=>[$k_{+3}$][$k_{-3}$] 3X_1}\:\: ; \: \quad \ce{ X_1  <=>[$k_{+4}$][$k_{-4}$] Y_4 \;,}
\end{align}
 made up of two dynamical species $\text{X}_1$ and $\text{X}_2$ and four chemostated species $\text{Y}_y$ (with concentrations denoted $c_{\text{Y}_y}$). The model has 2 emergent cycles $C^1=(1,0,0,1)$ and $C^2=(0,1,1,0)$ with corresponding cycles affinities $\mathcal{f}_{nc}^1=\log \frac{k_1 k_4 c_{Y_1} }{k_{-1}k_{-4} c_{Y_4}}$ and $\mathcal{f}_{nc}^2=\log \frac{k_2 k_3 c_{Y_2} }{k_{-2} k_{-3} c_{Y_3}}$. The dynamics is detailed balanced when the chemical potentials of all $c_{Y_i}$'s are equal, so that the cycle affinities are zero.
 
\begin{figure}
\center
\includegraphics[width=0.5\textwidth]{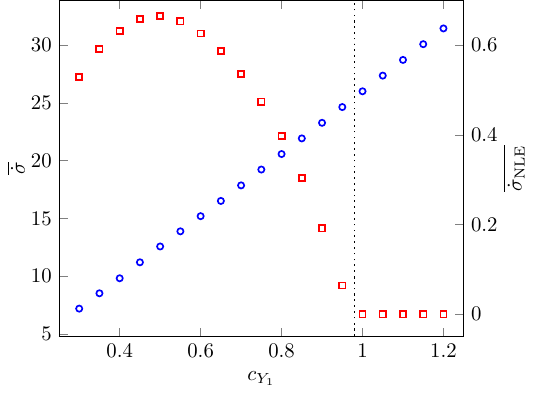}
\caption{Mean scaled entropy production rate averaged over one period in the infinite system-size limit: Exact expression  corresponding to \eqref{epr_macro} (circles $\textcolor{blue}{\circ}$) and the approximation based on the expression  \eqref{ep_nle_mean} given by the nonlinear Langevin equation (squares ${\square}$). The vertical dashed line shows the critical value $\tilde c_{Y_1} \simeq 0.97$ for the onset of the supercritical Hopf bifurcation. Parameters are $k_{1} = 1.00, k_{2}c_{Y_2} = 3.00, k_{3} = 1.00, k_{4} = 1.00, k_{-1} = 0.01, k_{-2}c_{Y_3} = 0.01, k_{-3} = 0.01, k_{-4}c_{Y_4}  = 0.01$.}
\label{fig:Brus}
\end{figure}

Choosing the concentration of $Y_1$ as a control parameter, the deterministic rate equation \eqref{rate_eq_chem} displays a supercritical Hopf bifurcation, namely, a  transition from a single stable fixed point to a stable limit cycle with period $t_\mathrm{p}$. This is seen from a linear stability analysis of the deterministic rate equations,
\begin{align}\label{rate_eq_brus}
\begin{aligned}
&d_t \mathcal{x}_{1} =r_1 -r_{-1} - r_2 +r_{-2} + r_3 -r_{-3} - r_4 +r_{-4} , \\
&d_t \mathcal{x}_{2} = r_2 -r_{-2} - r_3 +r_{-3}, 
\end{aligned}
\end{align}
with macroscopic scaled rates
\begin{align}
\begin{aligned}
&r_1=k_1 c_{Y_1},\quad \quad r_{-1}=k_{-1} c_1, \\
&r_{2}= k_2 c_1 c_{Y_2}, \quad r_{-2}= k_{-2} c_2 c_{Y_3},\\
&r_3=k_3 c_{1}^2 c_2, \quad \, \, r_{-3}=k_{-3} c_1^3,\\
& r_{4}= k_4 c_1 , \quad \quad \; r_{-4}= k_{-4} c_{Y_4}.
\end{aligned}
\end{align}
The eigenvalues of the Jacobian of the drift field evaluated at the fixed point, $\partial_c F(\mathcal{x}^*)$, are a complex pair that develops a positive real part above some critical value $\tilde c_{Y_1}$ and a stable periodic trajectory appears. 
Here we point out the failure of the Langevin approximation in the description of the system thermodynamics. The 
macroscopic entropy production rate associated to the chemical Langevin equation \eqref{ep_nle_mean} time-averaged over a period reads
\begin{align}\label{epr_periodic_NLE}
\begin{aligned}
&\overline{\dot \sigma_\text{NLE}}:= \frac{1}{t_\text{p}} \int_0^{t_\text{p}} dt \lim_{V \to \infty } \langle \dot \sigma_\text{NLE} \rangle \\
&= \frac{1}{t_\text{p}}  \int_0^{t_\text{p}} dt F(\mathcal{x}^*(t))  \cdot D^{-1}(\mathcal{x}^*(t)) \cdot F(\mathcal{x}^*(t)) ,
\end{aligned}
\end{align}
with $\mathcal{x}^*(t)=\mathcal{x}^*(t+t_\text{p})$ the periodic solution of \eqref{rate_eq_brus} 
%and diffusion coefficient
%\begin{align}
%D(c)= \frac 1 2 
%\begin{pmatrix}
%\sum_{\rho=1}^4 (r_\rho+r_{-\rho}) & -\sum_{\rho=2}^3 (r_\rho+r_{-\rho}) \\
%-\sum_{\rho=2}^3 (r_\rho+r_{-\rho}) & \sum_{\rho=2}^3 (r_\rho+r_{-\rho})
%\end{pmatrix}
%\end{align}
In Fig. \ref{fig:Brus}, we show, by integrating numerically the rate equation \eqref{rate_eq_chem}, that the exact period-averaged entropy production rate 
\begin{align}
%\begin{aligned}
&\overline{\dot \sigma}:=\frac{1}{t_\text{p}} \int_0^{t_\text{p}} dt \lim_{V \to \infty }  \langle \dot \sigma \rangle \\
&= \frac{1}{t_\text{p}}  \int_0^{t_\text{p}} dt \sum_{\rho>0} [r_\rho(\mathcal{x}^*(t)) - r_{-\rho}(\mathcal{x}^*(t))] \ln \frac{r_\rho(\mathcal{x}^*(t)) }{r_{-\rho}(\mathcal{x}^*(t))}  ,\nonumber
%\end{aligned}
\end{align}
is largely underestimated by the approximation \eqref{epr_periodic_NLE}, and even its qualitative behavior as a function of $c_{Y_1}$ is incorrect. In particular, the latter is identically zero for $c_{Y_1} > \tilde c_{Y_1}$, i.e. in the absence of limit cycle, since the drift $F$ is by definition null on the fixed point, as shown in general in \eqref{ep_nle_mean_stationary}. Note that the Langevin approach is often used to estimate the energetic cost of biochemical oscillations \cite{Xiao08,Cao20}.

\subsection{Electronic systems}\label{sec:circuit}

We consider an arrangement of ideal conductors whose state is determined by the charge vector $q$  and electrostatic potential vector $V$ \cite{Freitas2021ec}. The potentials are measured with respect to some reference (ground) potential and are linearly related to the charges, $q= C \cdot V $, through the capacitance matrix $C$. The internal energy of the system is thus
\begin{align}\label{energy_circuit}
E= \frac 1 2 q \cdot V.
\end{align}
The system is open when $N_Y$ conductors have fixed potentials imposed by external voltage sources, so that the state is specified by $N$ charges. Elementary charges with value $q_e$ are transported  between pairs of conductors by two-terminal devices. Each of these channels is modeled as a bidirectional Poisson process labelled by $\pm \rho$ with rate $R_{\pm \rho}(q)$, which changes the system state as $q \to q \pm q_e \Delta_{\rho}$. This allows one to describe many  relevant devices, e.g., tunnel junctions, diodes, MOS transistors in subthreshold operation \cite{Freitas2021ec}. The matrix with entries $\Delta^i_\rho$, encoding the circuit network, is the analogue of the stoichiometric matrix of chemical reactions in Sec. \ref{sec:ex1}. Its left (resp. right) null vectors identify the conserved quantities (resp. the cycles).  The stochastic dynamics is given by \eqref{eq_ndot} with $n q_e=q$ and $q_e J_\rho$ the electric current on each channel device. Heat conduction is assumed large enough such that the temperature in each conductor is constant and equal to $T$ (i.e. no self heating \cite{semenov2006impact}).

A closed circuit satisfies the detailed balance condition
\begin{align}\label{ldb_cicrcuits}
T \ln \frac{R_\rho(q)}{R_{-\rho}(q+q_e\Delta_\rho)} = - E(q+q_e \Delta_\rho)+E(q),
\end{align}
with the quadratic energy \eqref{energy_circuit} (because $S_\mathrm{int}=0$),
and relaxes to the equilibrium distribution at fixed value of the conserved quantities $L(q)$,
\begin{align}\label{Gibbs_circuit}
P_\text{eq}(q) \propto e^{-E(q)/T} \delta(L(q)-L).
 \end{align}
For open circuits, the Massieu potential is obtained by subtracting from the electrostatic energy the contribution exchanged with the regulated conductors
\begin{align}\label{free_ene_circuit}
 \Phi (q)=E (q)  -  \tilde V \cdot q,
\end{align}
with an appropriate voltage vector $\tilde V$ that is constructed from the quantities $L(q)$ that are no longer conserved when the system is opened.
Very analogously to chemical reaction networks, the forces $a_\rho$ are differences between the voltages of regulated conductors in the connected components of the channel $\rho$ \cite{Freitas2021ec}.

The explicit form of the transition rates can be obtained from the $I-V$ curve that gives the macroscopic average electric current $\mean{I}(V)$ through a two-terminal device as a function of the applied voltage $V$ across it. In the case of a constant voltage $V$, the mean current  
\begin{align}\label{I-V_curve}
\mean{I}(V)= q_e (R_{+}-R_{-})
\end{align}
and the local detailed balance 
\begin{align}\label{ldb_V_constant}
T \ln \frac{R_+}{R_{-}}= q_e V
\end{align}
are two independent equations that allow one to determine $R_{\pm}$. In the general case of a device embedded in a circuit, the transition rates $R_{\pm \rho}$ are obtained replacing the constant $V$ in \eqref{I-V_curve} and \eqref{ldb_V_constant} with the average of the voltage difference before and after the transition $q \to q \pm q_e \Delta_\rho $ \cite{Freitas2021ec}. In closed circuits, choices different from such midpoint rule would lead to a stationary probability density different from the equilibrium \eqref{Gibbs_circuit}, thus entailing the possibility to indefinitely extract energy (see Brillouin paradox \cite{brillouin1950}). 

We focus on those devices in which the characteristic capacitance $C \to \infty$ and the transition rates $R_{\pm \rho}$ (thus the charges $q \to \infty$) scale with the system size. Hence, in the macroscopic limit we   consider the elementary voltage $v_e = q_e/C \to 0$ negligible in comparison to all other voltage scales of the circuit and equal to the inverse of the large parameter $V\equiv 1/v_e \to \infty$. 

Open electronic circuits under detailed balance conditions cannot be used to store information nor to generate signals with specific frequencies. These are prevented by the fact that the the Massieu potential  \eqref{free_ene_circuit} has a unique minimum $\mathcal{x}_\mathrm{eq}=0$ and the spectrum of the generator of the (overdamped) dynamics is real (see Sec. \ref{sec:multi_meta}), respectively.
Multistability and oscillations require nonconservative forces created by voltage differences. In practice, they can be realized connecting 2 inverters, i.e. NOT gates, in a loop (SRAM cell \cite{Rezaei2020}) and an odd number of inverters in a chain (ring oscillator \cite{Hajimiri1999Jun}), respectively.

\subsubsection{Dissipative logical states: CMOS SRAM cell}\label{sec:ex2_CMOS}

\begin{figure*}
\center
\includegraphics[width=0.26\textwidth]{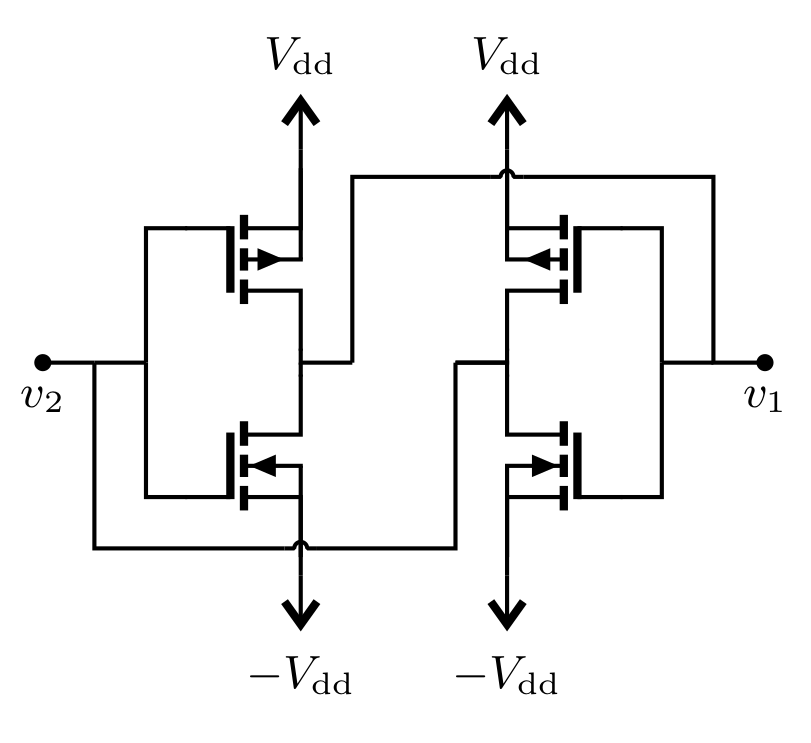}
\includegraphics[width=0.35\textwidth]{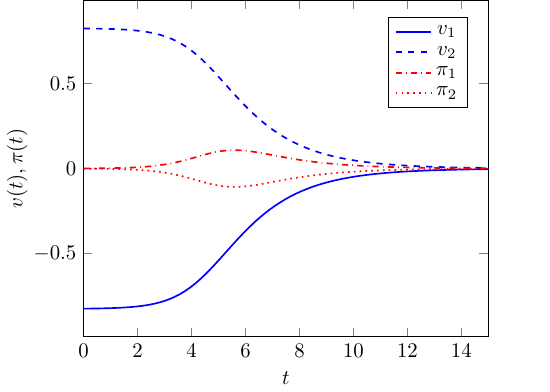}
\includegraphics[width=0.35\textwidth]{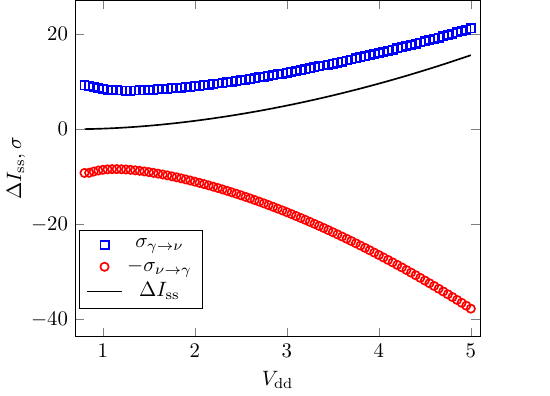}
\caption{\emph{Left}: Circuit diagram of a CMOS SRAM cell. 
\emph{Right}: Instanton obtained by the minimum action method \cite{zakine2023minimum} evolves at $v_1=-v_2$ and $\pi_1=-\pi_2$. Parameter values are $(I_0/C) e^{-V_\mathrm{th}/V_\mathrm{T}}$, $\mathrm{n=1}$, $V_\mathrm{T}=1$, $V_\mathrm{dd}=0.9$. 
\emph{Right}: The height of the quasi-potential barrier obtained from \eqref{Iss_CMOS} (solid) as a function of $V_\mathrm{dd}$, and its upper and lower bounds, i.e. entropy production of the instanton (square) and negative entropy production of the relaxation (circle), respectively. The latter ones are obtained according to \eqref{local_epr_CMOS} with a small parameter $\epsilon=0.02$. }
\label{fig:CMOS}
\end{figure*}

Low-power static random access memory (SRAM) cells are usually implemented by connecting two CMOS inverters. Each inverter is composed of a $p$MOS and a $n$MOS transistor, see Fig.~\ref{fig:CMOS}, and is powered by a voltage bias $2V_{dd}$. Each $p$MOS (resp. $n$MOS) transistor can be
modeled as a conduction channel between drain and source terminals, with associated transition rates $R^{p}_{\pm}$ (resp. $R^{n}_{\pm}$) \cite{Freitas2021ec}. The voltages $v_1$ and $v_2$ at the output of each inverter are the two independent degrees of freedom, given by $v=n q_e/C$ with $n$ the number vector of charges and $C$ the capacitance characterizing the device. Following the general procedure introduced in the previous section based on the I-V curve and the local detailed balance, with Massieu potential $\Phi(v_1,v_2)=C(v_1^2+v_2^2)/2+ CV_\mathrm{dd}^2$, one obtains 
\begin{align}
\begin{aligned}
R^{p}_+(v_1,v_2)&= (I_0/q_e)  e^{(V_\mathrm{dd}-V_\mathrm{th}-v_2)/(\mathrm{n}V_\mathrm{T})} \\
R^{p}_-(v_1,v_2)&= R^{p}_+(v_1,v_2)  e^{-(V_\mathrm{dd}-v_1)/V_\mathrm{T} -v_e/(2V_\mathrm{T})}
\end{aligned}
\end{align}
and $R^{n}_{\pm}(v_1,v_2)=R^{p}_{\pm}(-v_1,-v_2)$ \cite{freitas2022reliability}, where $I_0$, $V_\mathrm{th}$ and $\mathrm{n}$ are parameters characterizing the transistor. The non conservative work of the transitions is $a_{\pm \rho}= \pm q_e V_\mathrm{dd}$.

The macroscopic limit is obtained as $v_e \to 0$ with the bias voltage $V_\mathrm{dd}$ and the thermal voltage $V_\mathrm{T}=k_\mathrm{B} T/q_e$ kept finite. The corresponding macroscopic transition rates read
\begin{align}
\begin{aligned}
r^{p}_+(v_1,v_2)&= (I_0/C) e^{(V_\mathrm{dd}-V_\mathrm{th}-v_2)/(\mathrm{n}V_\mathrm{T})} \\
r^{p}_-(v_1,v_2)&= r^{p}_+(v_1,v_2)  e^{-(V_\mathrm{dd}-v_1)/V_\mathrm{T} },
\end{aligned}
\end{align}
and yield the deterministic dynamics of voltages
\begin{align}
\begin{aligned}\label{det_dyn_CMOS}
 d_t \mathcal{v}_1&=I_p(\mathcal{v}_1,\mathcal{v}_2) -I_n(\mathcal{v}_1,\mathcal{v}_2)\\
 d_t \mathcal{v}_2 &= I_p(\mathcal{v}_2,\mathcal{v}_1) -I_n(\mathcal{v}_2,\mathcal{v}_1),
\end{aligned}
\end{align}
in terms of the difference of currents between the two transistors, $I_{p/n}(\mathcal{v}_1,\mathcal{v}_2)=r^{p/n}_+(\mathcal{v}_1,\mathcal{v}_2)-r^{p/n}_{-}(\mathcal{v}_1,\mathcal{v}_2)$. Equations \eqref{det_dyn_CMOS} admit a single stable fix point $\mathcal{v}_1^*=\mathcal{v}_2^*=0$ for small bias $V_\mathrm{dd} \leq V_\mathrm{T}\ln2  $ and two stable (symmetric) fixed points $\pm \mathcal{v}^*=\pm(\mathcal{v}_\mathrm{bit},-\mathcal{v}_\mathrm{bit})$ with ($\mathrm{n}=1$)
\begin{align}
\mathcal{v}_{\mathrm{bit}}=V_\mathrm{dd} + V_\mathrm{T} \ln \left ( \frac 1 2  + \sqrt{1/4-e^{-2V_\mathrm{dd}/ V_\mathrm{T} }} \right),
\end{align}
separated by a saddle in the origin for $V_\mathrm{dd} > V_\mathrm{T}\ln2  $. These explicit formulae hold for $\mathrm{n}=1$, although more lengthy expressions can also be derived for $\mathrm{n} \neq 1$. The pitchfork bifurcation undergone by the dynamical system \eqref{det_dyn_CMOS} at the critical value $\tilde V_\mathrm{dd}:=V_\mathrm{T}\ln2$ corresponds to the emergence of bistability in the deterministic dynamics and makes the system usable as a volatile memory. Initializing the voltages in a given basin of attraction allows one to store 1 bit of information for a mean time equal to the inverse of the escape rate $\kappa \asymp e^{-[I_\mathrm{ss}(0)-I_\mathrm{ss}(\mathcal{v}^*)]/v_e}$. The energy dissipated to maintain such a logical state is $T \mean{\sigma_{\pm 1}}$, where the mean entropy production rate \eqref{ep_metastable_states} reads
\begin{align}
\mean{\dot \sigma_{\gamma=\pm 1}} &=  V_\mathrm{dd} q_e \sum_\rho \mean{ j_\rho }= 4V_\mathrm{dd} I_0 e^{-V_\mathrm{th}/V_\mathrm{T}} .
\end{align}

The quasi-potential cannot be obtained analytically. Nevertheless, if only rare fluctuations leading to memory losses are of interest, we can restrict the analysis to the linear space $c=(v_1-v_2)/2$. Due to the symmetry of the rates, relaxation and instanton trajectories that connect the fixed points to the saddle lie entirely on the reaction coordinate $c$ at $v_1+v_2=0$, see Fig. \ref{fig:CMOS}. The determination of $I_\mathrm{ss}(c)$ reduces to a one dimensional problem very analogous to the one of Sec.~\ref{sec:ex1_SM}. Writing \eqref{HJ0} in terms of the new variable $c$ and setting $v_1=-v_2$, $\pi_1=-\pi_2=\partial_c I_\mathrm{ss}(c)$ 
gives \eqref{one_step}, whose integral is
\begin{align}
%\begin{aligned}
 I_\mathrm{ss}(c) &= \frac{c^2}{V_\mathrm{T}} +\frac{\mathrm{n}}{ V_\mathrm{T}(\mathrm{n}+2)} [\text{L}(-c,V_\mathrm{dd})-\text{L}(c,-V_\mathrm{dd}) \nonumber \\
& \quad
+\text{L}(c,V_\mathrm{dd}) -\text{L}(-c,-V_\mathrm{dd})] +\mathrm{const},\label{Iss_CMOS}
%\end{aligned}
\end{align}
with $\text{L}(c,V_\mathrm{dd})= \mathrm{Li}_2(-\exp((V_\mathrm{dd}+c(1+2/\mathrm{n}))/V_\mathrm{T})$ and $\mathrm{Li}_2(.)$ the polylogarithm function of second order. Equation \eqref{Iss_CMOS} was derived in \cite{Freitas2021reliability} and is reported here in an equivalent form that shows explicitly the symmetry $I_\mathrm{ss}(c)=I_\mathrm{ss}(-c)$.

For $V_\mathrm{dd} \to 0$ \eqref{Iss_CMOS} reduces to the macroscopic Massieu potential on the subspace $v_1=-v_2$,
\begin{align}
\phi(c):= \lim_{v_e \to 0} v_e \Phi(c,-c)= \frac{c^2}{V_\mathrm{T}}.
\end{align}
The quasi-potential barrier takes the simple expression 
\begin{align}
I_\mathrm{ss}(0)-I_\mathrm{ss}(\mathcal{v}_\mathrm{bit}) \simeq \frac{2}{2 + \mathrm{n}} \frac{V_\mathrm{dd}^2}{V_\mathrm{T}}
\end{align}
in the far-from-equilibrium regime $V_\mathrm{dd} \gg \tilde V_\mathrm{dd}$, and  leads to the error rate due to thermal noise
\begin{align}
-\lim_{v_e \to 0} v_e \ln \kappa \propto  \mean{\dot \sigma_{\gamma =\pm 1}}^2.
\end{align}
Remarkably, the stability of the electronic memory grows proportionally to the square of the macroscopic dissipation of the device. This direct relation between the two quantities is made possible by the fact that the macroscopic current $I_0 e^{-V_\mathrm{th}/V_\mathrm{T}}$ is constant in the bistable regime.

As for the Schl\"{o}gl model of Sec.~\ref{sec:ex1_SM}, relaxation and instanton dynamics along $c$ are related 
by time reversal, sharing the same local speed $\dot c(c)$.
%\begin{align}
%\dot c(c)=\frac{I_0}{C} \left(e^{\frac{2 c}{\text{VT}}}-1\right) \left(e^{\frac{2 c}{\text{VT}}}-e^{\frac{x+1}{\text{VT}}}+1\right) e^{-\frac{\text{Vth}+2 c}{\text{VT}}}
%\end{align}
Yet, the corresponding local entropy production rates differ, leading in general to
\begin{align}\label{local_epr_CMOS}
\sigma_{\nu \to \gamma} = \int_\mathcal{D} dc \frac{\dot \sigma_{\nu \to \gamma}(c)}{\dot c(c)} \neq -\int_\mathcal{D} dc \frac{\dot \sigma_{\gamma \to \nu}(c)}{\dot c(c)} = -\sigma_{ \gamma \to \nu},
\end{align}
when the integration over the domain $\mathcal{D}=[1/\sqrt{\Omega},\mathcal{v}_\mathrm{bit} - 1/\sqrt{\Omega}]$ is performed. Equality in modulus is achieved only at linear order in $V_\mathrm{dd}$,
\begin{align}
\frac{\dot \sigma_{\nu \to \gamma}(c)}{\dot c(c)}&=-\frac{2 c}{V_\mathrm{T}} +\frac{2 V_\mathrm{dd}}{V_\mathrm{T}} \tanh \left(\frac{(\mathrm{n}+2) c}{2 \mathrm{n} V_\mathrm{T}}\right)+ O(V_\mathrm{dd}^2),
\end{align}
whose integral gives the near-equilibrium approximation \eqref{linear_resp} of the rate function \cite{freitas2021linear}. 
Since bistability does not appear at vanishing bias the thermodynamic bounds \eqref{upper_lower_bounds} are quite loose due to large values of the adiabatic entropy production, see Fig.~\ref{fig:CMOS}. Nevertheless, the bounds \eqref{upper_lower_bounds} turn into an equality by replacing the thermodynamic entropy production by the coarse-grained quantity \eqref{ep_info_schlogl} along $c$ with the lumped rates $\tilde r_\pm(v_1,v_2)= r^p_\pm(v_1,v_2) + r^n_{\mp}(v_1,v_2) $ \cite{Freitas2021NatCom}. Note that close to the bifurcation point the quasi-potential barrier vanishes and the regions of typical fluctuations around the fixed point overlap. The escape rate between attractors is dominated by sub-exponential terms neglected in \eqref{trans_saddle} and \eqref{upper_lower_bounds} becomes obsolete -- see Sec. \ref{sec:ex3} for a discussion of the attractor stability in this regime.

\subsection{{Driven Potts models}}\label{sec:ex3}

We consider a system that consists of $V \equiv M$ all-to-all interacting identical units subject to the same nonconservative force $a$ and identically coupled to a thermal bath of temperature $T$. Each unit is made of $N \equiv q$ identical states on a ring coupled to their nearest neighbors \footnote{We use the letter $q$ to denote the number of accessible mesoscopic states, as customary for the Potts model \cite{wu82potts}}. Due to the mean field nature of the interactions {between units}, all configurations with the same number $n_i$ of units in states $i=1, \dots, q$ are equivalent.
The internal entropy of the system resulting from such a degeneracy is \cite{Herpich2020Jun}
\begin{align}\label{S_int_Potts}
S_\mathrm{int}(n)= \ln \frac{M!}{\prod_{i=1}^q n_i!} ,
\end{align}
and the energy accounts for the interactions between all units in the same state $i$
\begin{align}\label{energy_Potts}
U= \frac{1}{2M} \sum_{i=1}^q u_i n_i (n_i-1),
\end{align}
where the prefactor $1/M$ makes the mean-field energy extensive, according to Kac's prescription \cite{kac1963van}. 
{The system has $q$ forward and $q$ backward mesoscopic transitions each connecting two successive single-unit states.} {The stochastic transitions $\rho=\pm i$ change the populations of nearby states $i$ and $i+1$ as $(n_i, n_{i+ 1}) \to  (n_i \mp 1 , n_{i+ 1} \pm 1)$, that is $\Delta^i_{\rho}= -\delta_{ \rho ,i}+\delta_{-\rho, i} + \delta_{ \rho, i-1} - \delta_{ -\rho, i-1}=-\Delta^i_{-\rho}$, where $i$ and $\rho$ are meant modulo $q$.}

\begin{figure}
    \centering
    \includegraphics[width=0.48\textwidth]{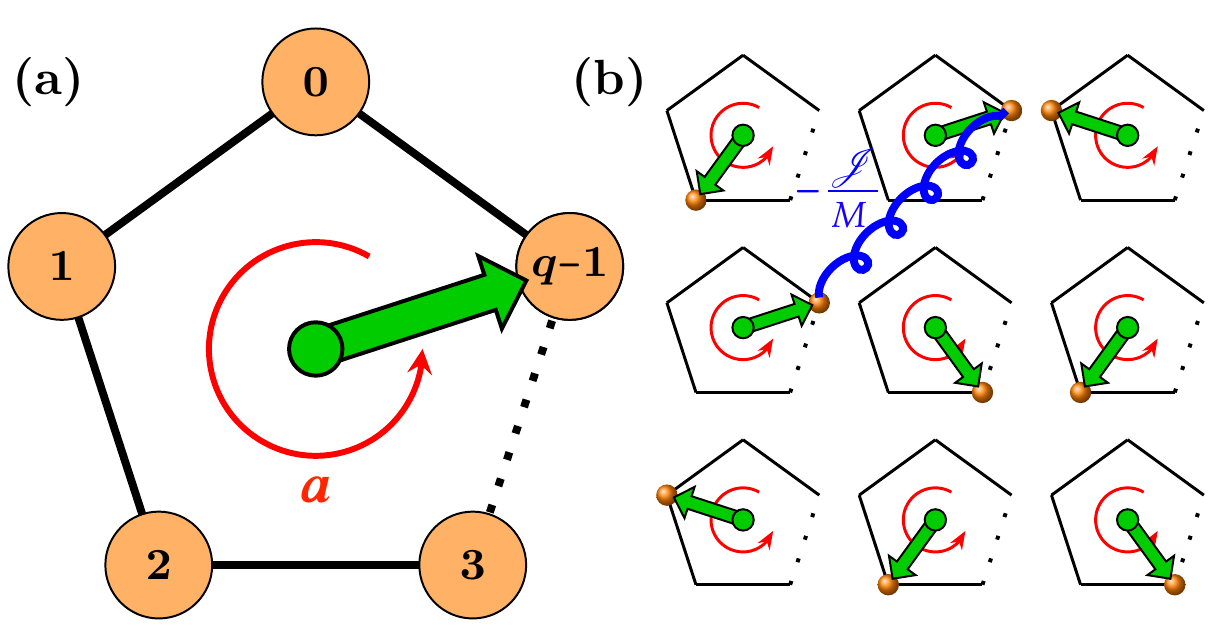}
    \caption{{Sketch of the driven Potts model. (a) A single unit is composed of $q$ states, identified by a single spin configuration. Counterclockwise transitions are favored by a nonconservative thermodynamic force $a>0$. (b) Interactions, entering the transition rates \eqref{rates_Potts}, are between the same state $i$ of all $M$ units.}}
    \label{fig:Potts}
\end{figure}

In the limit $M \to \infty$, we obtain the scaled internal entropy (using Stirling's approximation)
\begin{align}\label{S_int_Potts}
s_\mathrm{int}(c)= -\prod_{i=1}^q  c_i \ln c_i ,
\end{align}
{constituting the macroscopic system entropy \eqref{macro_s_sys}},
and the scaled energy 
\begin{align}\label{energy_Potts}
u(c)= \frac{1}{2} \sum_{i=1}^q u_i c_i^2,
\end{align}
in terms of the occupation densities $c:=n/M$.
Hence, the macroscopic entropy production \eqref{ldb_macro} for transitions {$|\rho|=i$ between single-unit states $i$ and $i + 1$ takes the form
\begin{align}\label{macro_epr_rho_Potts}
\sigma_\rho(c)= \mp[(\partial_{c_{i+1}}-\partial_{c_i}) \phi(c)] \pm a ,
\end{align}}
%Hence, the macroscopic entropy production \eqref{ldb_macro} for transitions between single-unit states $i$ and $i \pm 1$ takes the form
%\begin{align}\label{macro_epr_rho_Potts}
%\sigma_\rho(c)= -[(\partial_{c_{i\pm1}}-\partial_{c_i}) \phi(c)] \pm a 
%\\&=\frac 1 T (u_i c_i-u_{i \pm 1} c_{i \pm 1}) + \ln\frac{c_i}{c_{i \pm 1}} \pm  a \;,
%\end{align}
where $\phi(c)=u(c)/T-s_\mathrm{int}(c)$ is the scaled Helmholtz free energy divided by the temperature $T${, as introduced in \eqref{helmoltz}}, and the nonconservative part $a_{\pm \rho} = \pm a $ favors {unidirectional} rotation along the single-unit states.
%A common choice of the transition rates that respect the local detailed balance \eqref{micro_ldb} is the so-called Arrhenius form that is characterized by a constant kinetic factor \cite{}.
%\begin{align}\label{rates_Potts}
%R_\rho(n) = \Gamma e^{-\frac{1}{2T} [\Phi(n+\Delta_\rho) -\Phi(n) -T a_\rho] },
%\end{align}
With attractive interactions $u_i = -\mathcal{J} >0$ equal for all state $i$, this class of systems includes the driven Curie-Weiss model, i.e. the mean-field Ising model ($q=2$) -- in which $c_1-c_2=m$ is the magnetization -- and the ferromagnetic mean-field Potts model ($q \geq 3$).
In all cases, the state of a unit can be identified with a spin orientation{, see Fig. \ref{fig:Potts}}.

A possible choice of single-unit transition rates that respects the local detailed balance is the so-called Arrhenius form that is characterized by {an exponential form and a global time scale $1/\gamma$ \cite{meibohm2024small}}.
In this case, the scaled macroscopic transition rates \eqref{r_rho} for  transitions {$\rho=i$ (resp. $\rho=1-i$) between single unit states $i$ and $i+1$ (resp. $i-1$) reads} 
\begin{align}\label{rates_Potts}
r_\rho(c) = \gamma c_i e^{\frac{1}{2} [ \frac {1 \mp \xi } {T} (u_{i \pm 1} c_{i \pm 1}-u_i c_i) \pm  a] } ,
\end{align}
with $-1 \leq \xi \leq 1$. {Note that the logarithm of $r_i(c)/r_{-i}(c)$ equals precisely \eqref{macro_epr_rho_Potts} with the change of internal entropy $s_\text{int}$ given by $\ln(c_i/c_{i +1})$.}
Extensions have been studied that include a constant energy bias for each state $i$, interactions between different single-unit states and coupling to baths at different temperatures \cite{Herpich2020Jun}. 

In the detailed balance Curie-Weiss model in contact with a single thermal bath ($q=2$ and $a_\rho=0$), the emergence of a dynamical phase transition was observed upon an instantaneous disordering quench from low to high temperature, i.e. the appearance at the critical time $t_\mathrm{c}$ of a kink in the time-dependent rate function of the magnetization $I(m,t)$ \cite{meibohm2022finite}. This corresponds to a sudden change in the optimal fluctuations driven by the competition in \eqref{P_action_initialI} between the statistical weights of {a trajectory and of its initial condition, which are $\mathcal{A}$, as given in \eqref{action}, and $I(m(0),0)$, respectively}.
An analogous dynamical phase transition occurs in the time-dependent rate function of the exchanged heat, similarly due to a change of dominance in the most likely trajectories \cite{meibohm2023landau}. For the Curie-Weiss model in contact with two different thermal baths (i.e. two sets of transitions \eqref{rates_Potts} each associated with a bath at a different temperature), a different dynamical phase transition appears as a kink in the steady state cumulant generating function of the exchanged heat \cite{Herpich2020Jun}. In this case, {as we have described in Sec. \ref{sec:currents_ss}}, the singularity stems from the competition between the multiple stationary solutions of the tilted Hamiltonian equations \eqref{hamiltonian_eqs_z}, and represents a signature of the bistability of the system in the current statistics.

The driven Potts model ($q \geq 3$) has been extensively studied, e.g. as a model for synchronization in work-to-work converters \cite{Herpich2018Sep,Herpich2019}. The deterministic dynamics can be examined thoroughly. As $T/\mathcal{J} \to \infty $ the system becomes effectively detailed balanced and the dynamics is entropy-dominated with a single stable, disordered fixed point $\mathcal{x}^*=(1/q,\dots,1/q)$. Conversely, as $T/\mathcal{J} \to 0$ the system is energy-dominated with $q$ {stable, ordered} fixed points characterized by all units in the same state, i.e. $\mathcal{x}_{\gamma=1}^*=(1,0, \dots, 0)$ and {$\mathcal{x}_{\gamma >1}^*$ is obtained by cyclic permutations of the elements of the vector $\mathcal{x}_{\gamma=1}^*$}. Above a critical coupling strength $\mathcal{J}_\mathrm{c}$, the disordered fixed point becomes unstable. For $|a| \to 0$, independently of the choice of transition rates (e.g. $\xi$ in \eqref{rates_Potts}), $\mathcal{J}_\mathrm{c} \to qT$ and the macroscopic system settles in one ordered fixed point. When detailed balance is broken, $\mathcal{J}_\mathrm{c}$ depends on the specific class of transition rates \eqref{rates_Potts}, and for $\mathcal{J}> \mathcal{J}_\mathrm{c}$ the system can display time-dependent attractors $\mathcal{x}_\gamma^*(t)$ corresponding to the synchronized motion of all the units. The macroscopic deterministic dynamics close to the bifurcation can be solved exactly by transforming to Fourier modes $\hat c_k := \sum_{j=0}^{q-1} c_j e^{\mathrm{i}2 \pi  j k/q }$ and deriving normal form equations for their amplitude $\mathcal{r}_k$ and phase $\varphi_k$ \cite{meibohm2024minimum,meibohm2024small}. 
The emerging phase diagram -- as a function of $q$ and the dynamical class, e.g. $\xi$ in \eqref{rates_Potts} -- is extremely rich. One finds regions with a finite number of active modes $|S_\mathrm{a}| =1, 2, \dots$, i.e. modes $k$ such that $\mathcal{r}_k >0$. In the region $|S_\mathrm{a}|=1$, several single-mode oscillating states can be simultaneously stable. For $|S_\mathrm{a}| \geq 2$ a unique multi-mode oscillating state exists, which is in general a quasi-periodic attractor since the different modes show frequencies with irrational ratios.

The coexisting limit cycles at $|S_\mathrm{a}|=1$ become metastable at large but finite $M$. In particular, very close to the bifurcation, the system switches between them thanks to typical fluctuations, not via rare instantons. 
Indeed, the quasi-potential barrier separating {the stochastic limit cycles} vanishes, $\Delta I_\mathrm{ss} \to 0$, and the escape rate is dominated by the subexponential prefactor neglected in \eqref{trans_saddle}.
The life time of such competing states is decided by the phase space contraction rate $\lambda$ defined in \eqref{contraction_rate}, whose magnitude corresponds to the macroscopic limit of the inflow rate \eqref{inflow_rate} averaged over the attractor. In general, such connection between the Lyapunov stability of the deterministic dynamics and the typical fluctuations is given by \eqref{contraction_fluctuations}. Close to the bifurcation point, one can prove the stability-dissipation relation \cite{meibohm2024minimum,meibohm2024small}
\begin{align}\label{stability_dissipation}
\Delta \dot \sigma  \propto  \Delta \lambda,
\end{align}
where $\Delta \mean{\mathcal{O}}:= \lim_{N \to \infty}(\mean{\mathcal{O}}- \mean{\mathcal{O}}_{\mathcal{J}=0})/\mean{\mathcal{O}}_{\mathcal{J}=0}$ stands for the normalized variation of the average values of the observable $\mathcal{O}$ between interacting and noninteracting {(i.e. $\mathcal{J=0}$)} conditions. The macroscopic mean, $\lim_{N \to \infty}\mean{\mathcal{O}}$, is in general different on each attractor.
Therefore, the relation \eqref{stability_dissipation} implies that the least dissipative attractor is the most stable one. This constitutes the first derivation of a minimum entropy production principle valid far-from-equilibrium.
{It is crucial to note that the entropy production rate entering \eqref{stability_dissipation} is a thermodynamic quantity, linked to the microscopic energetics of the system by the local detailed balance \eqref{ldb_macro}, not an abstract definition of dissipation based on information theory arguments \cite{Daems1999contraction}.}

The above Potts model can be embedded in a $d$-dimensional physical space. To this end, we consider a lattice in which each node contains a large number of $q$-state units described in the macroscopic limit by a local density $c_{i,x}$ (see \ref{subsec:continuous}). 
Units can not only change their internal state, i.e. rotate the spin, but also jump to nearby lattice sites with rates belonging to the sets $R$ and $C$, respectively.
The scaled entropy production rate reads
{\begin{align}\label{macro_epr_rhoR_Potts_space}
\sigma_\rho(\{c_{i,x}\})= \mp[(\partial_{c_{i+1,x}}-\partial_{c_{i,x}}) \phi] \pm a^\mathrm{rot} ,
\end{align}
for transitions $\rho \in R$ between states $i$ and $i+1$} and
\begin{align}\label{macro_epr_rhoC_Potts_space}
\sigma_\rho(\{c_{i,x}\})= -[(\partial_{c_{i,x'}}-\partial_{c_{i,x}}) \phi] + a^\mathrm{tra}_{\rho} ,
\end{align}
for transitions $\rho \in C$ from site $x$ to the neighboring site $x'$. Detailed balance is broken by the nonconservative contributions $\pm a^\mathrm{rot}$ and $a^\mathrm{tra}_{\rho}$ that force the spin to rotate and to drift in space, respectively. By choosing $a^\mathrm{tra}_{\rho}$ to be constant in magnitude and aligned with the local spin direction, one obtains a toy model for self-propelled particles.
In the following, we restrict ourselves to the case of active Ising spins, i.e. $q=2$, but the discussion can be seamlessly generalized to $q>2$.

Taking the continuous limit by sending the lattice spacing $\epsilon$ to zero as detailed in Sec.~\ref{subsec:continuous}, one gets a stochastic field theory for $c_{i,x} \to c_i(r,t)$,
\begin{equation}\label{SPDEs}
	\begin{aligned}
    & \partial_t c_i=-\nabla \cdot \bigg [  \underbrace{\gamma c_i \left( -\nabla  \frac{\delta u}{\delta c_i}+f_i \right)-\gamma T \nabla c_i + \xi_i}_{j_i[c]}   \bigg] +I_i.
    \end{aligned}
\end{equation}
Here $\xi_i$ are independent zero-mean Gaussian white noise fields with variance $2 \gamma T c_i/\Omega$ and $\Omega$ the system size, $I_1=-I_2$ is a white noise field with Skellam distribution \cite{skellam1946jrss} being the difference of Poissonian jumps in and out of the single-unit state $i$ (e.g. with rates \eqref{rates_Potts}), and {constant drift vector $f \equiv f_1=-f_2$ that derives from} the continuous limit of $a^\mathrm{tra}_{\rho}$.
Note that $\chi=\gamma c \mathbb{I}$ is the mobility matrix and the diffusive flux $- \gamma T \nabla c$ stems from the entropic term $- \chi \cdot \nabla \delta s_\mathrm{int}/ \delta c$.
Summing and subtracting the two equations in \eqref{SPDEs}, one can obtain closed stochastic equations for the local magnetization $m(r,t):=c_1(r,t)-c_2(r,t)$ and the local density of spins $c_\mathrm{tot}(r,t):=c_1(r,t)+c_2(r,t)$.

The macroscopic entropy production follows directly from the general expression derived in Sec. \ref{sec:entropy_continuous}. In particular, in a stationary state it reduces to the nonconservative contribution \eqref{wnc_continuous}, 
\begin{align}\label{wnc_Ising_active}
 \dot \sigma_\mathrm{nc}=\int dr \left [a^\mathrm{rot}\mean{I_1} + f \cdot (\mean{j_1}-\mean{j_2}) \right],
 \end{align}
which displays, in order, the contribution of the single-unit rotational driving and of the active drift in space. Using the explicit expressions of the currents $\mean{j_i}$ in \eqref{SPDEs},  {the entropy production of translation in \eqref{wnc_Ising_active} reads 
\begin{align}
\begin{aligned}\nonumber
 \dot \sigma_\mathrm{nc} &\underset{a^\mathrm{rot}=0}{=}\int dr f \cdot (\mean{j_1}-\mean{j_2}) \\
 &=  \gamma f^2 c_\mathrm{tot} +  \gamma f \cdot \int dr \left( c_1  \nabla \frac{\delta \phi}{\delta c_1}-c_2 \nabla \frac{\delta \phi}{\delta c_2}  \right) .
\end{aligned}
\end{align}
In particular, one finds
\begin{align}
\begin{aligned}\label{wnc_Ising_active_macro}
{ \dot \sigma_\mathrm{nc} \underset{a^\mathrm{rot}=0}{=}}   \gamma f^2 c_\mathrm{tot}   ,
\end{aligned}
\end{align}
in any homogeneous state or whenever the contribution of interfaces (between magnetized and disordered domains) is negligible compared to the bulk one.} Note that~\eqref{wnc_Ising_active_macro} is the same dissipation as that of a gas of driven non-interacting particles, and is thus independent of the system phase.

For $a^\mathrm{rot}=0$, one obtains a thermodynamically consistent version \cite{agranov2024thermodynamically} of the active Ising model $(q=2)$ \cite{solon2013revisiting,Solon2015flocking}, generalized to $q\geq 2$ in  \cite{mangeat2020flocking,Solon2022susceptibility}.
Therein, the energy variation due to the jumps of the units in space -- namely, the first term on the right hand side of \eqref{macro_epr_rhoC_Potts_space}  --
was neglected. Alternatively, one can interpret {the model in \cite{solon2013revisiting,Solon2015flocking} as satisfying local detailed balance with a fine-tuned field-dependent nonconservative part $\tilde a^\mathrm{tra}_\rho(\{c_{i,x}\})$ that exactly counterbalances the energy variations, thus yielding a constant drift in space. Here, instead, we discuss a thermodynamic consistent model in which the translational nonconservative force on each spin is constant while the ensuing drift is affected also by the interactions with nearby spins.}
The original model \cite{solon2013revisiting,Solon2015flocking}  predicts the emergence of flocking at sufficiently large density and small temperature, i.e. a state with $\mean{m} \neq 0$, which can be retrieved by the thermodynamically consistent version by adding nearest-neighbour interactions \cite{proesmans2024active,agranov2024thermodynamically}.
The model by \cite{solon2013revisiting,Solon2015flocking}   has been used in \cite{Yu2022energy} to evaluate the entropy production \eqref{epr_macro}, which was found to display a kink at the onset of flocking in disagreement with \eqref{wnc_Ising_active_macro}. In the light of the above discussion, such apparent entropy production is an artifact caused by neglecting the energy variations of translating spins. 

Finally, for active particles with internal rotation, $a^\mathrm{rot} \neq 0$, the entropy production cannot be inferred only from the spacial dynamics, as often assumed in active field theories \cite{nardini2017entropy}. Indeed, the first term on the right hand side of \eqref{wnc_Ising_active} requires explicit knowledge of the internal current between single-unit states and reduces to the quadratic form of linear irreversible thermodynamics \cite{Tomer2021active}, in which $\mean{I_1} \propto a^\mathrm{rot} $, only if $a^\mathrm{rot} \to 0$.

\section{Discussion}\label{sec:discussion}

\subsection{Critical summary}

\begin{figure*}
\center
\includegraphics[width=0.8\textwidth]{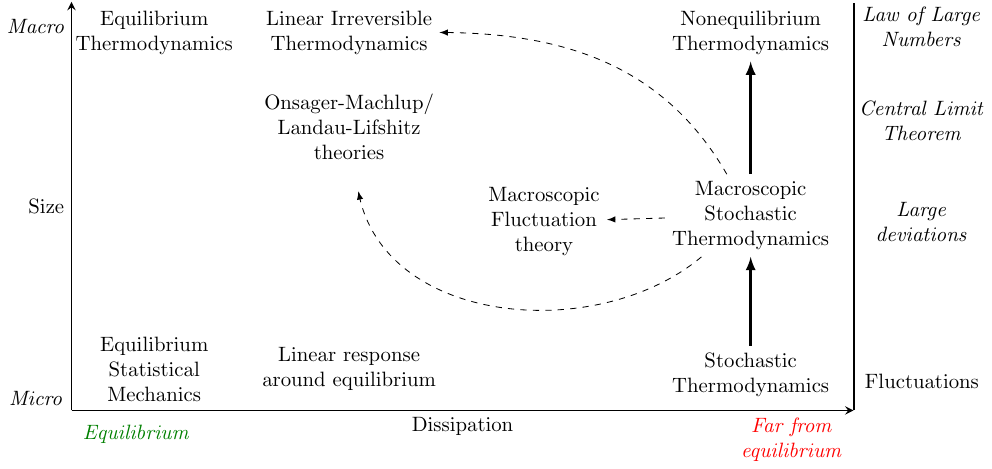}
\caption{Schematic classification of the thermodynamic theories in terms of the system size and distance from equilibrium. The present paper extends stochastic thermodynamics to macroscopic systems, and recovers Irreversible Thermodynamics (deterministic) and Onsager-Machlup/Laundau-Lifshitz theories (Gaussian fluctuations) in the near-equilibrium regime. It also comprises Macroscopic Fluctuation Theory wherein forces are linearly related to fluxes but are nonlinear functions of macroscopic variables, which thus have nonGaussian fluctuations. \label{fig:theories}}
\end{figure*}

We started from a description of an open system using a stochastic dynamics in terms of mesoscopic configurations specified by the occupation number of mesoscopic states, and we reviewed the standard procedure to build the superstructure of stochastic thermodynamics on it. 
In doing so, we took for granted that transitions rates between states, and thermodynamic observables associated to the states, can be expressed in terms of occupation numbers only. 
This means that we assumed that for each mesoscopic state, the internal degrees of freedom are at equilibrium and an equilibrium Massieu potential $\Phi(n)$ -- and thus an internal entropy $S_\text{int}(n)$ -- can be assigned to them that is solely a function of the occupation number of the mesostate. 
As a result, the entropy production can be expressed in terms of the mesoscopic transitions and the probability distribution over the mesostates, see \eqref{epr} and \eqref{ent_flow}. 
We note that even when these conditions are satisfied, obtaining such transitions rates in practice may not always be straightforward \cite{prinz2011markov}. 
%Given that, we showed that the entropy production can be expressed in terms of all mesoscopic transitions, see \eqref{epr} and \eqref{ent_flow}, and making reference to either the underlying equilibrium, see \eqref{nc_driv_epr}, or the stationary state, see \eqref{ad_nonad_epr}.

We then imposed the necessary constraints to obtain a deterministic macroscopic dynamics and an extensive thermodynamics, namely, extensive transition rates and Massieu potentials, together with nonconservative forces $a_\rho$ that are nonextensive.
In the resulting entropy production, the contribution due to the system entropy stemming from the uncertainty of the mesoscopic state (i.e. the Shannon entropy) vanishes, see \eqref{shannon0}, and only the internal entropy (if there is one) survives.
%entropy of the system is only determined by the entropy of the microscopic unresolved degrees of freedom (such as the kinetic energies).
For example, for systems in which the mesoscopic state $i$ is occupied by $n_i$ \emph{noninteracting} units (such as reacting chemicals), the internal entropy is just the Boltzmann entropy of the complexion number, $S_\text{int}(n) = \ln \frac{|n|!}{\prod_i n_i!}$, whose scaled macroscopic limit reads $s_\text{int}(c)= -\sum_i c_i \ln c_i $, see Sec \ref{sec:ex3}. Remarkably, this has the form of a Shannon entropy on the space of concentrations, but has nothing to do with the Shannon entropy in terms of the mesoscopic probabilities which, as we just mentioned, vanishes. Accordingly, it counts the uncertainty in the microscopic configurations within the mesoscopic states $i$, not the randomness in the distribution of the mesoscopic state. However, be aware that an extensive Shannon entropy \eqref{shannon0} can appear if a system has a number of (near) degenerate minima of the quasi-potential \eqref{global_QP} that scales exponentially with the size $V \to \infty$, similarly to glasses \cite{castellani2005pedestrians}.

It is worthwhile stressing that we only focused on thermal noise. We excluded systems whose dynamics remains noisy in the macroscopic limit, e.g., in the presence of a quenched disorder \cite{degiuli2022} or any kind of extrinsic noise \cite{Bressloff2017}. 
We also specialized the discussion of the macroscopic limit to dynamics characterized only by isolated, nondegenerate fixed points. However, the theory is expected to hold more generally \emph{mutatis mutandis} in presence of sufficiently regular time-dependents attractors, such as limit cycles that we mentioned in passing. The practical difficulty in this case is to determine the large fluctuations and the quasi-potential. 

We then considered a continuous-space limit that leads to a stochastic field theory. The conserved dynamics -- called model B in the standard classification \cite{hohember77} -- is the special case, extensively treated in \cite{Bertini2015Jun}, that we retrieved when all transitions scale diffusively. In this respect, we reframe and contextualize many results of macroscopic fluctuation theory within ST. 
In general, we connected with fluctuating hydrodynamic theories  \cite{landau1959fluid,Groot1984,dezarate06} -- in particular, nonlinear ones \cite{zubarev1983} -- although in our treatment the intensive thermodynamic quantities characterizing the equilibrium reservoirs are not dynamically coupled to the mesoscopic variable $c$ -- e.g. the temperature is unaffected by the dissipated heat. We thus achieved the goal of systematizing all well-established thermodynamics theories, from deterministic to statistical ones, within the overarching framework offered by ST (see Fig. \ref{fig:theories}).

{It is worth noting that while we have a priori assumed the existence of a well defined large scale parameter that controls the thermodynamic limit, this setting may not be the only one that leads to macroscopic deterministic dynamics. The low temperature limit (with respect to a typical energy scale) is a much studied example, especially in the context of Langevin dynamics \cite{graham1987macroscopic}. Moreover, stochastic dynamics in continuous space converge to deterministic ones within an appropriate hydrodynamic description that averages the microscopic observables over suitably small volumes \cite{Bertini2015Jun} -- without necessarily invoking that the number of particles per lattice site diverges, as we have effectively done in \ref{subsec:continuous}. }

Finally, in Section \ref{sec:emergent}, we coarse-grained the long-time dynamics as a Markov jump process over the attractors and discussed the resulting thermodynamics. The attractors provide a notion of states which emerge from the underlying dynamics and are generically out-of-equilibrium, thus overcoming a central starting assumption of ST. This section, together with the bounds on the macroscopic transition rates of Sec. \ref{sec:bounds_kappa}, creates a long-sought bridge between Freidlin–Wentzell theory of large deviations and thermodynamics \cite{freidlin,graham1985weak}. 
In this respect, we have worked at a formal level, in particular making use of generating functions conditioned on subsets of trajectories. While conditioning the entropy production on a single basin of attraction has recently received some attention \cite{Fiore2021current}, conditioning on transition paths -- and the underlying problem of calculating the statistics of transition times out of equilibrium -- remains an issue to be explored thoroughly. These tools are needed if one wishes to explicitly compute the emerging observables from the underlying mesoscopic dynamics. However, such bottom-up derivation is expected to be possible only in relatively simple models. 
The main goal of the coarse-graining in Section \ref{sec:emergent} is to identify the structure of the emerging thermodynamics in terms of nonequilibrium states, so as to guide the construction of effective models in complex systems, possibly informed by measurements of emerging observables.

In Sec. \ref{sec:applications} we have outlined three classes of models that comply with the general theory, namely, reaction networks of ideal chemicals, nonlinear electronic circuits and driven Potts models. 

\subsection{Outlook} 

We focused on the macroscopic limit of the core structure of ST, thus providing the basis for a systematic analysis of many specific aspects connected to the nonequilibrium thermodynamic limit and for further extensions to other classes of systems. 

\subsubsection{Thermodynamic uncertainty relations}

%omitting to analyze some more specialized, yet important results that have been recently considered in connection with the thermodynamic limit. Here we only mention (i) the thermodynamic uncertainty relation (TUR) and (ii) the information flows in partite systems.

In the simplest formulation, the thermodynamic uncertainty relation (TUR) asserts that the ratio between the squared mean and the variance of a current integrated over a time $\tau$ is bounded from above by half the entropy produced in the time $\tau$, see e.g.~\cite{Barato2015,macieszczak2018unified,gingrich2017inferring,hasegawa2019fluctuation,timpanaro2019thermodynamic,dechant2020,falasco2019unifying,van2020entropy,dechant2021improving,van2023thermodynamic}. For extensive observables of \ref{sec:macro_ft} introduced in \eqref{time_int_obs}, it is immediate to conclude that the TUR is  valid in terms of the scaled moments, as both sides of the inequality scale as $V$ \cite{koyuk2022thermodynamic}. Its explicit expression will dramatically depend on the observation timescale $\tau$, though. For systems initially prepared in an attractor $\gamma$, as a consequence of metastability, the TUR will involve only the entropy production of the attractor $\gamma$,
\begin{align}\label{eq_TUR}
\frac{\mean{\mathcal{o}}^2}{\mean{\mathcal{o}^2}-\mean{\mathcal{o}^2}} \leq \frac{ \mean{\dot \sigma_\gamma}\tau}{2},
\end{align}
if the integration time $\tau$ is much shorter than the minimum escape time $1/\kappa_\nu$. The  TURs \eqref{eq_TUR}, if used for inference, provide multiple bounds to the dissipation of each attractor rather than a single bound on the entropy production of the full system. Combining \eqref{eq_TUR} with the TUR evaluated at $\tau \gg 1/\kappa_\nu$, one can possibly bound the transition entropy production $\sigma_\nu$. 

{The macroscopic limit not only decomposes the TUR into multiple bounds for the different macroscopic attractors, as given by \eqref{eq_TUR}, but also generates tighter constraints on single degrees of freedom. Indeed, it has been shown in \cite{yoshimura2021thermodynamic} that the deterministic dynamics \eqref{macro_dcdt} is associated to the bound\footnote{{This result was obtained for deterministic chemical reaction networks but clearly holds for all systems described in this review that admit \eqref{macro_dcdt} as macroscopic dynamics.}}
\begin{align}\label{deterministic_TUR}
 |d_t \mathcal{x}_i|=   |F_i| \leq \sqrt{D_{ii}\dot \sigma_i},
\end{align}
where $D_{ii}$ is the diagonal component on the diffusion matrix \eqref{diffusion}, and $\dot \sigma_i:= \sum_{\rho>0: \Delta^i_{\rho}\neq0} [r_\rho(\mathcal{x}(t))-r_{-\rho}(\mathcal{x}(t))]\sigma_\rho(\mathcal{x}(t))$ is the scaled macroscopic entropy production associated to variations of the $i$th component of the configuration vector $\mathcal{x}$. Notice that $\dot \sigma_i \leq \mean{\dot \sigma}$ which means that \eqref{deterministic_TUR} cannot be obtained by the microscopic TUR, the latter being a global bound in terms of the total dissipation. As a corollary, integration of \eqref{deterministic_TUR}  yields a local speed limit which replaces -- with respect to its microscopic counterpart \cite{Shiraishi2018,Vo2020speed,VanVu2023optimal} -- mesoscopic probabilities $p(n,t)$ with macroscopic configurations $\mathcal{x}$.}

It remains to be explored how recent approaches based on the statistics of residence and return times behave in the macroscopic limit~\cite{marsland2019,skinner2021estimating, skinner2021improved,van2022thermodynamic,harunari2022learn}.

\subsubsection{Phase transitions}

Macroscopic ST provides a unifying language to describe nonequilibrium phase transitions, not only dynamically but also from an energetic point of view. 

\begin{figure}[t]
    \centering
    \includegraphics[width=0.48\textwidth]{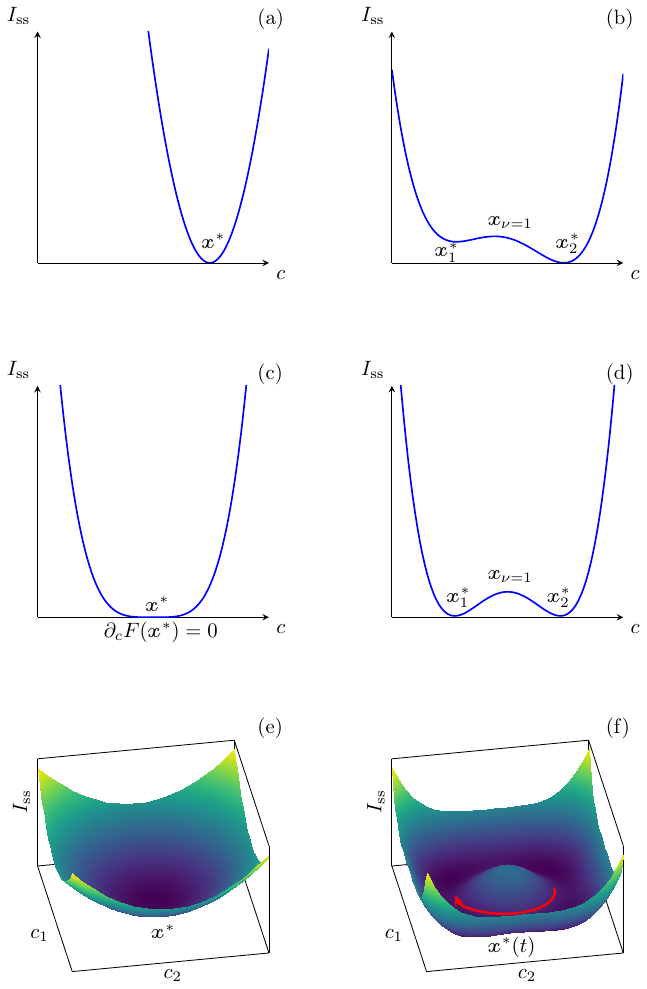}
    \caption{{
    Pictorial representation of the qualitative changes in the quasi-potential $I_\text{ss}$ at phase transitions, obtained from variations of an external parameter. The corresponding bifurcations in the macroscopic dynamics \eqref{macro_dcdt} are saddle-node bifurcation (from (a) to (b)), pitchfork bifurcation (from (c) to (d)) and Hopf bifurcation (from (e) to (f)) in which a finite circulation is present. While the first two are analogous to first and second order phase transitions at equilibrium, respectively, the last one has a genuine dissipative character.}}
    \label{fig:phase_transitions}
\end{figure}

On one hand, bifurcation theory can be employed to study the deterministic dynamics upon changing some external parameter \cite{strogatz2015nonlinear}, for instance the intensive thermodynamic variables of the reservoirs or the system's kinetic factors.
Nonconservative forces give rise to a far richer scenario than that encountered in equilibrium phase transitions, where only fix points are present and their statistical weight can be changed. {In the macroscopic limit, bistability can emerge as a consequence of a saddle-node (Fig. \ref{fig:phase_transitions} (a) and (b)) or a pitchfork bifurcation (Fig. \ref{fig:phase_transitions} (c) and (d)) in processes similar to equilibrium first or second order phase transitions, respectively. A critical point, corresponding to a locally flat quasi-potential (see Fig. \ref{fig:phase_transitions} (c)), is accompanied by diverging fluctuations so that the Langevin approach of Sec.~\ref{sec:Onsager_Machlup} breaks down \cite{vanKampen}. A higher order expansion of the master equation captures the critical slowing down of the mean dynamics and the different scaling in $V$ of the fluctuations \cite{dekker1980critical,remlein2024nonequilibrium}. This phenomenology is akin to phase transitions in systems with detailed balance dynamics, with the quasi-potential $I_\text{ss}$ playing the role of the Massieu potential $\phi$. A transition that has no counterpart at equilibrium is the emergence of a limit cycle, e.g., through a Hopf bifurcation (see Fig. \ref{fig:phase_transitions} (e) and (f)). In this case, the minimum of the quasi-potential lies on a flat manifold, on which a finite dissipative drift $v_\text{ss}(\mathcal{x}^*(t))$ is present and determines the probability distribution along the cycle as $p(\mathcal{x}^*(t)) \propto |v_\text{ss}(\mathcal{x}^*(t))|^{-1} $    \cite{vance1996fluctuations}.  }
On the other, large deviations theory is instrumental in identifying dynamical phase transitions. These are singularities in the  rate function of dynamical observables appearing either at long times \cite{GarrahanLecomtePrl2007,garrahan2009first,hurtado11,espigares2013,gingrich14,nyawo2017} (see Sec. \eqref{sec:currents_ss}) or developing at some critical finite time \cite{meibohm2022finite,meibohm2023landau,blom2023global} {(see Sec. \ref{sec:ex3})}.

A significant research line concerns the possibility of characterizing the phase transitions and the phase coexistence that appear in the thermodynamic limit by means of the nonequilibrium thermodynamic quantities of the system. The entropy production rate and the Massieu potential (or their derivatives) can display singularities at the critical points, in ways that seem to depend on the specificity of the model under scrutiny \cite{Falasco2018a,noa2019entropy,nguyen2020exp,rana2020,martynec2020entropy,Fiore2021current}. 
Nonequilibrium intensive quantities, such as chemical potential \cite{guioth2018large,guioth2019nonequilibrium,guioth2019lack}, pressure \cite{Takatori2014swim,Solon2015pressure,Joyeux2016pressure} and surface tension \cite{Bialke2015negative,zanike2020surface}, have been studied to understand whether they can predict the chemical and mechanical stability, especially of active systems. For pressure in particular, a generalized equation of state for generic diffusive systems was derived in \cite{falasco2016virial} that displays the dissipated heat alongside the standard virial terms. For some notable models, such as active Brownian particles \cite{Solon2015pressure}, such formula yields a pressure that is only a function of bulk variables -- independent of the details of confinement, as in equilibrium. We believe that the macroscopic ST outlined here can help unifying these somewhat detached research efforts.

\subsubsection{Finite size effects}
 
Within the framework described above, interesting effects often appear at large yet finite system size. Going beyond the asymptotic regime considered here, requires to retain subleading corrections in the thermodynamic and kinetic variables, Eqs. \eqref{ldb_macro} and \eqref{lambda}. The resulting modification of the large deviations form for probability distributions can be found by means of WKB expansions \cite{caroli1979diffusion,caroli1980wkb,proesmans2019large,assaf2017wkb}. For instance, \eqref{rateI} is replaced by
\begin{align}
p(c,t) = \left (A_{1}(c,t)+O(V^{-1}) \right) e^{-V I(c,t)}
\end{align}
with $A_{1}$ satisfying a transport equation obtained by expanding the master equation \eqref{me}. In particular, the long-time limit of $A_{1}$ would allow one to compute subleading corrections to the transition rates \eqref{trans_saddle}, i.e. the nonequilibrium generalization of the Kramers formula \cite{hanggi90kramers}. To the best of our knowledge this has been done only for systems with Gaussian noise \cite{Maier1993effect,maier1997limiting} where $A_{1}$ can be linked to the instanton dynamics \cite{bouchet2016generalisation}, {and presents an additional contribution due to the phase space contraction rate for systems lacking detailed balance.}
Moreover, subextensive terms should appear in the deterministic dynamics \eqref{macro_dcdt}. Such corrections to the drift field are often computed by adding a small fluctuation of order $O(V^{-1/2})$ to the deterministic solution $\mathcal{x}$ and averaging it. {This procedure leads to the deterministic evolution}
\begin{align}
\dot{\mathcal{x}}= F(\mathcal{x}) + \frac{1}{2V} \text{Cov} :\partial_c\partial_c F(\mathcal{x}),
\end{align}
with the covariance $\text{Cov}:=\mean{(c-\mathcal{x})(c-\mathcal{x})}$ computed (at the leading order) by means of \eqref{linear_langevin}.
Such modified drift may better predict the phase diagram of finite systems -- e.g. changing the order of phase transitions in extended systems as the active Ising model of Sec. \ref{sec:ex3} \cite{Solon2015flocking} -- but it is not guaranteed to respect the thermodynamic structure of the theory. Even more subtly, at large but finite $V$ the very definition of metastable states cannot be reduced to the sole notion of deterministic attractors, which should be identified by the spectral analysis of the Markov generator \cite{gaveau1998,kurchan2009}.

Finally, subextensive contributions play a key role in the information flows within a system \cite{Parrondo2015Feb}. For systems with a (at least) bipartite structure -- i.e. with two sets of transitions, each affecting only one set of states, as in section \ref{subsec:continuous} -- it is possible to quantify the entropy exchanges via the rate of variation of mutual information between the two sets of states \cite{Horowitz14,hartich2014stochastic}. This framework allows one to describe \emph{inter alia} the thermodynamics of autonomous Maxwell demons, in which a part of the system creates correlations used by the rest, e.g., to drive a current \cite{freitas2021characterizing}. This mechanism, being based on the rectification of thermal fluctuations, breaks down in the macroscopic limit we considered. Nevertheless, it can be rescued by making the nonconservative force $a_\rho$ extensive in $V$, although at the expense of a vanishing efficiency \cite{Freitas2022maxwell,Freitas2023information}.

\subsubsection{Odd-parity variables and quantum systems}

Importantly, we focused on systems described only by state variables that are even under time reversal. Repeating the approach of this work with odd-parity variables, such as momenta, would lead to, e.g. underdamped fluctuating hydrodynamics \cite{nakamura2009derivation,Manacorda2017lattice}. The noiseless limit of such theory has been considered in \cite{Forastiere2023} with the conventional approach of closing moments hierarchy based on the Boltzmann equation -- while the asymptotic fluctuations should be derivable by using large deviations theory \cite{bouchet2020boltzmann}. 
{The theory can be extended to long-range interacting systems, such as those composed by self-gravitating particles \cite{chavanis2006hamiltonian,chavanis2019generalized}, whose energy can be made extensive, albeit not additive, under the Kac's prescription.}
Generalizing the theory to open quantum systems is also a research program to explore. 

\subsubsection{Response theory}

We did not discuss response theory around nonequilibrium states \cite{Baiesi13update}. The linear theory allows to calculate, e.g., thermodynamic susceptibilities \cite{bok11,bai14,fal16a,fal16b} and transport coefficients \cite{spe06,sei10,bai11,falasco2019unifying,Chun2021nonequilibrium}.
Arguably, the most general setup to highlight the connection with thermodynamics is the one based on path probabilities, which shows how purely kinetic factors -- the excess dynamical activity -- pair with the entropy production to produce the system's response \cite{Baiesi09response}. How the macroscopic limit affects the scaling of these two contributions remains to be seen.  Thermodynamics bounds on susceptibilities have also recently appeared \cite{Owen2020,aslyamov2023nonequilibrium}.
Nonlinear response, originally developed for thermostatted dynamical systems \cite{eva90,mar08,rue09},  has received comparatively less attention in the context of ST \cite{andrieux2007fluctuation,dechant2020,falasco2022beyond} despite its experimental importance in dissipative mesoscopic systems, such as in active microrheology \cite{muller2020properties}, mechanical resonators \cite{con13,Geitner2017low} and optical microscopy \cite{radunz2009hot}.

\subsubsection{Phenomenological laws}

Sections \ref{sec:bounds_kappa} and \ref{sec:emergent} present new results, emerging as natural consequences of the macroscopic limit of ST. We speculate that they may provide links to some phenomenological laws  that have been previously postulated.

On one hand, the maximum entropy production principle of Sec. \ref{sec:bounds_kappa} is fundamentally different from the usual statement which goes under the same name -- an empirical law with mixed success. The latter can be rephrased in the language of this review, by the postulate that the attractor with the largest dissipation $\mean{\dot \sigma_\gamma}$ is the one with the smaller escape rate, i.e. the longest life time \cite{martyushev2006maximum, nicolis2010stability}. 
The two statements can agree only under some peculiar conditions, for instance if most of the nonconservative entropy production is localized on a small neighborhood of the stable fixed points.
In this case, we should have for the relaxation
\begin{align}
    \sigma_{\nu \to \gamma}^\text{nc}= \int dt \sum_\rho a_\rho r_\rho(\mathcal{x}(t)) \simeq   \mean{\dot \sigma_\gamma} \tau^\text{rel}_\gamma,
\end{align}
where $\tau^\text{rel}_\gamma$ is the longest relaxation time of the linear dynamics \eqref{linear_db}, i.e. the inverse absolute value of the smallest eigenvalue of $\partial_c F (\mathcal{x}_\gamma^*)$. Similarly, we should have for the instanton
\begin{align}
 \sigma_{\gamma \to \nu}^\text{nc}= \int dt \sum_\rho a_\rho r_\rho(c(t)) e^{\Delta_\rho \cdot \partial_c I_\text{ss}(c(t))}\simeq   -\mean{\dot \sigma_\gamma} \tau^\text{inst}_\gamma,
\end{align}
where $\tau^\text{inst}_\gamma$ is the largest timescale of the linearized instanton dynamics 
\begin{align}
d_t c= (c-\mathcal{x}^*_\gamma) \cdot [\partial_c F(\mathcal{x}_\gamma^*)+2D(\mathcal{x}_\gamma^*)\partial_c^2 I_\text{ss}(\mathcal{x}_\gamma^*)],
\end{align}
i.e. the inverse absolute value of the smallest eigenvalue of $\partial_c F^\dagger (\mathcal{x}_\gamma^*)$.
In order for \eqref{upper_lower_bounds} to be tight, one needs $\tau^\text{rel}_\gamma \simeq \tau^\text{inst}_\gamma$, which is in general true only if $\mean{\dot \sigma_\gamma} \to 0$ (see \eqref{reversibility}), i.e. close to detailed balance dynamics. This does not exclude that in some specific systems the condition $\tau^\text{rel}_\gamma \simeq \tau^\text{inst}_\gamma$ can be met even far from detailed balance, but suggests that some fine tuning or peculiar symmetries are present. Nevertheless, it is important to remark that in dissipative stochastic systems a local criterion of stability, e.g. the life-time of an attractor, does not determine the global stability, e.g. the stationary probability of the latter. While detailed balance states that are surrounded by the highest energy barrier (and have hence the largest life-time) are also the most populated in the long time, dissipative states with small escape rate may have small probability \cite{maes2013heat}.

On the other hand, the emergent second law \eqref{macro_jump_epr}, when only baths at the same temperature are present, is an energy balance that counts the energy of maintenance of $\gamma$ and the energy of changing $\gamma$. Imagine a system with a very large number $M \to \infty$ of metastable states coordinatized by the continuous variable  $m:=\gamma/M $. In the particular case $\kappa_{-\nu} \ll \kappa_{\nu}$ for all $\nu>0$, the random walk along $m$ is almost surely increasing. Equation \eqref{macro_jump_epr} takes the general form postulated for the energetics of the resting metabolism of organisms \cite{lynch2015}, e.g. with $m$ being the (normalized) mass of an individual and $\mean{\dot \sigma_{\gamma}}$, the resting metabolic expenditure. This analogy highlights the possibility that the emergent thermodynamics of Sec. \ref{sec:emergent} can bridge over to phenomenological energy balances \cite{Yang2021}, helping to clarify the mesoscopic origin of empirically determined laws such as the  allometric scaling of $\mean{\dot \sigma_{\gamma}}$ \cite{ilker2019}.

\emph{Acknowledgments}---The authors thank D. Forastiere, N. Freitas and J. Meibohm for stimulating discussions.
GF is funded by the European Union - NextGenerationEU - and by the program STARS@UNIPD with the project “ThermoComplex”.
ME is supported by project ChemComplex
(No. C21/MS/16356329) funded by Fonds National de la Recherche Luxembourg (FNR), and by Project No. INTER/FNRS/20/15074473 funded by FNR and Fonds de la Recherche Scientifique Belgium.

\bibliography{bibliography} 

\begin{appendix}

\section{Derivation of the entropic upper bound on the transition rates}\label{app:bound}

%We want to obtain a bound on $\kappa_\nu$ that involves the mean entropy production on transition paths that start from the stable fixed point $\mathcal{x}^*_\gamma$ and reach the unstable fixed point  $\mathcal{x}_\nu$.
For autonomous systems, we start from the decomposition into adiabatic and nonadiabatic entropy production at the level of single trajectories, multiply both sides of the time-integrated version of \eqref{ad_nonad_epr_macro}  by $\delta(c(0)-\mathcal{x}^*_\gamma)\delta(c(t)-\mathcal{x}_\nu)$ and take the average. 
Hence, among all trajectories we sample only those that start from fixed points and reach a saddle.
For large $t$, the lefthand side can be written in terms of  the entropy production of the instantonic trajectory, defined as

\begin{align}\label{sigma_inst}
&\sigma_{\gamma \to \nu} := \\
& \hspace{0.8cm} \frac{1}{p^{(\gamma)}_\text{ss}(\mathcal{x}_\nu)}\lim_{t \to \infty}\lim_{V \to \infty} \mean {\sigma \delta(c(0)-\mathcal{x}^*_\gamma)\delta(c(t)-\mathcal{x}_\nu)}, \nonumber 
%-I^{(\gamma)}_\text{ss}(\mathcal{x}_\nu) 
\\
& \hspace{0.8cm} = \int_{c(0)=\mathcal{x}_\gamma^*}^{c(\infty)=\mathcal{x}_\nu} dt \,\textstyle{\sum_\rho} r_\rho(c(t)) \sigma_\rho(c(t)) e^{\Delta_\rho \cdot \pi(t)}, \nonumber
\end{align}
where the second equality is \eqref{currents_inst_rel}, $c(t)$ and $\pi(t)=\partial_c I(c(t))$ being solution of \eqref{hamilton_eq} with boundary conditions $c(0)= \mathcal{x}^*_\gamma$ and $c(\infty)=\mathcal{x}_\nu$.
Here, we divided by $p^{(\gamma)}_\text{ss}(\mathcal{x}_\nu)$  to get rid of the exponentially small probability of the instanton, see \eqref{trans_rate}, since we are only interested in the value of the entropy production along the trajectory -- not on its statistical weight.
Note that the macroscopic limit removes the Shannon entropy of the boundary states, so that $\sigma_{\gamma \to \nu}$ counts only the entropy flow of the instanton. 
%Note that we have removed the mean self-information of the final state (the one of the initial state, $I^{(\gamma)}_\text{ss}(\mathcal{x}^*_\gamma)$ is zero), so that $\sigma_{\gamma \to \nu}$ counts only the entropy flow of the transition trajectory. 

Next, we show that the conditional mean of the adiabatic entropy production is still positive. The proof goes as follows. First, we note that 
\begin{align}\label{ft_ad_epr}
\sigma_\text{ad}[\mathcal{X}]= - \frac{1}{V}\ln \frac{P^\dagger[\mathcal{X}]}{P[\mathcal{X}]} ,
\end{align}
where $P^\dagger[\mathcal{X}]$ denotes the path probability under the dual dynamics with transition rates $r^\dagger_\rho(c)$ given in \eqref{scaledDualRate}. 
%Such adjoint dynamics is known to preserve the stationary distribution, i.e. $I_\text{ss}(c)$ is the stationary rate function associated to $P^\dagger$. 
We therefore write the transition probability $P^\dagger (\mathcal{x}_\nu,t |\mathcal{x}^*_\gamma,0)$ from $\mathcal{x}^*_\gamma$ to $\mathcal{x}_\nu$ under the dual dynamics as the conditional average over paths,
\begin{align}\label{Jensen}
P^\dagger(\mathcal{x}_\nu,t |\mathcal{x}^*_\gamma,0) = \sum_{\mathcal{X}} P^\dagger[\mathcal{X}] \delta(c(0)-\mathcal{x}^*_\gamma)\delta(c(t)-\mathcal{x}_\nu) \nonumber \\
= \sum_{\mathcal{X}} P[\mathcal{X}]  e^{-V\sigma_\text{ad}[\mathcal{X}]} \delta(c(0)-\mathcal{x}^*_\gamma)\delta(c(t)-\mathcal{x}_\nu) 
 \nonumber \\
\geq P(\mathcal{x}_\nu,t |\mathcal{x}^*_\gamma,0) e^{-\frac{V}{P(\mathcal{x}_\nu,t |\mathcal{x}^*_\gamma,0)}  \mean {\sigma_\text{ad}\delta(c(0)-\mathcal{x}^*_\gamma)\delta(c(t)-\mathcal{x}_\nu)}}\;.
\end{align}
In the last line we used Jensen's inequality applied to the normalized conditional average, $\mean{\dots \delta(c(0)-\mathcal{x}^*_\gamma)\delta(c(t)-\mathcal{x}_\nu)}/  P(\mathcal{x}_\nu,t |\mathcal{x}^*_\gamma,0)$,
in the same spirit as \cite{falasco2020dissipationtime}.
Second, we use the fact that for large $V$ and large $t$ the transition probability approaches the (local) stationary probability distribution at the saddle point $\mathcal{x}_\nu$:
\begin{align}
\lim_{t \to \infty} P(\mathcal{x}_\nu,t |\mathcal{x}^*_\gamma,0)   \asymp  e^{- V[I^{(\gamma)}_\text{ss}(\mathcal{x}_\nu)-I^{(\gamma)}_\text{ss}(\mathcal{x}^*_\gamma)] } =p^{(\gamma)}_\text{ss}(\mathcal{x}_\nu).
\end{align}
But since the dual dynamics preserves the steady state $p^{(\gamma)}_\text{ss}$, we have that  $P^\dagger (\mathcal{x}_\nu,t |\mathcal{x}^*_\gamma,0)= P(\mathcal{x}_\nu,t |\mathcal{x}^*_\gamma,0)$ for large $t$. Hence, \eqref{Jensen} gives

\begin{align}
%1 \geq e^{-\frac{V}{p^{(\gamma)}_\text{ss}(\mathcal{x}_\nu) }  \sigma_\text{ad}_{\gamma \to \nu}},
1 \geq e^{-V \sigma^\text{ad}_{\gamma \to \nu}},
\end{align}
which implies that the adiabatic entropy production of an instantonic trajectory is nonnegative:
\begin{align}\label{ad_epr_inst}
 & \sigma^\text{ad}_{\gamma \to \nu} :=  \\ 
 &\hspace{0.2cm} \frac{1}{p^{(\gamma)}_\text{ss}(\mathcal{x}_\nu)} \lim_{t \to \infty}\lim_{V \to \infty} \mean {\sigma^\text{ad}\delta(c(0)-\mathcal{x}^*_\gamma)\delta(c(t)-\mathcal{x}_\nu)} \geq 0 . \nonumber
\end{align}
As in \eqref{sigma_inst}, we have factored out the exponentially small probability $p^{(\gamma)}_\text{ss} \geq 0$ of the instanton.

At this point we go back to the conditional average of the entropy production decomposition \eqref{ad_nonad_epr_macro}, and use \eqref{ad_epr_inst} and \eqref{sigma_inst} to write
\begin{align} \label{nonad_bound_inst}
\sigma_{\gamma \to \nu} &\geq - \int_{\mathcal{x}(0)=\mathcal{x}^*_\gamma}^{\mathcal{x}(\infty)=\mathcal{x}_\nu} d_{\tau} I^{(\gamma)}_\text{ss}(\mathcal{x}(\tau)) d\tau \nonumber \\
 &= -I^{(\gamma)}_\text{ss}(\mathcal{x}_\nu) + I^{(\gamma)}_\text{ss}(\mathcal{x}^*_\gamma).
\end{align}
Multiplying by $V$ and exponentiating, we finally obtain the upper bound \eqref{upper_lower_bounds} on the transition rate $\kappa_\nu$.

We conclude by pointing out that integrating the entropy production between fixed points gives a diverging quantity -- unless detailed balance holds -- because relaxation (and instantonic) trajectories have infinite duration. This is traced back to the Laplace approximation \eqref{hamilton_eq}, which is valid for states $c$ at which the noise is small in comparison to the drift field $F(c)$. This assumption breaks down on fixed points, in which the drift field is null.
Clearly, one should consider boundary conditions that belong to the boundary of a ball centered on the fixed point with a radius of the order of the  Gaussian fluctuations $O(V^{-1/2})$. Inside any such ball the weak-noise limit breaks down\footnote{A rigorous treatment that includes subexponential corrections would require the use of a boundary layer approximation \cite{schuss2009theory}}. Changing the integration boundaries in this way introduces in the left hand side of \eqref{nonad_bound_inst} a correction of order $O(1/V)$ (since  $\partial_c I^{\gamma}_\text{ss}$ is zero on every fixed point), hence a subexponential correction to $\kappa_\nu$, but makes $\sigma_{\gamma \to \nu}$ and $\sigma_{\nu \to \gamma}$ finite {by discarding the entropy produced at the fixed point. Using \eqref{info_epr}, for which all the above derivations can be repeated, rather than the thermodynamic entropy production is an alternative approach to remove the divergence (without tuning the integration boundaries) and thus tighten the bounds.}

\section{Derivation of the emerging transition rates }\label{app:transition_rate}

We give a formal derivation of the relation \eqref{trans_saddle} along the lines of the proof developed in \cite{bouchet2016generalisation} for diffusive dynamics. 
We consider an autonomous system initially localized in a basin of attraction $\gamma$, and place absorbing boundary conditions in the neighborhood of the saddle points $\mathcal{x}_\nu$. For times much smaller than the escape time from the basin (which we wish to determine) the system probability density $p_\text{abs}(c,t)$ coincides with $p_\text{ss}^{(\gamma)}(c) \asymp e^{-V I_\text{ss}^{(\gamma)}(c)}$ for all $c$ in the basin but on a tiny boundary layer, of linear size $O(V^{-1/2})$, around the saddles $\mathcal{x}_\nu$. To quantify how probability mass gets removed we thus introduce the survival probability, the monotonic decreasing function \cite{red01}
\begin{align}\label{survival_prob}
\mathcal{P}_\text{surv}(t) = \int dc \, p_\text{abs}(c,t) ,
\end{align}
that we wish to express in terms of the total escape rate $\kappa$ from the attractor. 
In general the survival probability depends on all eigenvalues of the master equation operator with absorbing boundary conditions \cite{freitas2022reliability}. But in presence of metastability (see Sec. \ref{sec:deterministic_dyn}) only the largest contributes, $-\kappa$, unless $t \to 0$.
%we take $\kappa$ to be constant since 
This means that escape events are exponentially distributed in the limit $V \to \infty$, i.e., $\mathcal{P}_\text{surv}(t)= e^{-\kappa t}$ \cite{day83}. Moreover, we assume that the saddles are well separated and so contribute independently  to the escape rate  as $\kappa=\sum_\nu \kappa_\nu $. The evolution of the survival probability can be deduced from the master equation 
\begin{align}\label{me_abs}
\partial_t p_\text{abs}(c,t) =- \partial_c \cdot \mathfrak{j}_\text{abs}(c,t)
\end{align}
with absorbing conditions $p_\text{abs}(\mathcal{x}_\nu,t)=0$. Integrating \eqref{me_abs} over the basin of attraction, using \eqref{survival_prob} and the divergence theorem, we obtain
\begin{align}\label{me_p_surv}
\mathcal{P}_\text{surv}(t) \sum_\nu k_\nu =  \sum_\nu \mathfrak{j}_\text{abs}(\mathcal{x}_\nu,t) \cdot \hat{e}(\mathcal{x}_\nu) \Delta \mathcal{S}
\end{align}
where $\Delta \mathcal{S}$ is a small surface element of the boundary of the basin and $\hat{e}(\mathcal{x}_\nu)$ is the local outward unit vector. 
Using the definition of probability current $\mathfrak{j}_\text{abs}(c,t)=\mean{\dot c(t) \delta(c(t)-c)}$, we can write \eqref{me_p_surv} for $V \to \infty$  
\begin{align}
\mathfrak{j}_\text{abs}(\mathcal{x}_\nu,t) = \dot c(t) p_\text{abs} (\mathcal{x}_\nu,t),
\end{align}
with $\dot c(t)$ the velocity of the typical escape path in a neighborhood of the saddle.
Equation \eqref{me_p_surv} is valid in particular at times much shorter than $1/\kappa$ when $\mathcal{P}_\mathrm{surv}(t \ll 1/\kappa) \simeq 1$ and $p_\text{abs}(c,t \ll 1/\kappa) \asymp p^{(\gamma)}_\text{ss}(c)$ is indistinguishable from the local stationary probability \eqref{trans_rate} at leading order \cite{bouchet2016generalisation}.
Therefore, discarding subexponential contributions, \eqref{me_p_surv} reads
\begin{align}
 \sum_\nu k_\nu  \asymp \sum_\nu e^{-V [I^{(\gamma)}_\text{ss}(\mathcal{x}_\nu)-I^{(\gamma)}_\text{ss}(\mathcal{x}^*_\gamma)]} ,
\end{align}
which gives \eqref{trans_saddle} under the hypothesis of independent escape paths through each separate saddle point.
\end{appendix}

\end{document}